\documentclass{aa}
\usepackage{amsmath}
\usepackage{cases}
\usepackage{txfonts}
\usepackage{graphicx}
\usepackage{natbib}
\usepackage{epstopdf}
\usepackage{nicefrac}
\usepackage{hyperref}
\usepackage{threeparttable}
\usepackage[dvipsnames]{xcolor}

\begin{document}

\title{J-PLUS: The Star Formation Main Sequence and Rate Density at $d \lesssim 75$ Mpc}
\subtitle{}
\author{G. Vilella-Rojo\inst{1}
\and R.~Logro\~no-Garc\'{\i}a\inst{1} 
\and C.~L\'opez-Sanjuan\inst{2} 
\and K.~Viironen\inst{1}
\and J.~Varela\inst{2}
\and M.~Moles\inst{1}
\and A.~J.~Cenarro\inst{2}
\and D.~Crist\'obal-Hornillos\inst{1}
\and A.~Ederoclite\inst{3}
\and C.~Hern\'andez-Monteagudo\inst{2}
\and A.~Mar\'{\i}n-Franch\inst{2}
\and H.~V\'azquez Rami\'o\inst{2}
\and L.~Galbany\inst{4}
\and R.~M.~Gonz\'alez Delgado\inst{5}
\and A.~Hern\'an-Caballero\inst{1}
\and A.~Lumbreras-Calle\inst{1}
\and P.~S\'anchez-Bl\'azquez\inst{6,7}
\and D.~Sobral\inst{8,9}
\and J.~M.~V\'{\i}lchez\inst{5}
\and J.~Alcaniz\inst{10}
\and R.~E.~Angulo\inst{11,12}
\and R.~A.~Dupke\inst{10,13,14}
\and L.~Sodr\'e Jr.\inst{3}
}

\institute{Centro de Estudios de F\'{\i}sica del Cosmos de Arag\'on, Plaza San Juan 1, 44001 Teruel, Spain\\ \email{gvilellarojo@gmail.com} 
\and
Centro de Estudios de F\'{\i}sica del Cosmos de Arag\'on, Unidad Asociada al CSIC, Plaza San Juan 1, 44001 Teruel, Spain
\and
Instituto de Astronomia, Geof\'{\i}sica e Ci\^encias Atmosf\'ericas, Universidade de S\~ao Paulo, 05508-090 S\~ao Paulo, Brazil
\and
Departamento de F\'{\i}sica Te\'orica y del Cosmos, Universidad de Granada, 18071 Granada, Spain
\and
IAA-CSIC, Glorieta de la Astronom\'{\i}a s/n, 18008 Granada, Spain
\and
Departamento de F\'{\i}sica de la Tierra y Astrof\'{\i}sica, Universidad Complutense de Madrid, 28040, Madrid, Spain
\and
IPARCOS, Facultad de CC F\'{\i}sicas, Universidad Complutense de Madrid, 28040, Madrid, Spain
\and
Department of Physics, Lancaster University, Lancaster, LA1 4YB, UK
\and
Leiden Observatory, Leiden University, P.O. Box 9513, NL-2300 RA Leiden, The Netherlands
\and
Observat\'orio Nacional, Rua General Jos\'e Cristino, 77 - Bairro Imperial de S\~ao Crist\'ov\~ao, 20921-400 Rio de Janeiro, Brazil
\and
Donostia International Physics Centre (DIPC), Paseo Manuel de Lardizabal 4, 20018 Donostia-San Sebastián, Spain
\and
IKERBASQUE, Basque Foundation for Science, 48013, Bilbao, Spain
\and 
University of Michigan, Department of Astronomy, 1085 South University Ave., Ann Arbor, MI 48109, USA
\and
University of Alabama, Department of Physics and Astronomy, Gallalee Hall, Tuscaloosa, AL 35401, USA
}

\date{Received 11 August 2020 / Accepted 7 January 2021}

\abstract{}{Our goal is to estimate the star formation main sequence (SFMS) and the star formation rate density (SFRD) at $\mathrm{z} \leq 0.017$ ($d \lesssim75$ Mpc) using the Javalambre Photometric Local Universe Survey (J-PLUS) first data release, that probes $897.4$ deg$^2$ with twelve optical bands.}
{We extract the H$\alpha$ emission flux of 805 local galaxies from the J-PLUS filter $J0660$, being the continuum level estimated with the other eleven J-PLUS bands, and the dust attenuation and nitrogen contamination corrected with empirical relations. Stellar masses ($M_{\star}$), H$\alpha$ luminosities ($L_{\mathrm{H}\alpha}$), and star formation rates (SFRs) were estimated by accounting for parameters covariances. Our sample comprises 689 blue galaxies and 67 red galaxies, classified in the $(u-g)$ {\it vs} $(g-z)$ color-color diagram, plus 49 AGN.}
{The SFMS is explored at $\log M_{\star} \gtrsim 8$ and it is clearly defined by the blue galaxies, with the red galaxies located below them. The SFMS is described as $\log \mathrm{SFR} = 0.83 \log M_{\star} - 8.44$. We find a good agreement with previous estimations of the SFMS, especially those based on integral field spectroscopy. The H$\alpha$ luminosity function of the AGN-free sample is well described by a Schechter function with $\log L_{\mathrm{H}\alpha}^{*} = 41.34$, $\log \phi^{*} = -2.43$, and $\alpha = -1.25$. Our measurements provide a lower characteristic luminosity than several previous studies in the literature.}
{The derived star formation rate density at $d \lesssim 75$ Mpc is $\log \rho_{\rm SFR} = -2.10 \pm 0.11$, with red galaxies accounting for 15\% of the SFRD. Our value is lower than previous estimations at similar redshift, and provides a local reference for evolutionary studies regarding the star formation history of the Universe.}
 

\titlerunning{J-PLUS. The Star Formation Main Sequence and Rate Density at $d \lesssim 75$ Mpc}
 
\maketitle

\section{Introduction}\label{sect:intro}
The star formation rate (SFR) of a galaxy accounts for the mass of gas that is transformed into stars per unit time. It is common to use units of solar masses $M_{\odot}$ per year to describe it. Through the observation and understanding of stellar population properties, one can unveil the history of a galaxy, and get better insight into its current state. To better comprehend the formation and evolution of galaxies, we aim to relate the SFR to other galaxy properties, such as its total stellar mass $M_{\star}$, morphology, environment, gas content, or nuclear activity. By doing so, we look for those parameters that might play a major role in the rate at which galaxies form stars. 

Since the pioneering study of \citet{Gallego1995}, where the star formation rate density (SFRD) in the local Universe is measured, a large amount of studies have traced the SFRD evolution with redshift up to $\mathrm{z} \sim 8$ \citep[e.g.][and references therein]{Lilly1996, Hopkins2006, cucciati12, Sobral2013, Madau2014, novak17, Maniyar2018, driver2018}. The current consensus is that the SFRD increases with look-back time to $\mathrm{z} \sim 2-3$, then decreases towards cosmic dawn.

More recently, and in addition to the SFRD, the fundamental relation between the SFR and the stellar mass, often called star formation main sequence (SFMS, \citealt{Brinchmann04, Noeske2007}) was introduced, and is now well established and supported by observations, showing correlation with other properties such as galaxy morphology \citep[see][and references therein]{RosaGD2016}. The linear relation in logarithmic scale between these two basic galaxy properties has been analyzed at $\mathrm{z} = 0-6$, suggesting a slope $a$ in the range $0.5 \lesssim a \lesssim 1$, and an evolving normalization $b$ that mirrors the SFRD behavior \citep[e.g.][and references therein]{Whitaker2012, Speagle14, Salmon2015, Lee2015, Tomczak2016, Santini2017, Popesso2019b, Leslie2020}.

In this framework, it is clear that a robust anchorage of these magnitudes and relations at $\mathrm{z} \sim 0$ is needed to set the current star formation properties of the Universe, and to provide a local reference for evolutionary studies. 

To analyze the SFR at a given cosmological epoch, a proxy for the star formation activity and a representative sample of galaxies are both needed. In the first case, we look for consequences of star formation processes, and we refer to these as SFR indicators. These can be classified in two main families. The first one is based on the direct consequences of the SFR; the main indicator of this family is the ultraviolet (UV) light emitted by young, short-lived, O and B stars. These stars have short life spans ($30$ Myrs, \citealt{Calzetti2013}), and are the most straightforward tracer of the SFR. However, this indicator suffers from two major drawbacks. The first one is dust attenuation in the star-forming region. Photons trying to escape it will interact with dust grains, causing a loss of the UV radiation that we receive. The other one is the Earth's atmosphere, which shields us from the UV radiation.

The second big family of SFR tracers are indirect. In this family, we include the dust-processed light, which is emitted by dust grains heated by the UV radiation, and the recombination lines that appear on top of the stellar continuum spectral energy distribution (SED) of the star-forming regions. Among these recombination lines, we have the H$\alpha$ line. This emission occurs when an electron in a hydrogen atom is ionized by the UV field of the young, massive stars, and then returns back to the fundamental state. During this process, the electron may transit from the $3$rd to the $2$nd energy level, emitting as a consequence a photon with rest-frame wavelength $6562.8\,\AA$. We refer to this transition as the H$\alpha$ line, as it is the least energetic of all the transitions that end in the $2$nd energy level (i.e., the Balmer series).

To build a representative sample of galaxies and derive the global SFR in a cosmological epoch, there are also different, complementary approaches. On the one hand, spectroscopic observations are those that provide the most accurate measurements of emission or absorption line fluxes. This is at the cost of large samples, or aperture corrections. While integral field spectroscopic (IFS) surveys are an excellent choice to overcome the spatial coverage limitation, galaxies require a target pre-selection. Single-fiber spectroscopy, despite being suitable to gather information of many galaxies in the same observing time, requires aperture corrections that dilute spatial information. 

On the other hand, photometric studies that count with well adapted narrow-band filters are able to retrieve the largest samples, with no need for aperture correction. This is at the cost of much lower spectral precision, that is translated into a loss in precision when measuring emission-line fluxes. However, in the investigation of properties of galaxies in the nearby Universe, a statistically meaningful sample is crucial. Cosmic variance may introduce biases that cannot be compensated for target pre-selection. To understand the properties of galaxies around ours, a blind, homogeneous study is the best tool. In this regard, the Javalambre Photometric Local Universe Survey (J-PLUS\footnote{\url{http://j-plus.es/}}, \citealt{Cenarro2019}) becomes ideal. By blindly surveying the Northern sky with twelve optical filters (seven narrow and five broad band), it gathers information about stars, galaxies, and objects in the Solar System. In particular, the J-PLUS filter $J0660$, of $14$ nm width, is centered at rest-frame H$\alpha$ emission and it is able to trace the SFR up to a distance of $d \sim 75$ Mpc ($\mathrm{z} \leq 0.017$).

In this Paper, we analyze the J-PLUS first data release (DR1) to derive the SFMS and SFRD at $\mathrm{z}\sim0$. Additionally, we compute the stellar mass function and H$\alpha$ luminosity function in our sample. This work, in combination with the forthcoming paper of \citet[][in prep.]{Logrono2020}, are the culmination of two previous studies: the first is \citet[][VR15 hereafter]{mio}, in which we present the best method to extract the H$\alpha$ flux using J-PLUS synthetic data, including corrections for dust attenuation and [\ion{N}{ii}] contamination. In the second paper, \cite{Logrono2019}, the methodology presented in VR15 is tested using real J-PLUS data and common star-forming regions observed spectroscopically by the Sloan Digital Sky Survey (SDSS, \citealt{sdss}) and the Calar Alto Legacy Integral Field Area (CALIFA, \citealt{Sanchez2012}) survey. The comparison between the spectroscopic H$\alpha$ flux and the one derived from J-PLUS data reveals that the photometric measurement is not biased and presents a minimum uncertainty of $20$\%.

This paper is organized as follows: in Sect.~\ref{sect:SampleSelection}, we explain how we select our sample of nearby galaxies ($\mathrm{z}\leq0.017$, or $d \lesssim 75$ Mpc under the assumed cosmology) and the measurement of their distances, stellar masses, H$\alpha$ luminosities, and SFRs. We analyze this data to obtain the SFMS and SFRD in the local universe in Sect.~\ref{sect:SFjplus}. In Sect.~\ref{sect:Discussion}, we present a discussion of our findings in the context of the current literature, and a summary of the main results is provided in Sect.~\ref{sect:Conclusions}. 

This paper makes use of the AB system \citep{Oke83} of magnitudes, a \cite{Salpeter55} initial mass function (IMF), and a flat Universe cosmology with $\Omega_{\rm m} = 0.3$, $\Omega_{\Lambda}=0.7$, and $h = 0.7$.

\begin{figure*}
   \centering
    \includegraphics[width=0.33\textwidth]{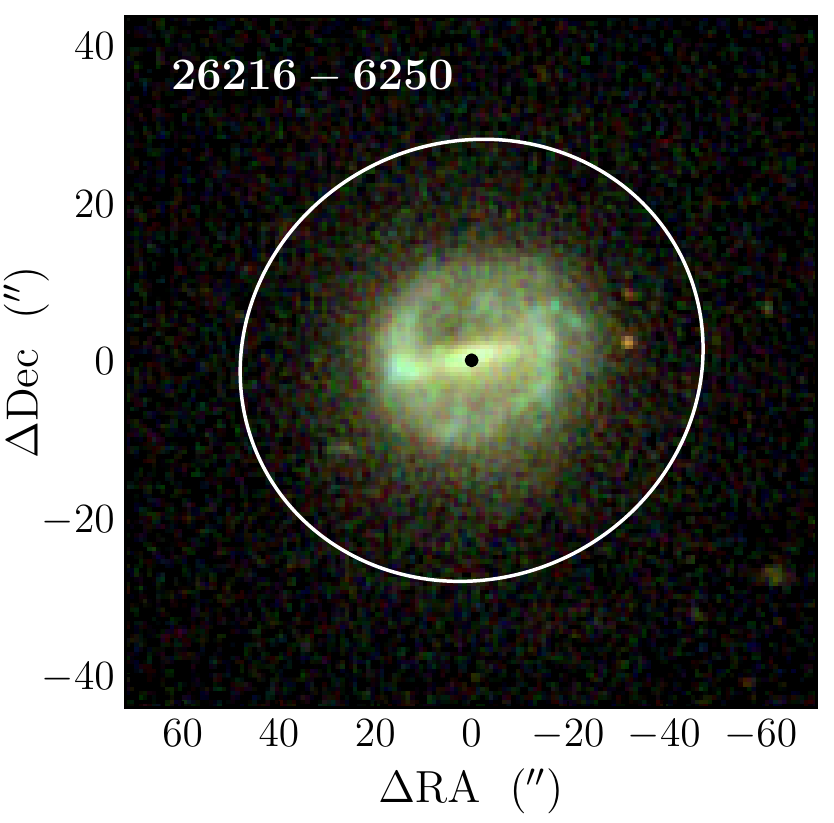}
    \includegraphics[width=0.33\textwidth]{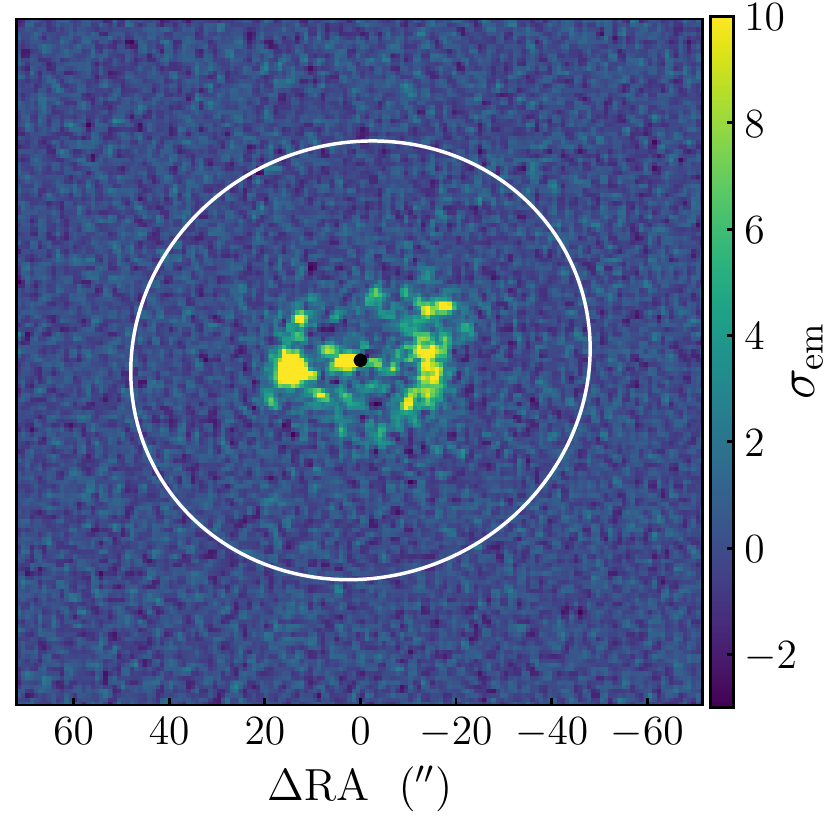}
    \includegraphics[width=0.33\textwidth]{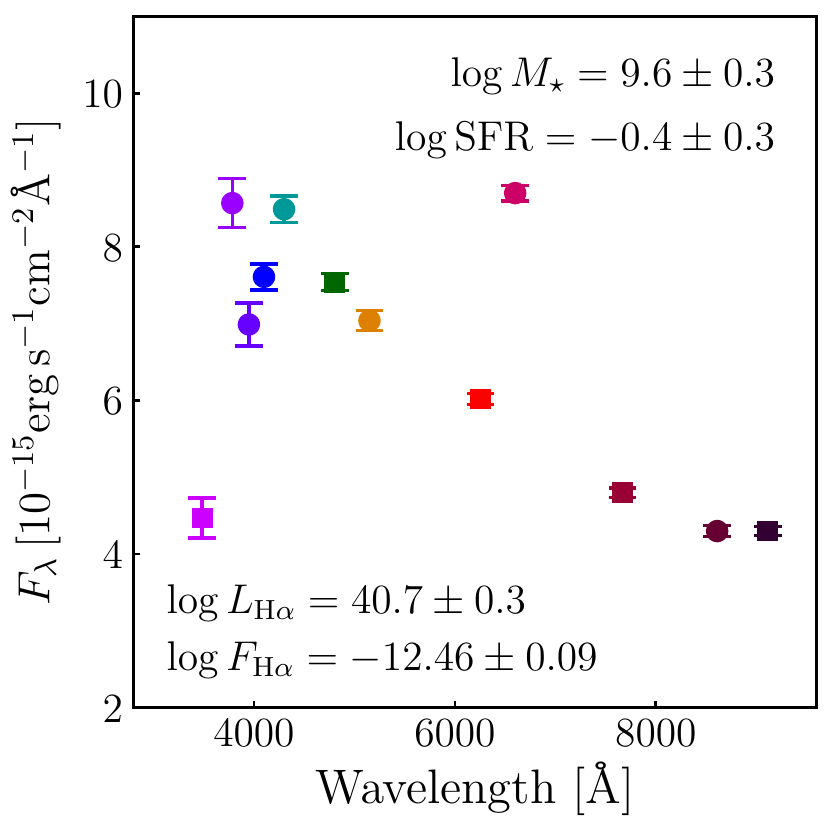}\\
    \includegraphics[width=0.33\textwidth]{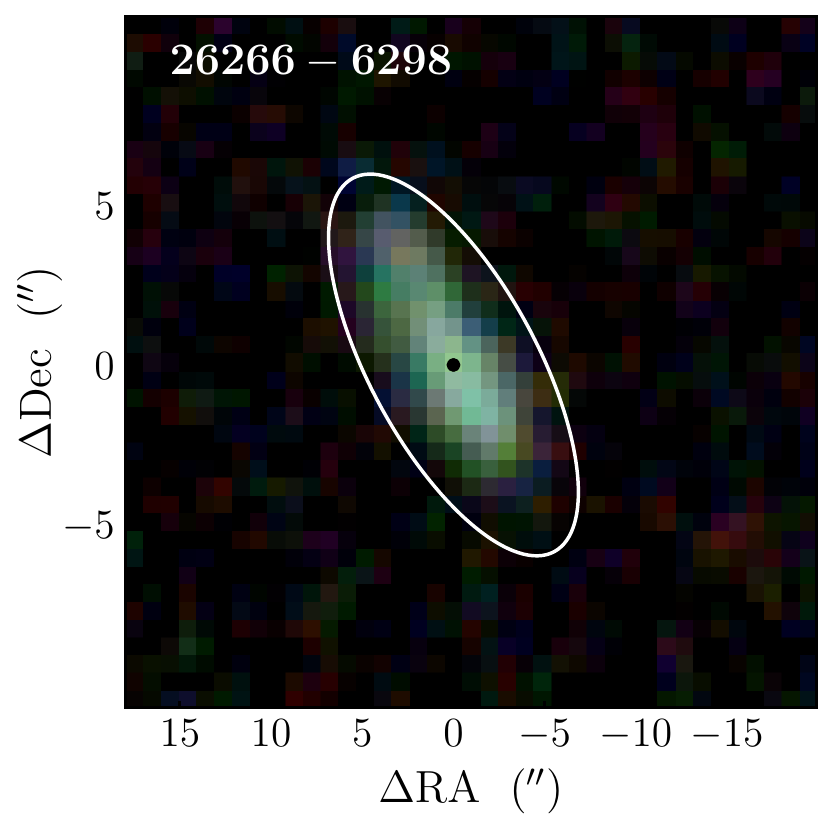}
    \includegraphics[width=0.33\textwidth]{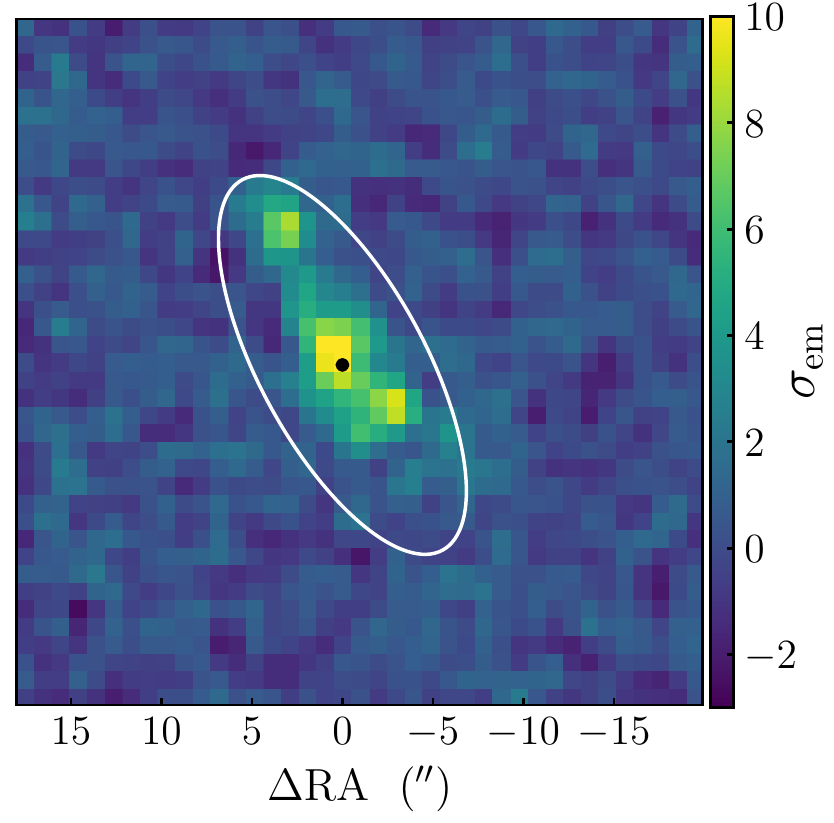}
    \includegraphics[width=0.33\textwidth]{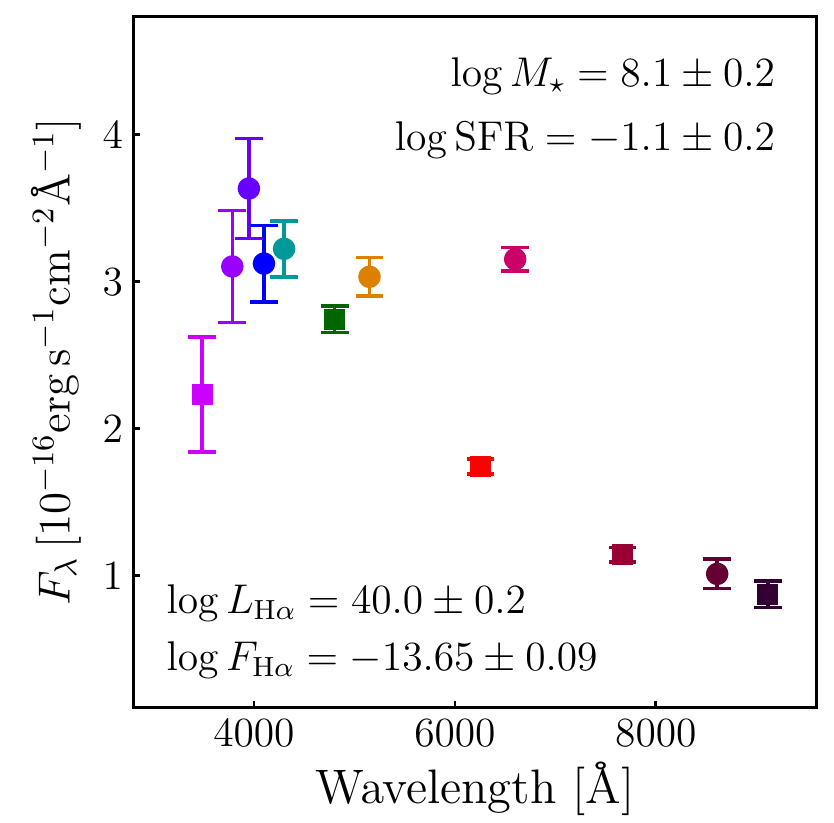}
    \caption{Illustrative examples of the J-PLUS DR1 sample of local galaxies, presenting sources 26216-6250 (${\rm RA} = 227.1397$, ${\rm Dec}= 52.2960$; UGC 9741 at $z_{\rm spec} = 0.0083$) and 26266-6298 (${\rm RA} = 231.1995$, ${\rm Dec}= 55.1277$; there is no spectroscopic redshift for this source in the analyzed databases, see Sect.~\ref{sect:excess_selection}). The sky location of the sources is marked with a bullet in the {\it left} and {\it central panels}, and the white ellipse marks the three effective radii contour for the sources. {\it Left panels}. Color composite of the galaxies, obtained from the $gri$ J-PLUS images. {\it Central panels}: $J0660$ emission in units of $\sigma_{\rm em}$, the dispersion of the pixels in the image (Sect.~\ref{sect:FluxExtraction}), as scaled in the color bar. This emission image has been obtained by applying the 3F methodology in VR15 to the $r$, $J0660$, and $i$ J-PLUS images. Several star-forming knots are apparent. {\it Right panels}. J-PLUS twelve-band photometry (the SED) of those pixels with $J0660$ emission larger than $\sigma_{\rm em}$ (Sect.~\ref{sect:FluxExtraction}) and inside three effective radii. Squares show the five SDSS-like filters ($ugriz$), and circles the seven medium- and narrow-band filters ($J0378$, $J0395$, $J0410$, $J0430$, $J0515$, $J0660$, and $J0861$). The flux in $J0660$, $J0515$, and $J0378$ is raised with respect to the continuum level due to the H$\alpha$ + [\ion{N}{II}], [\ion{O}{III}], and [\ion{O}{II}] emission, respectively. The derived physical properties of the galaxies (Sects.~\ref{sect:Halphaflux} and \ref{sect:LMSFR}) are labeled in the panels. 
   }
    \label{fig:galaxy_G0}
\end{figure*}

\section{Data and sample of local emission-line galaxies}\label{sect:SampleSelection}
In this Section, we describe how we retrieved H$\alpha$ emitters within our searching interval in redshift ($\mathrm{z} \leq 0.017$). First, we summarize the main characteristics of J-PLUS and its first data release in Sect.~\ref{sect:jplus}. Then, we explain the criteria for a source to be considered a local galaxy at $\mathrm{z} \leq 0.017$ (Sect.~\ref{sect:ha_emitters}). We assign distances, H$\alpha$ luminosities, stellar masses, and star formation rates to J-PLUS local emitters in Sect.~\ref{sect:MonteCarlo}. Finally, we show how galaxies distribute within this parameter space in Sect.~\ref{sect:SampleProperties}.

\subsection{J-PLUS photometric data}\label{sect:jplus}
J-PLUS is an imaging survey project intending to cover a fraction of the Northern sky from the Observatorio Astrof\'{\i}sico de Javalambre (OAJ\footnote{\url{https://oajweb.cefca.es/}}, \citealt{cenarro14}) using an 83-cm diameter telescope (JAST/T80) equipped with a 9.2k x 9.2k pixel camera (T80Cam, \citealt{t80cam}) providing a 2 deg$^2$ field of view. The system is equipped with a set of twelve, purpose designed, photometric filters, the five SDSS ($ugriz$) filters and seven medium or narrow-band filters primary designed to classify stars. They are placed on key stellar features covering around the $4\,000\ \AA$ break region ($J0378$, $J0395$, $J0410$, $J0430$), the magnesium doublet ($J0515$) and the calcium triplet ($J0861$), plus the filter on the H$\alpha$ line ($J0660$). Thus, J-PLUS covers the whole optical wavelength range, enabling different kind of studies in stellar astrophysics \citep{Bonatto2019, Whitten2019, Solano19}, galaxies at several redshift ranges \citep{Logrono2019, SanRoman2019, Nogueira-Cavalcante19} or in clusters \citep{Molino2019, JimenezTeja2019}, and extreme Lyman-$\alpha$ emitters at ${\rm z} > 2$ \citep{Spinoso20}.

The first J-PLUS data set was released in July 2018. The J-PLUS DR1\footnote{\url{https://www.j-plus.es/datareleases/data\_release\_dr1}} includes photometric information for $511$ pointings that cover an area of $897.4$ deg$^2$ after correcting for overlapping areas and masking optical artifacts. The catalogs, publicly accessible at the J-PLUS website, contains $\sim 13.4$ million objects detected in $r$ band using \texttt{SExtractor} \citep{Bertin1996} with $m_r \leq 21$ mag. The photometry in the twelve J-PLUS bands was performed with the \texttt{SExtractor} dual mode feature. Details on the reduction and calibration processes can be found in \cite{Cenarro2019} and \cite{Carlinhos_Calibracion}. 

\begin{figure*}
    \centering
    \includegraphics[]{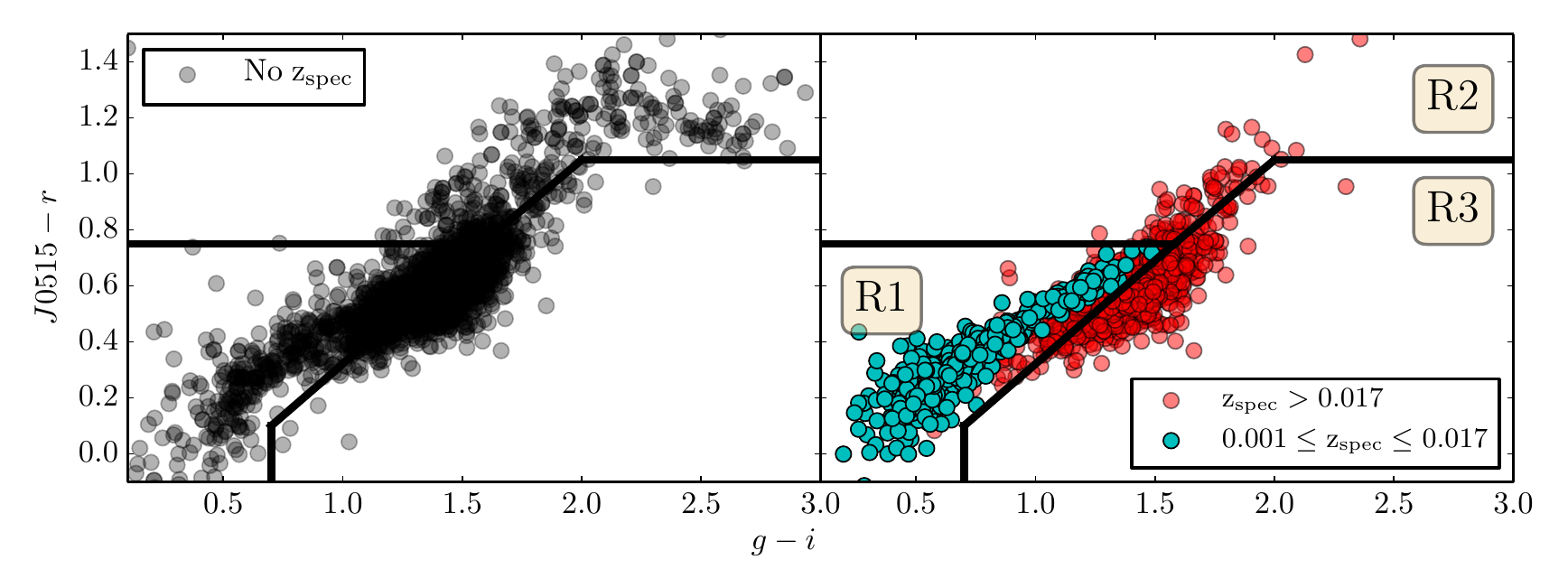}
    \caption{$(J0515-r)$ {\it vs} $(g-i)$ color-color diagram from \texttt{ISO\_GAUSS} magnitudes of $m_r \leq 18$ sources with significant $J0660$ excess. {\it Left panel}: Sources with no spectroscopic redshift in SDSS. {\it Right panel}: $J0660$ emitters, with spectroscopic redshift in SDSS. Cyan dots represent sources within our redshift of interest, while red dots are sources with higher redshift. The three regions defined to isolate low-z galaxies (R1), blended stars (R2), and high-z galaxies (R3) are labeled in the panel.}
    \label{fig:color-color-diagrams-emisores}
\end{figure*}

\subsection{Definition of the local sample}\label{sect:ha_emitters}

\subsubsection{Initial selection of spectroscopic sources}\label{sect:initial_selection}
We started from the general J-PLUS catalogs and retrieved all the detected sources with an apparent $\texttt{AUTO}$ magnitude $m_r \leq 18$ and high-quality flags in the surveyed area. This excluded sources that are at the edge of the images, near to a bright star, or affected by optical artifacts. To get rid of stars, we applied the stellarity parameter\footnote{Accessible at the ADQL table \texttt{StarGalClass} of the J-PLUS data base} derived by \cite{Carlinhos2019} and imposed \texttt{total\_prob\_star} $< 0.5$. This is a Bayesian classification that takes into account the morphological features of a source, and \emph{a priori} information from the {\it Gaia} DR2 parallaxes \citep{gaia_dr2}.

This initial selection has several benefits. It provides a completeness higher than $95$\% for stellar masses larger than $10^{9}$ M$_{\odot}$, and a surface brightness limit of $\mu_r \sim 24$ mag arcsec$^{-2}$ for nearby, well resolved galaxies (Sects.~\ref{sect:GlobalDistributions} and \ref{sect:SMF}). It also introduces a cut in the number of faint emitters at higher redshift that would contaminate our sample ([\ion{O}{iii}] + H$\beta$ at $\mathrm{z} \sim 0.33$, or [\ion{O}{ii}] at $\mathrm{z} \sim 0.77$; see \citealt{IzquierdoVillalba2019}); and finally, it covers a well defined, non pre-selected area to compute volume densities. 

The initial selection yielded $109\,815$ sources. We cross-correlated these sources with other databases, named NED\footnote{\url{https://ned.ipac.caltech.edu/}}, Simbad\footnote{\url{http://simbad.u-strasbg.fr/simbad/}} \citep{simbad}, and SDSS, to assign a spectroscopic redshift ($\mathrm{z}_{\rm spec}$) to them. After cleaning from duplicates and merging regions of unique, large galaxies wrongly classified by \texttt{SExtractor} as individual sources, we ended with $684$ galaxies at $0.001 < \mathrm{z}_{\rm spec} \leq 0.017$. A representative example is shown in the {\it upper panels} of Fig.~\ref{fig:galaxy_G0}.

At this stage, the completeness of the $z_{\rm spec}$ sample is unknown and a large fraction of local galaxies with $m_r \leq 18$ but without spectroscopic information could be hidden in J-PLUS data. To deal with this issue, we searched for high-confidence H$\alpha$ emitters without $z_{\rm spec}$.

\subsubsection{Selection of local galaxies with excess in $J0660$}\label{sect:excess_selection}
The H$\alpha$ emission of local galaxies is traced by the J-PLUS $J0660$ filter up to $\mathrm{z} = 0.017$. In combination with the emission of [\ion{N}{ii}], it causes the flux inside $J0660$ to be above the expected continuum. It is precisely this feature that we were looking for. 

For each source passing the initial magnitude ($m_r \leq 18$) and quality selection (Sect.~\ref{sect:initial_selection}), we queried the \texttt{ISO\_GAUSS} magnitudes in the $r$, $J0660$, and $i$ filters, along with their respective photometric errors. The \texttt{ISO\_GAUSS} magnitudes in the J-PLUS database were estimated as the usual \texttt{ISO} magnitudes in \texttt{SExtractor} over images convolved with a Gaussian kernel of $1.5$ arcsec. This reduces the differences due to the point spread function (PSF) variations among passbands with a well defined kernel.

To these fluxes, we applied the equations of the three filters (3F) method, as described in VR15. This algorithm is similar to the classical color - narrowband magnitude diagram that has been widely used in the literature \citep[e.g.][]{Bunker1995}, but instead of relying in a $0$-color for non-emitters, it assumes a featureless linear continuum for the line in the wavelength range of the emission. In our case, we relied on $r$ and $i$ to trace the continuum inside $J0660$. This was done using a Monte-Carlo approach, in which the flux inside the three aforementioned bands were perturbed within their respective error $j$ times, and in each iteration we retrieved an inference of the excess inside $J0660$, which we refer to as $F_{J0660,j}$. We did this $3\,000$ times. In the end, we computed the median of all the $F_{J0660,j}$, which we refer to as $\left\langle F \right\rangle$, and considered a source to have a significant $J0660$ excess if
\begin{equation}
 \frac{ \left\langle F \right\rangle  }{\mathrm{NMAD}\left( F_{J0660,\,j}\right)} \ge 3 \,,
\end{equation}

\noindent where NMAD denotes the normalized median absolute deviation \citep{nmad}. A total of $3\,426$ objects with $J0660$ excess were selected.

Within these sources, some of them are contaminants without H$\alpha$ emission. The assumption of a linear continuum without features produces that some non-emitting astrophysical objects exhibit a $J0660$ excess unrelated with star formation. To discern the nature of the selected sources, we analyzed them in a color-color diagram. After inspected all the available combinations, we chose the J-PLUS colors $(J0515-r)$ and $(g-i)$ as the best ones to discriminate between different objects (Fig.~\ref{fig:color-color-diagrams-emisores}).

\begin{figure}
    \centering
    \includegraphics[]{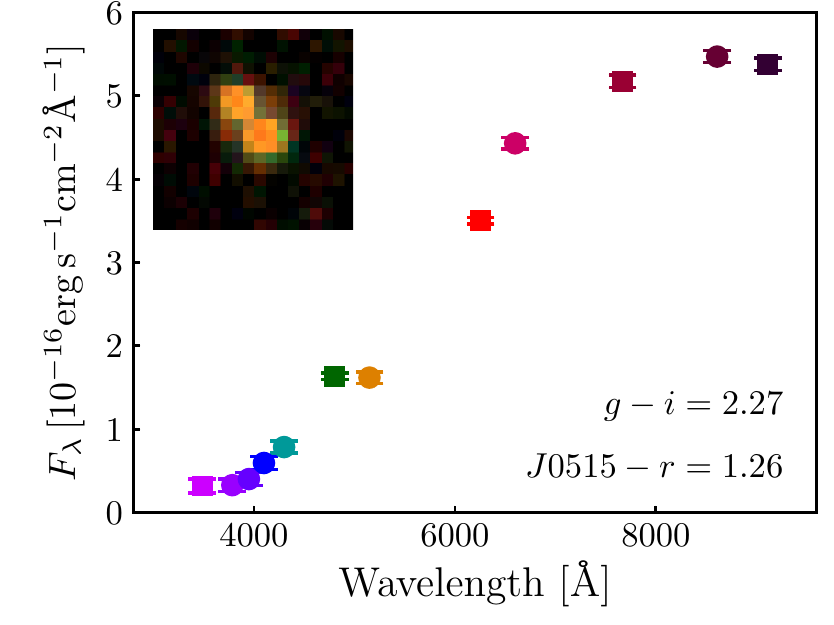}\\
    \includegraphics[]{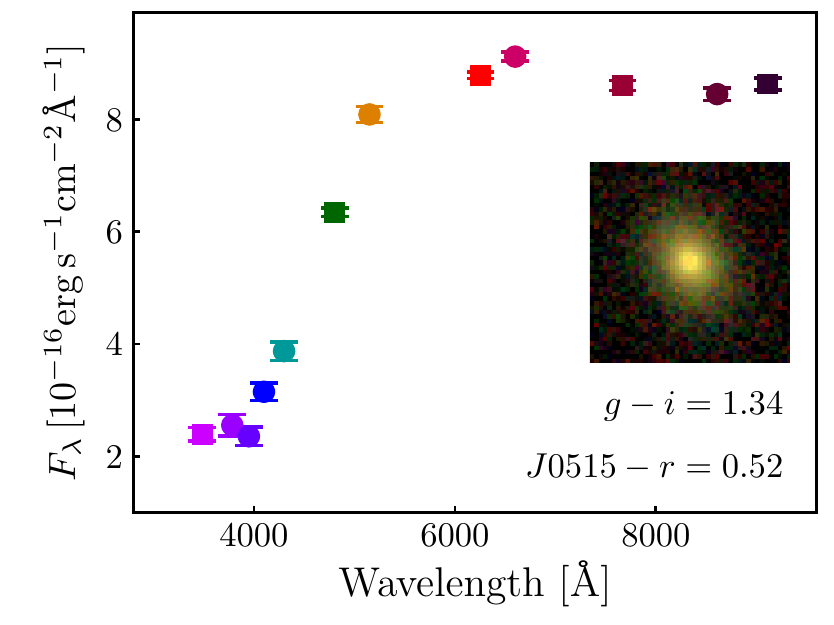}
    \caption{Representative example of a source in the region R2 ({\it upper panel}) and the region R3 ({\it bottom panel}) of the $(J0515-r)$ {\it vs} $(g-i)$ color-color diagram. In both panels, the J-PLUS \texttt{ISO\_GAUSS} photometry of the source in shown, with symbols as in the {\it right panels} of Fig.~\ref{fig:galaxy_G0}. A color composite of the sources, obtained from the $gri$ J-PLUS images, is presented in the insets.}
    \label{fig:regions_example}
\end{figure}

Three main populations were found in this color-color diagram. There is a concentration of sources at $(g-i) \sim 1.5$ and $(J0515 - r) \sim 0.6$, a sequence of sources extending bluewards down to $(g-i) \sim 0.5$ and $(J0515 - r) \sim 0.2$, and a set of redder sources with $(J0515 - r) \gtrsim 0.8$. We inspected the properties of these three populations, finding that they are dominated by different astrophysical sources. In this study, J-PLUS photometry was complemented with the available spectroscopic information. We found that the redder population is mainly composed by double (physically or in projection) stars that were detected as an unique, extended source by \texttt{SExtractor}. The SED of these red stars has a local maximum at $J0660$ in the J-PLUS filter system, being selected as objects with $J0660$ excess ({\it upper panel} in Fig.~\ref{fig:regions_example}). The main concentration of sources comprises early-type galaxies dominated by old stellar populations and located at $z \sim 0.1$. As in the previous case, this galaxy population exhibits a $J0660$ excess on its SED ({\it bottom panel} in Fig.~\ref{fig:regions_example}). Finally, the genuine H$\alpha$ emitting galaxies of interest are located in the bluest population.

From the analysis above, we defined three regions in the $(J0515-r)$ {\it vs} $(g-i)$ diagram. The region R1 ($1540$ sources) aims to isolate the desired H$\alpha$ emitters; region R2 ($351$ sources) is mostly populated by double stars; finally, region R3 ($1535$ sources) selects ${\rm z} \sim 0.1$ galaxies. The definition of region R1 was based on completeness, and it includes all the $466$ sources with $J0660$ excess and $0.001 < {\rm z}_{\rm spec} \leq 0.017$. As consequence, a large contamination of galaxies at higher redshift is also present, with $485$ sources having ${\rm z}_{\rm spec} > 0.017$ (Fig.~\ref{fig:color-color-diagrams-emisores}). Thus, there are $589$ sources in region R1 without spectroscopic information that can be located at low or high redshift.

We visually checked those sources, and classified them as low-redshift or high-redshift ($\mathrm{z} > 0.017$) galaxies. We looked for characteristic SED features of low-redshift galaxies, such as the [\ion{O}{iii}] emission traced by the $J0515$ filter at $0.007 < \mathrm{z} \leq 0.017$ ({\it bottom right panel} in Fig.~\ref{fig:galaxy_G0}), the [\ion{O}{ii}] emission traced by the $J0378$ filter ({\it top right panel} in Fig.~\ref{fig:galaxy_G0}), and a clear blue overall color at $\lambda \gtrsim 4500\ \AA$ (Fig.~\ref{fig:galaxy_G0}). These properties contrast with the typical red colors and the lack of other emission features displayed by the interlopers (Fig.~\ref{fig:regions_example}). We summarize the result of this exercise in Table~\ref{tab:numeros_tercer_sector}. We selected $158$ local emitters without spectroscopic information, and an illustrative example is presented in the {\it bottom panels} of Fig.~\ref{fig:galaxy_G0}.

As a summary, we gathered a sample of 842 nearby galaxies, out of which 684 have spectroscopic redshift information. From these, 466 present a $3\sigma$ excess in $J0660$ and 218 have lower emission significance. The remaining 158 galaxies that complete the sample do not have spectroscopic redshift information but have a significant excess in $J0660$. The next step was to derive the physical properties of these galaxies.

\begin{table}
\caption{Redshift classification of the S/N $\geq 3$ emitters in $J0660$ that lie within the region R1 inside the $(J0515-r)$ {\it vs} $(g-i)$ color-color diagram. There are $218$ extra galaxies with S/N $< 3$ in emission and ${\rm z}_{\rm spec} \leq 0.017$.}\label{tab:numeros_tercer_sector}
\begin{tabular}{@{\extracolsep{8pt}}cccc@{}}
\hline\hline\noalign{\smallskip}
\multicolumn{2}{c}{With z$_{\rm spec}$}   & \multicolumn{2}{c}{Without z$_{\rm spec}$}      \\
\noalign{\smallskip}\cline{1-2}\cline{3-4}\noalign{\smallskip}
Low-z           & High-z          & Low-z candidates & High-z candidates   \\ \noalign{\smallskip}\cline{1-1}\cline{2-2}\cline{3-3}\cline{4-4}\noalign{\smallskip}
$466$             & $485$             & $158$              & $431$         \\
\hline
\end{tabular}
\end{table}       
         
\subsection{Physical properties of the local sample}\label{sect:MonteCarlo} 
We describe in this Section the extraction of the J-PLUS photometry of the $842$ local galaxies in our sample (Sect.~\ref{sect:FluxExtraction}), and the estimation from these information of their H$\alpha$ flux ($F_{\mathrm{H}\alpha}$, Sect.~\ref{sect:Halphaflux}),  distance ($d$, Sect.~\ref{sect:Distancias}), H$\alpha$ luminosity ($L_{\mathrm{H}\alpha}$), stellar mass $M_{\star}$, and SFR (Sect.~\ref{sect:LMSFR}).

\subsubsection{J-PLUS photometry of local emitters}\label{sect:FluxExtraction}
Here, we detail the measurement of the photometric flux in the twelve J-PLUS filters for the galaxies in our sample with the final goal of providing the best possible estimation of their H$\alpha$ flux, as described in Sect.~\ref{sect:Halphaflux}.

We started by isolating galaxies in the twelve J-PLUS images. For this, we use ten effective radii as estimated by \texttt{SExtractor}, creating a cropped image from the original J-PLUS data. These images were homogenized to a common PSF in advance, as described in \citet{Logrono2019}. We run again \texttt{SExtractor} on the $r-$band image with a configuration optimized for large galaxies, obtaining the final structural parameters of the source (i.e., effective radius and the elliptical Kron aperture).

Then, a emission flux image was computed. We used the $r$ and $i$ images to define a linear continuum for each pixel, subtracting it from the observed $J0660$ image. The targeted H$\alpha$ + [\ion{N}{II}] emission is also included in the $r$ filter, so we used the equations described in VR15 to deal with this effect\footnote{We refer the reader to VR15 for a complete explanation of this procedure, which is called as Three Filter Method}. 

For each galaxy emission image, we computed the pixel flux histogram. This histogram resembles a Gaussian distribution, which is described by an average flux, $\mu_{\mathrm{em}} \sim 0$, and a dispersion, $\sigma_{\mathrm{em}}$. We note that the emission image encloses an area that extends up to ten effective radii of the galaxy, making the number of emitting pixels small compared to the whole number of pixels of the image. In addition, the dispersion $\sigma_{\mathrm{em}}$ was estimated from the 16-to-84 percentiles of the distribution, minimizing the impact of the emission pixels located in the positive tail. We estimated that the contribution of the emission pixels is therefore negligible when measuring the background flux. At this stage, we can classify any pixel in the emission image according to its signal in terms of $\sigma_{\mathrm{em}}$ ({\it middle panels} in Fig.~\ref{fig:galaxy_G0}).

Now, for each galaxy in the sample, we combined the flux of all the pixels that have an emission flux larger than $N$ times the dispersion $\sigma_{\mathrm{em}}$ with the following process:
\begin{enumerate}
    \item We selected a confidence level $N$.
    \item All pixels with emission flux larger than $N\sigma_{\mathrm{em}}$ and within three effective radii were tagged.
    \item The tagged pixels were selected in each of the twelve J-PLUS images.
    \item The signal in these pixels was added, creating a flux and an observed magnitude, in a given J-PLUS band.
    \item The fluxes and magnitudes were de-reddened from Milky Way extinction using the color excess from \texttt{Bayestar17}\footnote{\url{http://argonaut.skymaps.info}} \citep{bayestar17} at galaxy position and the extinction coefficients reported in \citet{Whitten2019}.
\end{enumerate}

In the end, eleven photometric catalogs were derived, where the confidence level $N$ belongs to the interval $\left[ -5,5 \right]$ with unity steps. We refer to these as \texttt{pix\{pm\}N} catalogs, where \texttt{N} refers to the positive (\texttt{p}) or negative (\texttt{m}) confidence level $N$ used to extract the photometry. It is important to remark that, even though these catalogs are created using the pixels that fulfill an emission criteria, the \texttt{pixm5} catalog \textit{de facto} contains the total apparent fluxes and magnitudes of each galaxy in each J-PLUS band. We note that each \texttt{pix\{mp\}N} photometry provide different information about the analyzed galaxies and their star formation properties, and we benefit of this extra spatial information along the paper.

The \texttt{pix\{mp\}N} catalogs were used to derive the physical properties of the $842$ local galaxies in our sample, as described in the next sections.

\subsubsection{Estimation of the H$\alpha$ emission flux}\label{sect:Halphaflux}
First, we measured the H$\alpha$ emission flux of each galaxy, $F_{\mathrm{H}\alpha}$. To do so, we used the SED-fitting methodology that is fully described in VR15, along with the statistical dust and [\ion{N}{ii}] corrections estimated in the same work. This methodology relies on the full SED of the galaxy to estimate the continuum level at the $J0660$ filter, including H$\alpha$ stellar absorption, so the twelve J-PLUS fluxes were used in the process. For completeness, we did this for each \texttt{pix\{pm\}N} catalog.

We remind here the empirical dust and [\ion{N}{ii}] corrections that we used. We estimated the color excess of the selected emitting area as
\begin{equation}\label{EcPolvo}
  E(B-V) =  0.206\left(g-i\right)^{\:1.68} -0.0457.
 \end{equation}
With this, the relation between the observed emission flux obtained from the SED-fitting routine, $F_{\rm obs}$, and the intrinsic, dust-free flux comprising the H$\alpha$ and [\ion{N}{ii}] emission, $F_{\mathrm{H}\alpha+\mathrm{[\ion{N}{ii}]}}$, is given by
 \begin{equation}
  F_{\mathrm{H}\alpha+\mathrm{[\ion{N}{ii}]}} = F_{\rm obs} 10^{0.4\,E(B-V)\,k'}\ \ \ \ {\rm [erg\,s^{-1}\,cm^{-2}]},
 \end{equation}
where $k'$ is a polynomial that depends on wavelength and the selective-to-total excess ratio $R_{V}$. We estimate $k' = 3.33$ from the \cite{Calzetti2000} parametrization for $R_{V} = 4.05$. This corrects the internal extinction of the analysed source, and the Milky Way extinction was accounted for in the estimation of the \texttt{pix\{pm\}N} catalogs.

 Once we have corrected the H$\alpha+$ [\ion{N}{ii}] emission from dust reddening, we removed the [\ion{N}{ii}] contribution. To this aim, we used the following empirical relation derived by VR15:
 \begin{equation}\label{EcNitrogenos}
  \log F_{\mathrm{H}\alpha} =  \begin{cases}
  0.989\log(F_{\mathrm{H}\alpha+\mathrm{[\ion{N}{ii}]}} )-0.193,  & \text{if } g-i  \leq 0.5, \\
  0.954\log(F_{\mathrm{H}\alpha+\mathrm{[\ion{N}{ii}]}} )-0.753,  & \text{if } g-i > 0.5.
  \end{cases}
 \end{equation}

The procedure above was observationally validated in \cite{Logrono2019} using the J-PLUS early data release (EDR). We selected a sample of 46 star-forming regions that had been observed with CALIFA or SDSS, but also with J-PLUS. With this we made sure that we had an spectroscopic measurement of the H$\alpha$ flux to compare with the photometric measurement. We conclude that our methodology provides an unbiased $F_{\mathrm{H}\alpha}$ with a minimum uncertainty of $20\%$, that is included in the final error budget together with the statistical error in the measurement. We further test the H$\alpha$ flux measurements in Sect.~\ref{sect:polvo}.

We were not able to extract an H$\alpha$ flux for $37$ galaxies in the local sample with the above procedure. There are low surface brightness galaxies with a low pixel-by-pixel significance in the emission. The final sample of local H$\alpha$ emitters used hereafter comprises $805$ galaxies.

\begin{figure}
    \centering
    \includegraphics{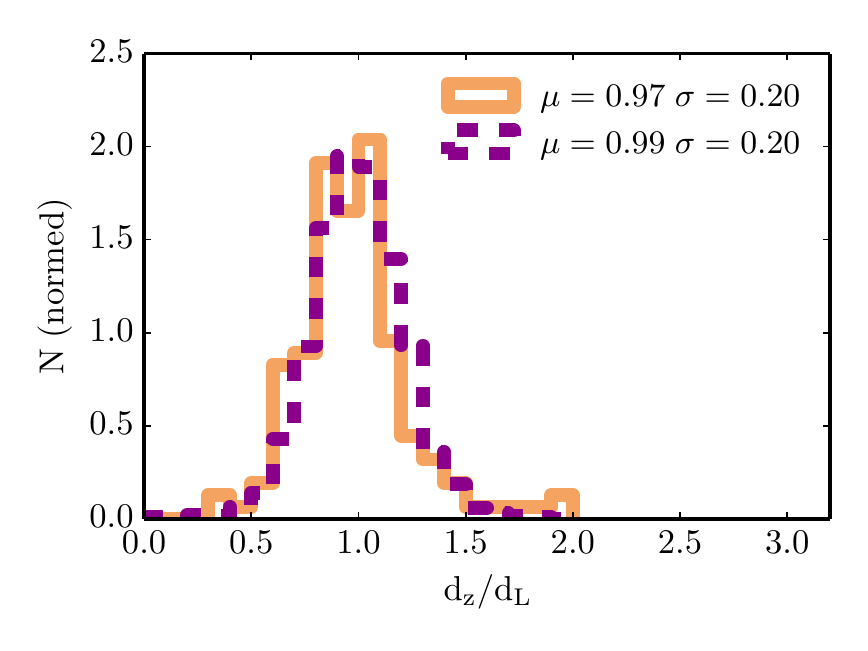}
    \caption{Histogram of the ratio between the redshift-derived ($d_{\rm z}$) and the redshift-independent ($d_{\rm L}$) luminosity distance for galaxies in Sample~$G2$ (orange histogram). The purple histogram represents the same, but for a sample of simulated galaxies distributed according to a volume prior and with a spectroscopic redshift that has been perturbed with a peculiar velocity term of $v_{\rm peculiar} = 750\,\mathrm{km\,s^{-1}}$.}
    \label{fig:histo_velocidades}
\end{figure}

\subsubsection{Computing distances to local galaxies}\label{sect:Distancias}
In this section, we explain how we computed the distance to the galaxies in our sample. This is important to describe it in detail because the distance was used both in the calculation of the stellar mass and the H$\alpha$ luminosity (Sect.~\ref{sect:LMSFR}). This introduces a correlation between these two parameters that has to be accounted for. Additionally, it is well known that redshifts (either photometric or spectroscopic) may be strongly affected by peculiar velocities in the local Universe and are thus a degraded proxy for the cosmological distance.

We started by dividing the sample of $805$ local H$\alpha$ emitters in three sub-samples, according to the information that we have to infer a distance\footnote{The notation is meant to be mnemonic: the higher the number that describes the sample, the more information we have for the galaxies in it.}:
\begin{itemize}
    \item Sample~$G0$: Galaxies with neither redshift-independent distance nor spectroscopic redshift. There are $158$ galaxies in this category ($20\%$ of the total sample).
    \item Sample~$G1$: Galaxies without a redshift-independent measurement of the distance, but with a measured spectroscopic redshift. There are $487$ galaxies in this situation ($60\%$ of the total sample).
    \item Sample~$G2$: Galaxies with a redshift-independent measurement of the distance. Distances were retrieved from the NED data base, and obtained with different methods, such as the Tully-Fischer relation, or the tip of the red giant branch. They account for the remaining $160$ galaxies ($20\%$ of the total sample). All of them also have a spectroscopic redshift.
\end{itemize}

We obtained the distance modulus $(m-M)$ for galaxies in Sample~$G2$ from the NED. The distance to these galaxies was determined by the associated distance to the median $(m-M)$, in the case that more than one value was provided in the NED. If just one value of $(m-M)$ is available, we used it. In the same way, we also computed distance errors. We explain how we assigned errors to these distances, noted $\delta d$, in Appendix~\ref{sect:A_Distance_errors}.

Galaxies in Sample~$G1$ have a redshift-derived distance. This redshift is not fully coupled to the Hubble flow, meaning that peculiar velocities introduce an uncertainty to the luminosity distance derived from z$_{\rm spec}$. To test the impact of peculiar velocities in the distances derived from spectroscopic redshifts, we used the redshift-independent information in Sample~$G2$.

We found that the ratio between the distances estimated from the Hubble flow ($d_{\rm z}$) and those estimated from redshift-independent methods ($d_{\rm L}$) in Sample~$G2$ is well described by a Gaussian with median $0.97$ and dispersion $0.20$ (Fig.~\ref{fig:histo_velocidades}). To find the best peculiar velocity that describes the observed relation for Sample~$G2$ galaxies, we generated a set of $5\,000$ synthetic galaxies distributed up to $\mathrm{z}=0.017$ according to a volume prior. To each of these galaxies, for which we know the real distance and the Hubble flow redshift, we added a term of peculiar velocity drawn from a Gaussian distribution, with $\sigma=v_{\mathrm{peculiar}}$. We then re-computed the luminosity distance that we would have obtained if we had used this perturbed z$_{\rm spec}$ to measure it. We did this for several $v_{\mathrm{peculiar}}$, and we found that $v_{\mathrm{peculiar}} = 750\,\mathrm{km\,s^{-1}}$ generates a distribution of relative errors in distance that reproduces well the observed one with Sample~$G2$ galaxies, as shown in Fig.~\ref{fig:histo_velocidades}. This component includes the typical peculiar velocity of the local galaxies and the mean uncertainty on the redshift-independent distances. It must be interpreted therefore as a formal minimum error in the comparison between redshift-dependent and independent distances. Moreover, the assumed value of $v_{\mathrm{peculiar}}=750\,\mathrm{km\,s^{-1}} $ does not have a significant impact in the main scientific results of this work, as shown in Appendix~\ref{sect:A_Distance_Algorithm}, where we repeat all the analysis carried out in further sections with different $v_{\mathrm{peculiar}}$ to assess the impact of this assumption.

The result above also provides a limiting distance, $d_{\rm lim}$, at which errors in the current methods to obtain a redshift-independent distance are larger than the floor uncertainty imposed by $v_{\rm peculiar}$ in the z$_{\rm spec}$ distances. Hence, we imposed a cut in $d_{\rm lim} = 60$ Mpc. Galaxies in Sample~$G2$ with a distance measurement larger than $60$ Mpc will be assigned a distance according to their spectroscopic redshift, disregarding the redshift-independent measurement. In Appendix~\ref{sect:A_Distance_Algorithm}, we show that changing this criterion does not significantly affect the results of this paper.

The detailed analysis above justifies the distance assignment that is described in the next Section. 

\begin{figure*}
    \centering
    \includegraphics{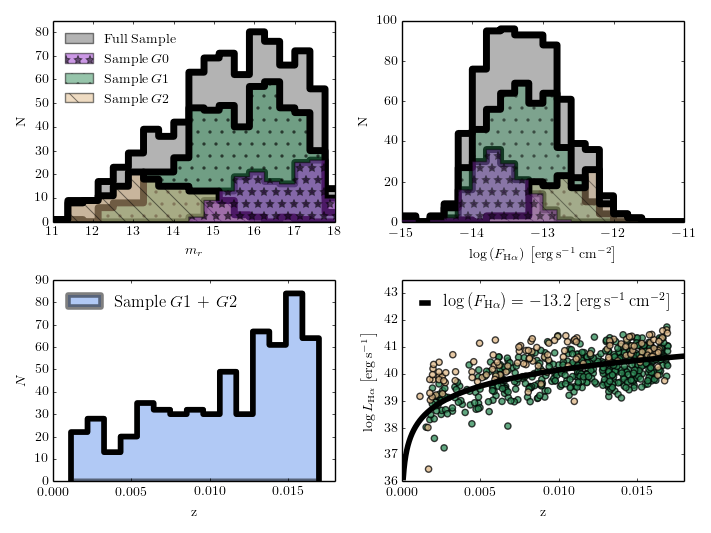}
    \caption{General properties of the J-PLUS DR1 local sample at $\mathrm{z} \leq 0.017$. {\it Upper left panel}: Distribution in the $r$-band apparent magnitude. The three sub-samples with different distance information are labeled in the panel (see Sect.~\ref{sect:Distancias}, for details). {\it Upper right panel}: Distribution in H$\alpha$ flux. {\it Lower left panel}: Redshift distribution of those galaxies in the sample with $\mathrm{z}_{\rm spec}$. {\it Lower right panel}: H$\alpha$ luminosity {\it vs}. redshift for those galaxies in the sample with $\mathrm{z}_{\rm spec}$. The solid line marks the luminosity for a galaxy with $\log F_{\mathrm{H}\alpha} = -13.2$ as reference.
    }
    \label{fig:panel_caracterizacion}
\end{figure*}

\begin{figure}
    \centering
    \includegraphics[]{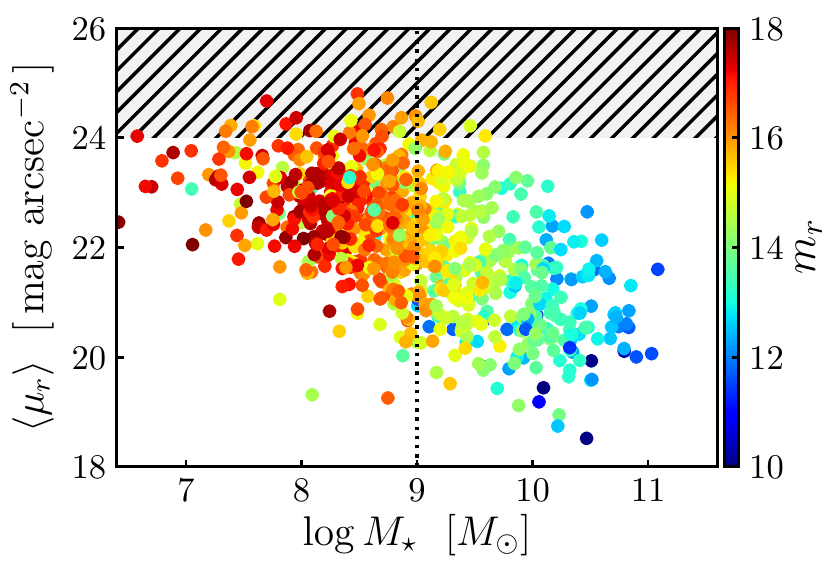}
    \caption{Mean $r-$band surface brightness within three effective radii, $\langle \mu_r \rangle$, as a function of the stellar mass for the local sample of H$\alpha$ emitters. The color scale shows the apparent magnitude of the sources. The hatched area marks a surface brightness larger than 24 mag arcsec$^{-2}$. The vertical dotted line marks a stellar mass of $\log M_{\star} = 9$.}
    \label{fig:mumag}
\end{figure}

\begin{figure*}
     \centering
     \includegraphics{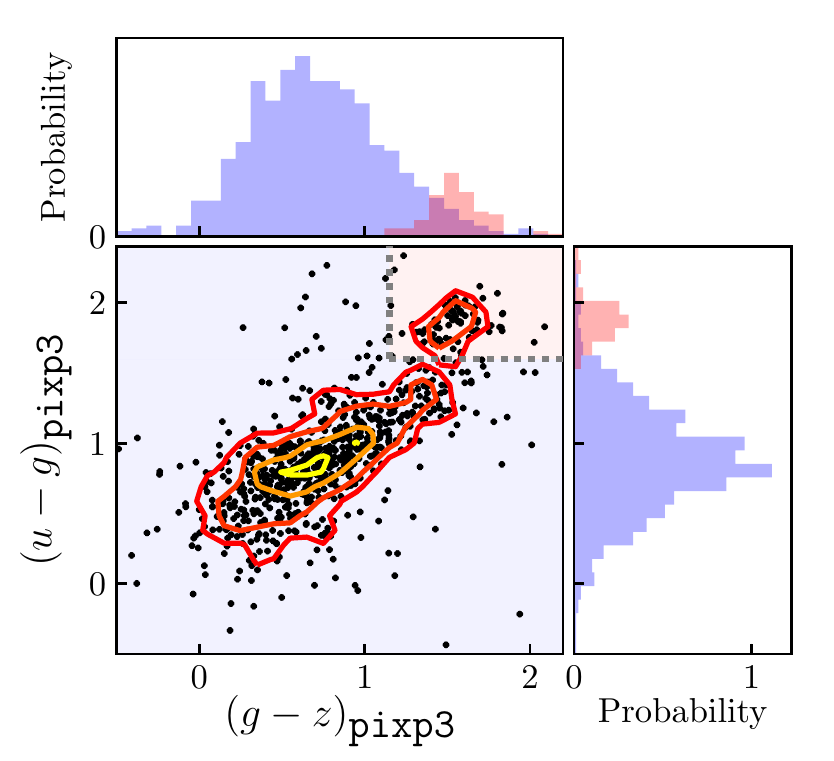}
     \includegraphics{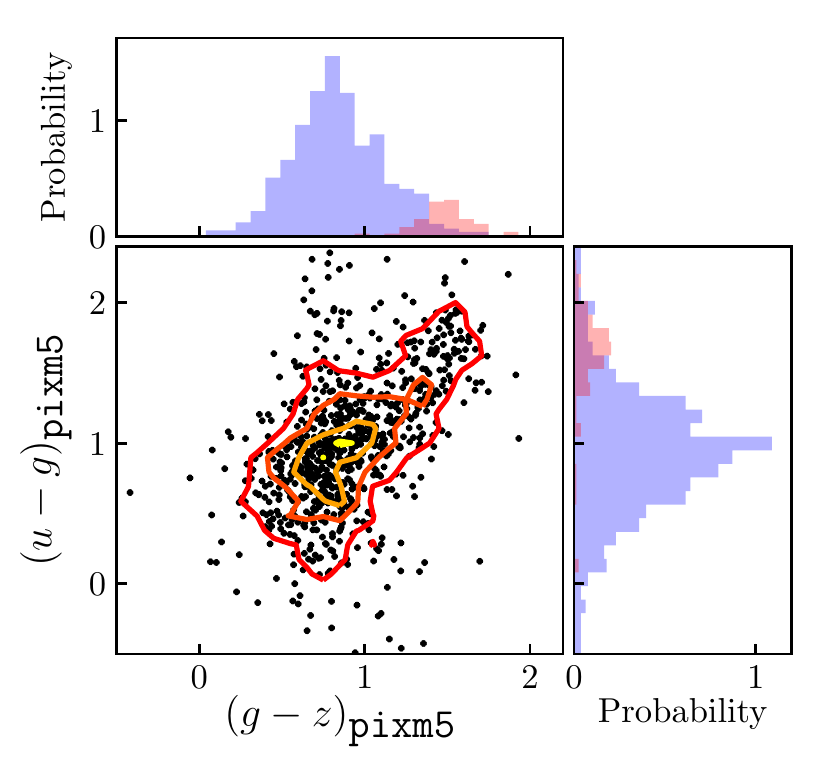}
     \caption{$(u-g)$ {\it vs}. $(g-z)$ color-color diagram of our local galaxy sample computed from pixels with signal-to-noise higher than 3 (\texttt{pixp3}; {\it left panel}) and higher than -5 (i.e., total colors , labeled as \texttt{pixm5}; {\it right panel}) in the H$\alpha$ detection images. The colored contours show the density of sources to highlight the presence of local maxima. The dotted lines in the {\it left panel} mark our selection for the Red Sample (red area), the Blue Sample being the complementary set (blue area). The upper- and right-side normalized histograms are the density projections in $(g-z)$ and $(u-g)$, respectively. The red and blue shaded histograms in both panels mark the Red and Blue samples selected with \texttt{pixp3} colors.}
     \label{fig:ug_gz}
\end{figure*}

\subsubsection{Stellar masses, H$\alpha$ luminosity, and SFR: The Monte-Carlo routine}\label{sect:LMSFR}
We now explain the Monte-Carlo routine that we used to assign an H$\alpha$ luminosity (hence, a SFR) and a $M_{\star}$ to the galaxies in our sample. This routine consists of perturbing the H$\alpha$ fluxes, and the distances, within their error bars. These two perturbations lead to a different value of $L_{\mathrm{H}\alpha}$, SFR, and $M_{\star}$ in each iteration. To start with, H$\alpha$ fluxes were always perturbed with a noise component drawn from a Gaussian distribution with $\sigma$ its error. Distances were perturbed within their errors depending on the information that we have. We used the following recipe:

\begin{itemize}
    \item Sample~$G0$: Galaxies with no distance information:
    \begin{itemize}
        \item In each iteration, we assigned a random distance according to a volume prior.
    \end{itemize}

    \item Sample~$G1$: Galaxies that do not have a z-independent measurement of the distance, but with a z$_{\rm spec}$:
    \begin{itemize}
        \item We started by assigning to each z$_{\rm spec}$ a perturbation $v'$, drawn from a Gaussian distribution, centered on zero and with $\sigma=750\,\mathrm{km\,s^{-1}}$;
            \begin{equation}
                {\mathrm z}_{\mathrm{perturbed}} = \left(1 + {\mathrm z}_{\mathrm{spec}} \right)\cdot \left( 1 + \frac{v'}{c} \right) - 1\,.
            \end{equation}    
        \item We computed the luminosity distance using $\mathrm{ z}_{\mathrm{perturbed}}$.
    \end{itemize}

    \item Sample~$G2$: Galaxies that have a collection of redshift-independent measurements of the distance: 
    \begin{itemize}
        \item If their median distance is smaller than $60$ Mpc, we used both the distance and the uncertainty from the redshift-independent methods.
        \item We perturbed the distance with a random error drawn from a Gaussian distribution with $\mu=0$ and $\sigma = \delta d\,\mathrm{Mpc}$. In case the final distance is negative, we took the absolute value. 
        \item If their median distance is larger than $60$ Mpc, we proceeded in the same way that with galaxies of Sample~$G1$.
    \end{itemize}
\label{distance_algorithm}    
\end{itemize}

To compute the stellar mass of the galaxies, expressed in Solar mass units, we used the mass-to-light vs. color relation (MLCR) for star-forming galaxies described in the work by \cite{carlinhos2019masas}. This relation is based on the observed (i.e., dust-attenuated) $g-i$ color and we scaled it to a Salpeter IMF:
\begin{equation}\label{eq:mass}
    \log M_{\star} = 1.626 + 0.212\left(g-i\right) +0.144\left(g-i\right)^{2} - 0.4M_{i},
\end{equation}
where $M_i$ is the absolute magnitude of the galaxy in the $i$ band. In this case, we did not perturb the apparent $g$ and $i$ magnitudes, as their error is negligible. However, by changing the distance, we were changing $M_{i}$. Finally, and to account for the intrinsic dispersion of this relation, we included an extra perturbation, drawn from a Gaussian distribution with $\sigma=0.07$ dex \citep{carlinhos2019masas}. We favoured the estimation of the stellar mass from a MLCR over a SED-fitting technique because both approaches provide similar accuracy \citep[e.g.][]{Taylor2011} and to minimize the correlation between the derived H$\alpha$ fluxes, based on SED-fitting, and the stellar masses.

We computed the H$\alpha$ luminosity as
\begin{equation}
    L_{\mathrm{H}\alpha} = 4\pi d^2\,C(q, T)\,F_{{\rm H}\alpha}\ \ \ {\rm [erg\,s^{-1}]},
\end{equation}

\noindent and the star formation rate as
\begin{equation}
    \mathrm{SFR} = 7.9\cdot10^{-42}\,L_{\mathrm{H}\alpha}\ \ \ {\rm [M_{\odot}\,yr^{-1}]},
\end{equation}

\noindent where the \citet{Kennicutt1998} relation between H$\alpha$ luminosity and SFR for a \citet{Salpeter55} IMF and case B recombination is assumed, and $C(q, T)$ is a statistical correction that accounts for the missing H$\alpha$ flux on inclined systems. This correction depends on the minor-to-major axis ratio of the galaxy, $q$, and its morphological type, $T$. The inclination correction is unity for 60\% of the galaxies in the sample and has a median value of $C = 1.22$ for the remaining 40\%, with a maximum correction of $C = 1.78$. The details about the estimation of this inclination correction are in the forthcoming paper by \citet{Logrono2020}, where the morphology of the local sample is derived and studied.

We repeated the process above $n = 5\,000$ times. In the end, for each galaxy, we had a collection of $n$ values for the distance, $M_{\star}$, $L_{\mathrm{H}\alpha}$, and SFR. The median of these values were used as our measurements in the rest of the paper, while the uncertainties were computed as the NMAD of the $n$ values of each parameter. We also computed the typical values for the covariance and the correlations, to find that the average correlation factor between $M_{\star}$ and the SFR is $\sim0.75$. These correlations were used in the statistical analysis of the sample.

We derived stellar masses, H$\alpha$ luminosities, and SFRs for each \texttt{pix\{pm\}N} catalog. In the following, stellar masses refers to \texttt{pixm5} (total) photometry, and $L_{\mathrm{H}\alpha}$ and SFR to \texttt{pixp1} photometry. We found that the latter are similar to the global ones from \texttt{pixm5} but with a higher signal-to-noise ratio.

\subsection{General properties of the sample}\label{sect:SampleProperties}
\subsubsection{Global distributions}\label{sect:GlobalDistributions}
In this Section, we characterize the sample of 805 local H$\alpha$ emitters. Distributions of $m_{r}$, redshift, H$\alpha$ flux, and the relation between $L_{\mathrm{H}\alpha}$ and redshift are plotted in Fig.~\ref{fig:panel_caracterizacion}. 

It is worth noting that galaxies in Sample~$G0$ appear to be the faintest in $m_{r}$ magnitude and H$\alpha$ flux, and their distribution in fluxes does not resemble the distribution of fluxes that combines Sample~$G1$ and Sample~$G2$. This can be due to two possibilities, either these galaxies are all particularly far (with a limit of $73$ Mpc), but have bright luminosities, or these galaxies are intrinsically less luminous and are spread all over our volume in a regular way.

To assess this problem, we recursively moved these galaxies through our volume, keeping the flux, but assigning the same distance to all of them and computing the luminosity distribution of these Sample~$G0$ galaxies as if all of them were at this distance. By doing this we found that, in order to reproduce the luminosity distribution of the rest of the sources with known distances, all galaxies in Sample~$G0$ should be at $\sim80$ Mpc, which means that we should not see an excess in the $J0660$ filter. This leads us to believe that these galaxies are indeed properly classified as low-redshift galaxies, but with faint H$\alpha$ luminosities.

Finally, we present in Fig.~\ref{fig:mumag} the relation between stellar mass and $\langle \mu_r \rangle$, defined as the mean $r-$band surface brightness within three effective radii. We found a limiting surface brightness of $\langle \mu_r \rangle \sim 24.0$ mag arcsec$^{-2}$, in agreement with the limit of the general J-PLUS catalog \citep{Cenarro2019}. The larger apparent size of the local sample with respect to the general galaxy population in J-PLUS makes the $m_r \leq 18$ selection equivalent in terms of surface brightness to the $m_r \sim 21$ limiting magnitude for the general population, minimizing the bias against low surface brightness galaxies. The trend between stellar mass and $\langle \mu_r \rangle$ that appears in Fig.~\ref{fig:mumag} suggests that the J-PLUS local sample is complete in stellar mass for $\log M_{\star} \gtrsim 9$ galaxies. This is confirmed in Sect.~\ref{sect:SMF}, where the stellar mass function of the sample is derived and analyzed.

\subsubsection{Active galactic nuclei}\label{sect:AGN}
The J-PLUS filter set does not allow us to discern active galactic nuclei (AGN) using known tools such as the BPT diagram \citep[][]{Baldwin1981}. To cope with this, we check the NED for information regarding the nuclear activity of our sources. From now on, we remove sources with any kind of nuclear activity from our study. After removing $49$ AGN from the $805$ sources in our catalog, a total of $756$ galaxies remain.

\subsubsection{Color properties of the sample}\label{sect:pix3color}
To provide further insights into the nature of our emitters, we study them in a $(u-g)$ \textit{vs}. $(g-z)$ color-color diagram. After revising the resulting diagrams with the available \texttt{pix\{pm\}N} photometries, we used the \texttt{pixp3} catalog. We recall that the photometry in this catalog contains information of the pixels with high emission flux. Hence, it encapsulates the information on the average properties of the star-forming regions. In contrast, if we use the \texttt{pixm5} catalog, we obtain the total colors of the galaxy. We compare the $(u-g)$ \textit{vs} $(g-z)$ obtained with both \texttt{pixp3} and \texttt{pixm5} catalogs in Fig.~\ref{fig:ug_gz}.

Interestingly, we found that the \texttt{pixp3} photometry unveils two different populations that are not clearly discernible if the total color of the galaxy, as traced by \texttt{pixm5}, is used. The separation between both populations increases as we move from \texttt{pixm5} to \texttt{pixp5} catalogue. However, not all sources have emission pixels above $5\sigma_{\mathrm{em}}$, and the colors are therefore not available. The \texttt{pixp3} catalog provides the optimum compromise between a clean separation of populations and minimum loss of sources that lack high-significance emission, providing a broadband colour measurement for most of the sample.

We separated the two populations by imposing a selection criteria on \texttt{pixp3} colors:

\begin{itemize}
    \item Blue Sample: Galaxies with $(u-g)\leq1.6 \cup (g-z)\leq1.15$.
    \item Red Sample: Galaxies with $(u-g)>1.6 \cap (g-z)>1.15$.
\end{itemize}

The $28$ galaxies without available \texttt{pixp3} colors were assigned to the Blue Sample. In the end, we selected $689$ galaxies in the Blue Sample and $67$ galaxies in the Red Sample. We call Full Sample the combination of the Blue and Red samples ($756$ galaxies). As we will demonstrate, the star formation properties of these two populations are remarkably different.

\subsection{Sample characterization: conclusions}
In this Section, we have explained the routine to retrieve low-z H$\alpha$ emitters, extract their J-PLUS photometry, and estimate their H$\alpha$ flux. This yielded a catalog of $805$ \textit{bona fide} galaxies. Then, we explained how we computed distances, and their uncertainties, to these galaxies. With this, we have presented the Monte-Carlo routine to obtain the H$\alpha$ luminosity, SFR, and $M_{\star}$. This allowed to perform a basic characterization of the main physical properties of these galaxies. In the end, after removing AGN, our final catalog contains $756$ galaxies, $689$ in the Blue Sample and $67$ in the Red Sample, which will be used in the following sections. The catalog with the information of the galaxies in the Full Sample is publicly available in the J-PLUS webpage\footnote{\url{http://j-plus.es/ancillarydata/dr1_halpha_local_galaxies}}. The study of the morphological properties of this sample is beyond the scope of the present work and it is addressed in the forthcoming paper by \citet{Logrono2020}.

\section{Star formation in the local Universe}\label{sect:SFjplus}
In this Section, we present the main scientific results of this paper. These are: the star formation main sequence (Sect.~\ref{sect:SFMS}), the two projections of this relation, i.e., the stellar mass function (SMF, Sect.~\ref{sect:SMF}) and the H$\alpha$ luminosity function (H$\alpha$LF, Sect.~\ref{sect:HaLF}), and the star formation rate density at $d \lesssim 75$ Mpc (Sect.~\ref{sect:SFRD}).

\begin{figure*}
    \centering
    \includegraphics[]{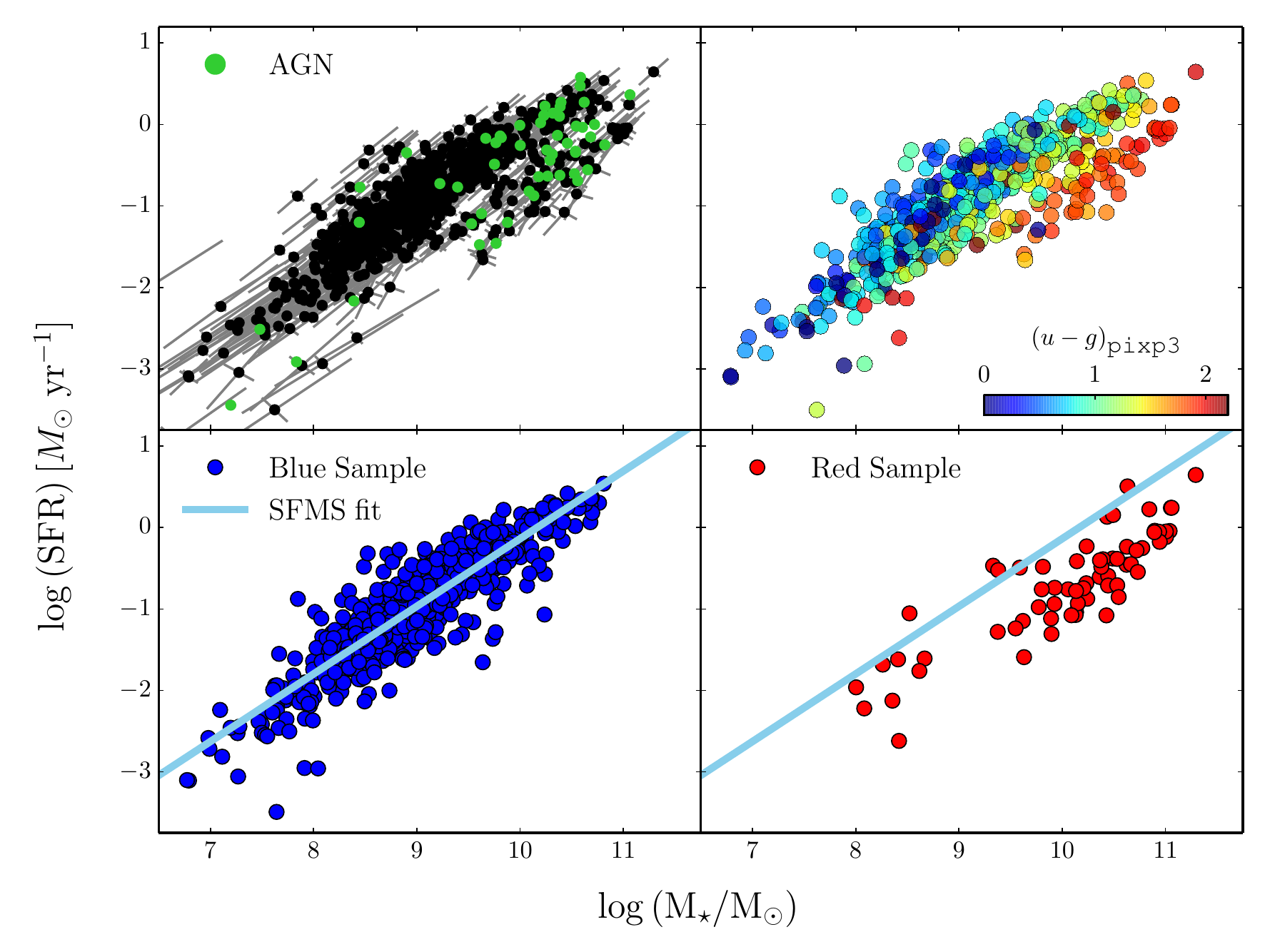}
    \caption{Star formation rate {\it vs}. stellar mass in the J-PLUS DR1 sample at $d \lesssim 75$ Mpc. {\it Upper left panel}: Relation for the Full sample ($756$ galaxies, bullets). The error bars provide the semi-major and semi-minor axis of the error ellipse, highlighting the covariance in the variables. The green dots mark the $49$ AGN in the sample; these are shown here only with illustrative purposes, and are not taken into account when computing the main results of this work. {\it Upper right panel}: Relation for the Full sample. The color of the points reflects the $(u-g)_{\texttt{pixp3}}$ color of the galaxy, as scaled in the inner bar. {\it Bottom left panel}: Relation for the Blue sample ($689$ galaxies). {\it Bottom right panel}: Relation for the Red Sample ($67$ galaxies). The solid line in both bottom panels is the best-fitting linear relation to the SFMS defined by the Blue Sample.}
    \label{fig:sfms_jplus}
\end{figure*}

\subsection{Star formation main sequence}\label{sect:SFMS}
In the {\it upper left panel} of Fig.~\ref{fig:sfms_jplus}, we present the relation between SFR and stellar mass for the Full Sample. Both the SFR and $M_{\star}$, as well as their errors, had been computed with the routine that is described in Sect.~\ref{sect:MonteCarlo}. We found that a main trend appears, spanning from $\log M_{\star} \sim 7.5$ up to $\log M_{\star} \sim 11$. This main trend is accompanied by a secondary, parallel sequence, with lower SFR that appears at $\log M_{\star} \sim 9.5$ up to $\log M_{\star} \sim 11$. Between these two sequences, there is an underpopulated gap. We highlight the impact of the correlated errors in the observed relation. The dispersion of the points is largely smaller than the computed uncertainties if both variables are assumed independent. However, the $\sim 0.75$ correlation between stellar mass and SFR, mainly driven by the shared distance, roughly moves the data along the observed relation and the dispersion is therefore dominated by the intrinsic variance in the properties of the galaxy sample, and not by the uncertainties in the measurements.

To provide more insight into the nature of the two observed sequences, we studied the $(u-g)$ color of the sources, computed using the \texttt{pixp3} catalog, in the {\it upper right panel} of Fig.~\ref{fig:sfms_jplus}. We found a clear gradient in color that is coupled to the gradient in stellar mass, being the most massive galaxies the reddest ones. At a fixed stellar mass of $\log M_{\star} \sim 10.5$, the average color of galaxies in the primary sequence is $(u-g) \sim 1.25$, and in the secondary sequence is $(u-g) \sim 1.80$. This suggests a link between the two observed sequences with the Blue and Red samples defined in Sect.~\ref{sect:pix3color}.

We found indeed that the Blue Sample defines the canonical star formation main sequence ({\it bottom left panel} in Fig.~\ref{fig:sfms_jplus}). The characterization of the SFMS is presented in the next Section. We also found that the secondary sequence is populated by the Red Sample ({\it bottom right  panel} in Fig.~\ref{fig:sfms_jplus}). This suggests a different origin for the star formation in these systems, as we will discuss latter.

\subsubsection{Fitting the star formation main sequence}
We fitted the SFMS defined by the Blue Sample while we explored the parameter space in the Monte-Carlo routine (Sect.~\ref{sect:MonteCarlo}). Each time we perturbed the SFRs and stellar masses, we fitted a linear relation to the resulting distribution. The form of the equation that we fitted is the following:
\begin{equation}
    \log \mathrm{SFR} = a\cdot\log M_{\star} + b.
\end{equation}

The fitting was done using \texttt{emcee} \citep{emcee}, a Markov Chain Monte-Carlo (MCMC) code. We used a set of $20$ walkers, $500$ steps, and a burn-in phase of $300$ steps. In each iteration of the Monte-Carlo process, we saved $100$ random values of \textit{a} and \textit{b} that the \texttt{emcee} walkers have sampled during the sampling phase. We emphasize that the fitting is not error-weighted. Instead, we fitted the linear relation using the resulting distribution of points after having been perturbed.

When the Monte-Carlo routine ends, we had $5\times10^5$ sampling points for each parameter, which we used to model the probability distribution function (PDF) of \textit{a} and \textit{b}, and the correlation between these two parameters. We present here the median and the NMAD values of the PDFs for \textit{a} and \textit{b}, which are used as error bars. The best-fitting parameters are:
\begin{eqnarray}
    a =  0.83 \pm 0.05,\\
    b = -8.44 \pm 0.50.  
\end{eqnarray}

The best-fitting SFMS is presented in the {\it bottom panels} of Fig.~\ref{fig:sfms_jplus}. We will put our results in the context of previous findings in Sect.~\ref{sect:SFMS_discusion}. The dispersion of the data with respect to the best-fitting line, $\sigma \left( \mathrm{SFR} - \mathrm{SFR_{fit}}  \right)$, is $0.25$ dex, comparable with previous work ($\sim 0.3$ dex, \citealt{Whitaker2012,Ilbert2015,Popesso2019a}). To conclude, the best~-~fitting values for the SFMS if the Red Sample is included are $a=0.73 \pm 0.05$ and $b=-7.6\pm 0.50$. As expected, the slope becomes flatter due to the presence of galaxies with low star formation rate at the high-mass end of the relation.

\subsubsection{Ionization source in red galaxies}\label{sect:pAGB}
We found that the Red Sample defines a lower SFR sequence with respect to the canonical SFMS traced by the Blue Sample. One can argue that the H$\alpha$ ionization source in these systems is not related with star-forming processes, and that AGN or post-AGB stars could account for the needed UV radiation field.

First, we had removed known AGN from our sample (Sect.~\ref{sect:AGN}), minimizing their possible impact. Second, the typical H$\alpha$ equivalent width (EW) of diffuse gas ionized by post-AGB stars is EW $\leq 3\ \AA$   \citep{CidFernandez2011, Kehrig2012, gomes16}. The instrumental setup from J-PLUS impose an EW limit of $12\ \AA$ (VR15), so the H$\alpha$ measurements are expected to be insensitive to post-AGB ionization.

The sequence of post-AGB ionized galaxies selected with H$\alpha$ EW $\leq 3\ \AA$ is studied in detail by \citet{CanoDiaz2019} using IFS data from the Mapping Nearby Galaxies at Apache Point Observatory (MaNGA, \citealt{manga}) survey. They find a clear sequence in the SFR {\it vs}. stellar mass space with $a = 1.09$, $b = -13.0$, and a dispersion of 0.18 dex. The near-unity slope is interpreted by the authors as a direct consequence of post-AGB stars being the main ionizing source, with the available UV flux scaling with the stellar mass of the galaxy. We checked that our Red Sample is more than 3$\sigma$ ($0.6$ dex) above the post-AGB ionization sequence in \citet{CanoDiaz2019}, reinforcing the star formation origin of our observed H$\alpha$ flux. This is also supported by the derived stellar mass function of the Red Sample (Sect.~\ref{sect:SMF}).

As a final remark, the star-forming and post-AGB ionized sources produce also two different populations in the spatially-resolved SFMS \citep[e.g.][]{Hsieh2017, CanoDiaz2019}. The J-PLUS DR1 sample analyzed in this paper is well suited for spatially-resolved studies, providing extra clues about the origin of the measured H$\alpha$ flux. This issue will be addressed in a future work.

We conclude that the origin of the H$\alpha$ flux observed in the Red Sample is compatible with star formation, and therefore we will include it in the estimation of the H$\alpha$ luminosity function (Sect.~\ref{sect:HaLF}) and the star formation rate density (Sect.~\ref{sect:SFRD}).

\begin{figure}
    \centering
    \includegraphics[width=0.5\textwidth]{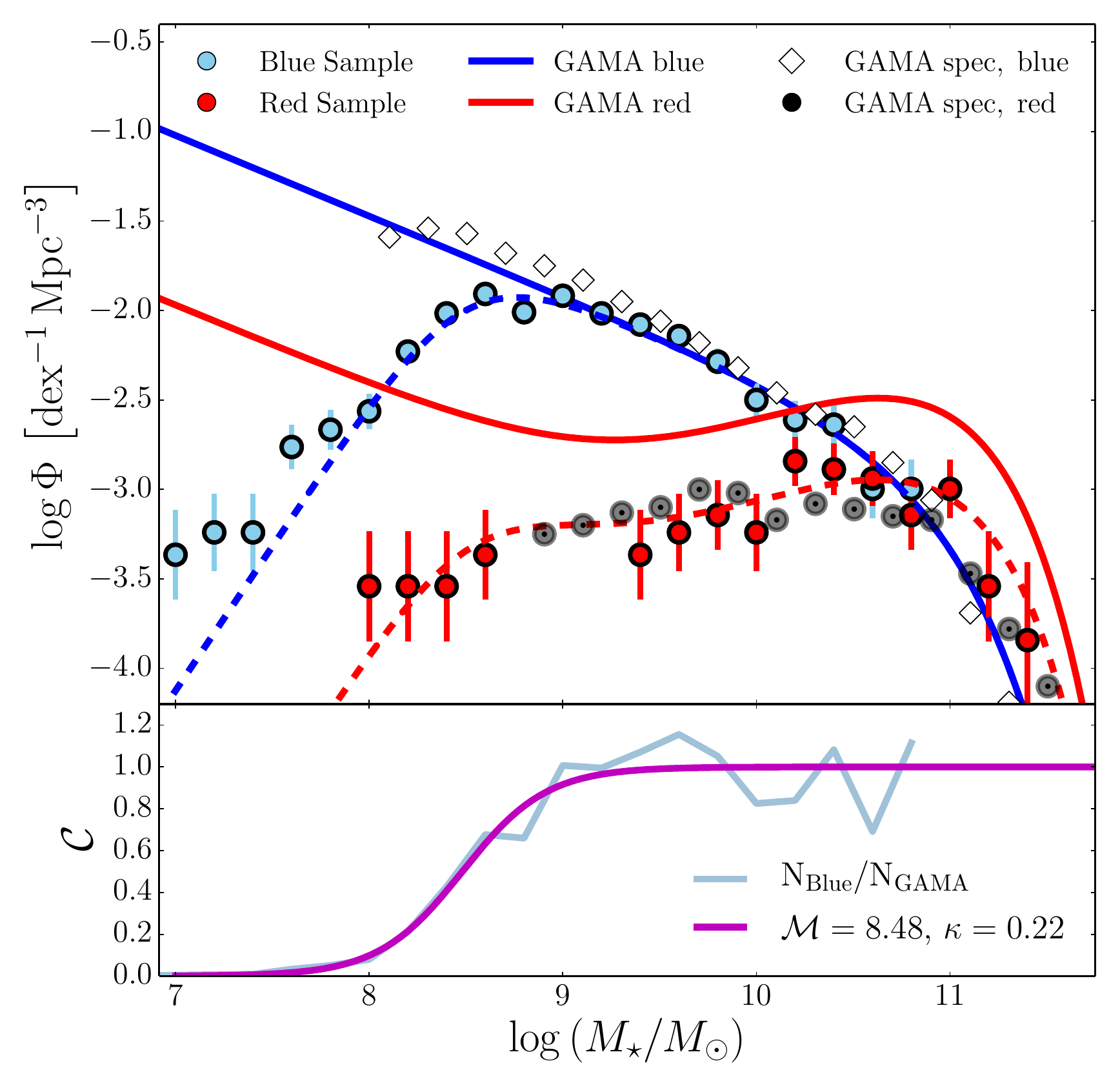}
    \caption{{\it Upper panel}: Stellar mass function for the Blue Sample (blue dots) and the Red Sample (red dots). Blue and Red solid lines represent the SMF for blue and red galaxies respectively, taken from \citet{Baldry2012}. White diamonds and black circles represent the SMF of spectroscopic H$\alpha$ emitters in GAMA, taken from \citet{Guna2015}. Dashed blue and red lines represent the GAMA blue and red SMF from \citet{Baldry2012}, but multiplied by the Blue Sample incompleteness, presented in the lower panel. An extra $0.35$ factor is also applied to the red SMF from \citet{Baldry2012} to match the observed fraction of red galaxies with spectroscopic H$\alpha$ emission from \citet{Guna2015}. {\it Bottom panel}: Stellar mass completeness estimated form the comparison between GAMA blue galaxies and Blue Sample SMF (cyan solid line). The purple line is the best-fitting Sigmoid function to the completeness, whose parameters are labeled in the panel.}
    \label{fig:Mass_function}
\end{figure}

\subsection{Stellar mass function}\label{sect:SMF}
We now study the two main projections of the SFMS along its axis. These are: the H$\alpha$ luminosity function, and the stellar mass function. The first one describes the number of sources that emit a given H$\alpha$ luminosity, per unit volume, and per unit luminosity. The second one describes the number of sources with a given stellar mass, per unit volume and unit mass. The estimations from our data are tabulated in Appendix~\ref{Sect:tablas}.

To estimate the SMF, we used the stellar masses computed during the Monte-Carlo routine described in Sect.~\ref{sect:MonteCarlo} and assumed a cosmological volume estimated from the unmasked area of 897.4 deg$^2$ surveyed by J-PLUS DR1 coupled with a maximum probed distance of $75$ Mpc. The reported errors only account for Poisson uncertainties in the counts.

The stellar mass function of the Blue and Red samples are presented in Fig.~\ref{fig:Mass_function}, and the data are provided in Table~\ref{tab:MF_values}. We found that the Red Sample dominates at masses higher than $\log M_{\star} \sim 10.5$, with the Blue Sample being larger in number density below this mass. The maximum of the Blue Sample SMF is reached at $\log M_{\star} \sim 8.9$, then the number density decreases towards lower stellar masses.

To interpret our results, we compared the SMF from J-PLUS with the SMF reported by \citet{Baldry2012} at $\mathrm{z} < 0.06$ in the GAlaxy and Mass Assembly (GAMA, \citealt{gama}) survey, both for red and blue galaxies as defined by the observed dichotomy in the GAMA $(u-g)$ {\it vs}. $M_{r}$ color-absolute magnitude diagram. We also included the SMF derived by \citet{Guna2015} in GAMA for spectroscopic star-forming H$\alpha$ emitters, both red and blue (Fig.~\ref{fig:Mass_function}). We find that the SMF of the Blue Sample is remarkably similar to blue SMFs from GAMA, both the general and the star-forming population. This suggests that our sample is probing a representative volume of the local Universe, and that the observed maximum in the stellar mass function reflects our completeness in the selection of blue, star-forming galaxies.

Given the good agreement between the J-PLUS and the GAMA SMFs, we used the \cite{Baldry2012} results to estimate the completeness of our sample. To do so, we divided the observed number counts inside each mass bin of our incomplete mass function by the predicted values from the \cite{Baldry2012} fitting. We found that the resulting distribution is well described by a Sigmoid function ({\it bottom panel} in Fig.~\ref{fig:Mass_function}),
\begin{equation}
    \mathcal{C} = \frac{1}{1+e^{-\frac{\log M_{\star}-\mathcal{M}}{\kappa}}},
\end{equation}
where $\mathcal{M}$ is the logarithm of the stellar mass at which the sample is $50\%$ complete. We find $\mathcal{M} = 8.48$ and $\kappa = 0.21$, meaning that our sample is more than 50\% complete in stellar mass at $\log M_{\star} \gtrsim 8.5$ and more than 95\% complete at $\log M_{\star} \gtrsim 9$. This completeness is in agreement with the expectations from the surface brightness analysis presented in Fig.~\ref{fig:mumag}. Later, we use this relation to infer the H$\alpha$ luminosity function weighting the number of sources by their incompleteness in stellar mass (Sect.~\ref{sect:HaLF_mass}).

The shape of the Red Sample is also similar to the red SMF reported by \cite{Baldry2012} but with a difference in the normalization. We qualitatively matched both SMFs by multiplying the red SMF in \cite{Baldry2012} by a factor of $0.35$ (Fig.~\ref{fig:Mass_function}). Moreover, the SMF of the Red Sample closely resembles the \citet{Guna2015} results for red star-forming, H$\alpha$ emitters. The shape and scale agreement with the spectroscopic H$\alpha$ emitters from \citet{Guna2015} further supports the star-forming origin of the H$\alpha$ emission for the Red Sample, as already discussed in Sect.~\ref{sect:pAGB}. In addition to the results from GAMA, \citet{Sobral2011} also find a significant population of red galaxies with H$\alpha$ emission at ${\rm z} = 0.84$. Our findings expand those in \citet{Sobral2011} and \cite{Guna2015} to the local volume.

Disentangling the physical origin of the Red Sample is beyond of the scope of the present paper, and we explore the morphological properties of the Red and Blue samples in the companion paper by \citet{Logrono2020} to get more clues on this regard.

\begin{figure*}
    \centering
    \includegraphics{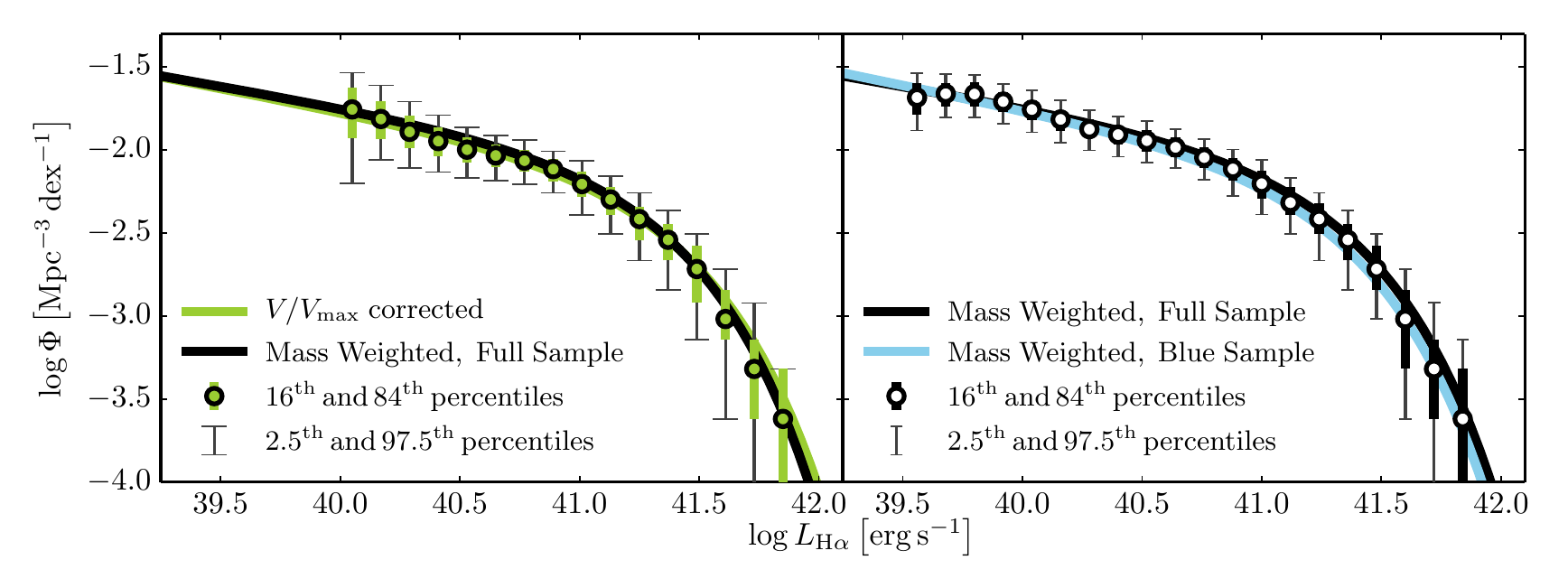}
    \caption{H$\alpha$ luminosity functions estimated from J-PLUS DR1 at $\mathrm{z} \leq 0.017$. {\it Left panel}: H$\alpha$LF estimated by the $V_{\rm int}/V_{\rm max}$ technique from the Full Sample (green dots), accompanied with the best-fitting Schechter distribution to these data (green line). The black line in this plot uses the same data, but a different weighting technique based on the Stellar Mass Function. {\it Right panel}: Empty dots are the H$\alpha$LF estimated using only the Blue Sample, together with the best-fitting Schechter distribution to these data, once weighted by their stellar mass (blue line). For comparison we show the best Schechter fitting using the mass-weighted Full Sample (black line, in common with left panel black line). We see that neither the weighting criteria nor the the sample have a major impact on the best-fitting distribution. However, the mass-weighted samples allow us to explore $0.5$ dex the faint-end slope of the distribution.}    
    \label{fig:HaLFs}
\end{figure*}

\subsection{H$\alpha$ luminosity function in the local Universe}\label{sect:HaLF}
In this Section, we present the star-forming H$\alpha$ luminosity function derived from J-PLUS DR1 data at $d \lesssim 75$ Mpc. We applied two different weighted schemes to account for volume and flux incompleteness, one based on the stellar mass function (Sect.~\ref{sect:HaLF_mass}) and the other on the $V_{\rm int}/V_{\rm max}$ technique (Sect.~\ref{sect:HaLF_vvmax}).

\subsubsection{Mass-weighted H$\alpha$ luminosity function}\label{sect:HaLF_mass}
Like in previous sections, we used a Monte-Carlo approach to infer the mass-weighted H$\alpha$LF. In this case, the root of the procedure remains the same than in Sect.~\ref{sect:MonteCarlo}. The main addition to the process is that, in each iteration of the Monte-Carlo sampling, and after computing the stellar masses of each galaxy, we fitted a Sigmoid function to the comparison between the resulting stellar mass distribution and the GAMA stellar mass function (Sect.~\ref{sect:SMF}). This fitting was used as a completeness proxy for that iteration of the process, and with it we computed a mass weight for each galaxy. These weights were used to create the mass-weighted distribution of H$\alpha$ luminosities. We note that the mass weight can be used as a proxy for the H$\alpha$ luminosity incompleteness because of the SFMS, that closely relate both quantities.

 We present the J-PLUS H$\alpha$LF at $d \lesssim 75$ Mpc in the {\it right panel} of Fig.~\ref{fig:HaLFs}, and the data are provided in Table~\ref{tab:FSMWHaLF}. We proved luminosities of $\log L_{\mathrm{H}\alpha} \gtrsim 39.5$. We compute it for the Blue Sample and the Full Sample. Even if we find that the star formation properties of the Red and Blue sample are different, in both cases we are probing the current star formation in the source (Sects.~\ref{sect:pAGB} and \ref{sect:SMF}). As expected from the results in Sect.~\ref{sect:SFMS}, the Blue Sample dominates the star-forming population at the faint end, and the Red Sample increases the density of star-forming galaxies at the bright end.

We fitted the observed H$\alpha$LFs with a \cite{Schechter76} distribution. This is expressed as
\begin{equation}\label{eq:schechter}
  \Phi \left(L_{\mathrm{H}\alpha}\ |\ L_{\mathrm{H}\alpha}^{*}, \phi^{*}, \alpha \right) \mathrm{d}L_{\mathrm{H}\alpha} = \phi^{*}\left(\frac{L_{\mathrm{H}\alpha}}{L_{\mathrm{H}\alpha}^{*}}\right)^{\alpha}e^{-\frac{L_{\mathrm{H}\alpha}}{L_{\mathrm{H}\alpha}^{*}}}\frac{\mathrm{d}L_{\mathrm{H}\alpha}}{L_{\mathrm{H}\alpha}^{*}}\, ,
 \end{equation}
where $L_{\mathrm{H}\alpha}^{*}\ \rm{[erg\,s^{-1}]}$, $\phi^{*}\ \rm{[Mpc^{-3}]}$, and $\alpha$ are the parameters that define the distribution.

This fitting was performed with \texttt{emcee}, using a sample of $20$ walkers, $2\,000$ sampling steps, and a burn-in phase of $1\,000$ steps. We did this $600$ times, and each time we stored $1\,000$ sampling points. In the end, we had $600\,000$ sampling points to draw the PDF of each parameter in the Schechter distribution. The approach that we have described is convenient to take into account the potential degeneracy between the parameters. The resulting best-fitting parameters for the Full Sample are:

\begin{eqnarray}
    \log L_{\mathrm{H}\alpha}^{*} = 41.34^{+0.12}_{-0.10},\\
    \log \phi^{*} = -2.43^{+0.11}_{-0.13},\\
    \alpha = -1.25 \pm 0.07.
\end{eqnarray}

We compare our values with previous work in the literature in Sect.~\ref{sect:HaLF_discussion}.

\subsubsection{$V_{\rm int}/V_{\rm max}$ H$\alpha$ luminosity function}\label{sect:HaLF_vvmax}
We re-computed the H$\alpha$LF from a different approach, and we compare the results to the previous one. In this approach, we did not use the stellar masses to asses the incompleteness of a galaxy, and we used instead the classical $V_{\rm int}/V_{\rm max}$ technique described in \cite{Schmidt1968} and \cite{Huchra1973}. We explain the idea behind this correction in detail in Appendix~\ref{sect:VintVmax}.

As a summary, for a given limiting H$\alpha$ flux, the $V_{\rm int}/V_{\rm max}$ technique accounts for the larger volume probed by the more luminous galaxies, that can be observed to larger distances. This effect is coupled with the completeness in the H$\alpha$ detection flux, that decreases the number of sources at faint fluxes. The stellar mass weight presented in Sect.~\ref{sect:SMF} and applied in the previous section is intended to account for both effects, and we test here this assumption.

We found that the distribution in H$\alpha$ flux has a \textit{plateau} that starts around $\log F_{\mathrm{H}\alpha} \sim -13$ (Fig.~\ref{fig:panel_caracterizacion}). For fainter emission fluxes, we are not able to recover all the sources. We set the limiting flux for the $V_{\rm int}/V_{\rm max}$ analysis in $\log F_{\mathrm{H}\alpha}^{\rm lim} = -13.2$ and varied it in the Monte-Carlo runs by $\pm 0.1$ dex and $\pm0.2$ dex, so that in the end, the distribution had been fitted $250$ times for each of the $5$ different values of $\log F_{\mathrm{H}\alpha}^{\rm lim}$. To fit the Schechter distributions, we used the same \texttt{emcee} walkers and steps. This time, we ended up with $1\,250\,000$ sampling points to describe the final best-fitting parameters, and their uncertainties.

The derived H$\alpha$LF for the Full Sample is presented in the {\it left panel} of Fig.~\ref{fig:HaLFs}, together with the mass-weighted results in the previous section. We found an excellent agreement between both methodologies down to $\log L_{\mathrm{H}\alpha} \sim 40$. The mass-weighted LF reach 0.5 dex fainter luminosities than the $V_{\rm int}/V_{\rm max}$ LF because of the completeness scheme used, that in addition to volume effects also include information about H$\alpha$ flux incompleteness. Therefore, fluxes below $\log F_{\mathrm{H}\alpha}^{\rm lim} = -13.2$ are used and fainter luminosities reached. As a reference, the average number of galaxies analyzed in the $V_{\rm int}/V_{\rm max}$ case is 325, that increases to 720 in the mass-weighted case.

We conclude that the mass-weighted results in previous section are similar with the well established $V_{\rm int}/V_{\rm max}$ technique, but reaching $0.5$ dex fainter luminosities. 

\subsubsection{Impact of AGN on the H$\alpha$ luminosity function}
The H$\alpha$LFs presented in previous sections only refers to star-forming galaxies. However, we could include in this observable the H$\alpha$ emission coming from galaxies targeted as AGN. In this case, the H$\alpha$LF did not account for the origin of the H$\alpha$ photons, being independent of the astrophysical process that causes them. Including 49 AGN in the analysis (Sect.~\ref{sect:AGN}), the robustness of the two methods would still hold, while our resulting best-fitting parameters in the mass-weighted case would have changed by $+0.03$ dex in $L_{{\rm H}\alpha}^*$, $+0.05$ dex in $\phi^*$, and $+0.04$ in $\alpha$. These changes are always compatible with the star-forming H$\alpha$LF at $1\sigma$ level.

\subsection{Star formation rate density}\label{sect:SFRD}
In this Section, we aim to infer the value of the star formation rate density at $d \lesssim 75$ Mpc. We defined the SFRD as \begin{equation}
\rho_{\mathrm{SFR}} = 7.9\cdot10^{-42} \mathcal{L}_{{\rm H}\alpha}\,\quad \mathrm{[M_{\odot}\,yr^{-1}\,Mpc^{-3}]},
\end{equation}
where $ \mathcal{L}_{\mathrm{H}\alpha}$ is the luminosity-weighted integral of the star-forming, mass-weighted H$\alpha$LF estimated in Sect.~\ref{sect:HaLF_mass},
\begin{equation}\label{integral_LF_SFRD}
     \mathcal{L}_{\mathrm{H}\alpha} = \int_{0}^{\infty}\!\!\!\Phi \left( L_{\mathrm{H}\alpha} \right) L_{\mathrm{H}\alpha}  {\rm d}L_{\mathrm{H}\alpha} = \phi^* L_{\rm{H}\alpha}^* \Gamma(\alpha + 2)\,\quad\mathrm{[erg\,s^{-1}\,Mpc^{-3}]},
 \end{equation}
where $\Gamma$ is the Gamma function. We integrated down to zero to facilitate the computation and because the estimated $\alpha$ value implies a small contribution to the total density from the lower luminosities. As an example, integrating down to $\log L_{\rm{H}\alpha} = 39.5$ or $37$ decreases the final density by 0.020 or 0.002 dex, respectively. In the end, we obtain 
\begin{equation}
    \log \rho_{\mathrm{SFR}} = -2.10 \pm 0.11.
\end{equation}
The retrieved error includes the statistical uncertainty ($\pm 0.04$ dex) and the impact of cosmic variance ($\pm 0.10$ dex), as detailed in the next section. We find that 15\% ($0.05$ dex) of the SFRD is located in the Red Sample. This is a significant contribution to the current star formation rate.

We discuss this result and the relation with previous determinations in Sect.~\ref{sect:SFRD_discussion}.

\subsubsection{Impact of comic variance}
The cosmic or sample variance (CV) is the excess variance with respect to a Poissonian process that emerges from the clustered nature of galaxies \citep[e.g.][]{somerville04,moster11}. The smaller the probed volume, the larger the impact of cosmic variance is, and measurements can significantly depart for the desired median value in the Universe. The volume at $\mathrm{z} \leq 0.017$ that is explored with the $897.4$ deg$^2$ of J-PLUS DR1 accounts for $34\,757.7$ Mpc$^3$. This is a relative small volume in a cosmological context, so the cosmic variance should impact our results.

The cosmic variance at a given luminosity is estimated as
\begin{equation}
    \sigma_{\rm CV}\,(L) = B\,(L) \, \sigma_{\rm dm},
\end{equation}
where $\sigma_{\rm dm}$ is the cosmic variance of the underlying dark matter distribution that is common to any galaxy, and $B\,(L)$ is the galaxy bias of the targeted population \citep{somerville04,moster11,robertson10,clsj15cv}. In addition, the cosmic variance imprints a large correlation between galaxies of different luminosities, leading to significant off-diagonal terms in the covariance matrix ($\Sigma$) of the luminosity function \citep{Smith2012}. Formally,
\begin{equation}
    \Sigma = \Sigma_{\rm P} + \Sigma_{\rm CV},
\end{equation}
where $\Sigma_{\rm P}$ accounts for the Poissonian uncertainties and
\begin{equation}
    \Sigma_{\rm CV}\,(L_1,L_2) = B(L_1)\,B(L_2)\,\sigma^2_{\rm dm}.
\end{equation}
The correlation between different luminosities due to the cosmic variance is typically $0.4 - 0.8$ \citep{Smith2012, clsj2017, kawin20}. We decided to not include the cosmic variance in the analysis of the H$\alpha$LF in Sect.~\ref{sect:HaLF} because the luminosity-dependent bias of H$\alpha$ emitters in the local Universe is currently unknown to our best knowledge, and a proper cosmic variance analysis is therefore unfeasible.

The luminosity-weighted integral in Eq.~(\ref{integral_LF_SFRD}) simplifies the problem, and the SFRD cosmic variance becomes equivalent to
\begin{equation}
    \sigma_{\rm SFRD,\,CV} = B_{\rm eff}\ \sigma_{\rm dm}, 
\end{equation}
where
\begin{equation}
     B_{\rm eff} = \frac{1}{\mathcal{L}_{\mathrm{H}\alpha}}\int_{0}^{\infty}\!\!\!B(L_{\rm{H}\alpha})\,\Phi \left( L_{\mathrm{H}\alpha} \right) L_{\mathrm{H}\alpha}  {\rm d}L_{\mathrm{H}\alpha}
 \end{equation}
is the number- and luminosity-weighted effective bias of the population under study. Because the peak of the H$\alpha$ luminosity density occurs at $\sim L_{\rm{H}\alpha}^*$ \citep{gama13}, the cosmic variance at this luminosity is a good proxy for the cosmic variance of the SFRD.

Following the reasoning above, we used the empirical prescription in \citet{driver10} as a proxy to estimate the impact of the cosmic variance in our determination of the SFRD, because it is based on $M_r = M_r^{*} \pm 1$ magnitude galaxies with $m_{r} \lesssim 18$ at ${\rm z} < 0.1$.  The cosmic variance is mainly driven by the probed cosmological volume, but the survey geometry also plays a relevant role. Assuming a squared, contiguous area, we estimated a $48\%$ ($0.17$ dex) cosmic variance in J-PLUS DR1, while the addition of $511$ independent square fields of $2$ deg$^2$ provides a 6\% ($0.02$ dex) cosmic variance. These are two extreme values, and the actual J-PLUS footprint has several, large contiguous areas, with a fraction of scattered fields (see \citealt{Cenarro2019}). To better estimate the cosmic variance in our study, we split the J-PLUS DR1 area in a series of contiguous, nearly independent pieces, and estimated the cosmic variance in each one after accounting by their geometry. Then, we combined them under independence assumption following \citet{moster11}. With this, we reached a $25\%$ ($0.10$ dex) uncertainty introduced by cosmic variance for the J-PLUS DR1 at $\mathrm {z} \leq 0.017$.

Compared with the statistical error of $0.04$ dex, the cosmic variance is the leading source of uncertainty in our SFRD measurement.

\section{Discussion}\label{sect:Discussion}
In the following sections, we compare our results about star formation in the local Universe with previous work in the literature. This will put our estimations in context, and permits the evaluation of possible systematic uncertainties in our measurements. 

\subsection{Systematic errors on H$\alpha$ flux estimation}\label{sect:polvo}
We start the discussion by looking at our H$\alpha$ measurements. The extraction of H$\alpha$ flux from J-PLUS photometry, including [\ion{N}{ii}] removal and dust de-reddening, is extensively tested in VR15 with synthetic data and in \citet{Logrono2019} with J-PLUS early data. The conclusion reached by these studies is that we are able to measure $F_{\mathrm{H}\alpha}$ without bias and with a minimum error of $20\%$, mainly driven by the statistical dust and [\ion{N}{ii}] corrections.

We expanded the analysis in the above papers with two extra tests. First, we repeated the study in \citet{Logrono2019} with the extended data provided by J-PLUS DR1. That added $99$ new comparison regions with SDSS spectroscopic measurements. The total number of spectroscopic regions analyzed was 
$125$ from SDSS and $20$ from CALIFA. The obtained results mirrors the initial findings by \citet{Logrono2019}, and we refer the reader to this work for extra information.

\begin{figure}
    \centering
    \includegraphics{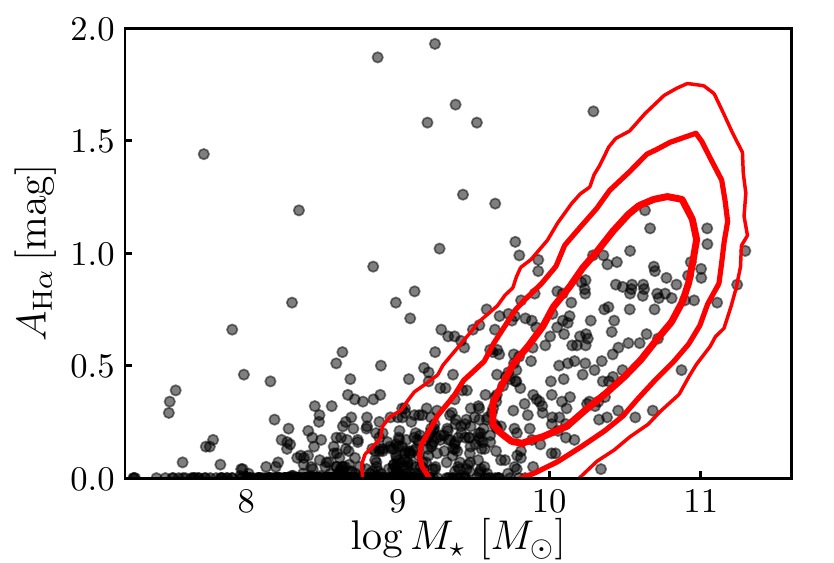}
    \caption{Relation between H$\alpha$ extinction, $A_{\mathrm{H}\alpha}$, and stellar mass in the Full Sample. The red contours depict the 1, 2, and 3$\sigma$ density of sources in SDSS spectroscopic data presented by \citet{DuartePuertas2017}.}
    \label{fig:Polvo_Duarte_Purtas}
\end{figure}

Second, we studied the distribution of H$\alpha$ extinction, $A_{\mathrm{H}\alpha}$, as a function of stellar mass \citep[see][]{GarnBest2010} in the Full Sample, as shown in Fig.~\ref{fig:Polvo_Duarte_Purtas}. This analysis was motivated by the large discrepancy in the median extinctions reported by \citet{Gallego1995}, \citet{Nakamura2004}, and \citet{DuartePuertas2017}; $A_{\mathrm{H}\alpha} \sim 0.85$, when compared with the median extinction in our sample, $A_{\mathrm{H}\alpha} \sim 0.15$. We find that our data closely follows the expected extinction {\it vs}. stellar mass relation estimated by \citet{DuartePuertas2017} using aperture-corrected SDSS spectroscopic data (see also \citealt{Gilbank2010, GarnBest2010}), and that the apparent discrepancy in $A_{\mathrm{H}\alpha}$ is just a reflection of the lower stellar masses probed by our sample. This result strengths the statistical dust correction presented in VR15.

We conclude that the initial results from VR15 and \citet{Logrono2019} about the reliability and accuracy of the J-PLUS estimation of H$\alpha$ fluxes have been reinforced with the analysis presented in this section.

\begin{figure*}
    \centering
    \includegraphics[width=0.95\textwidth]{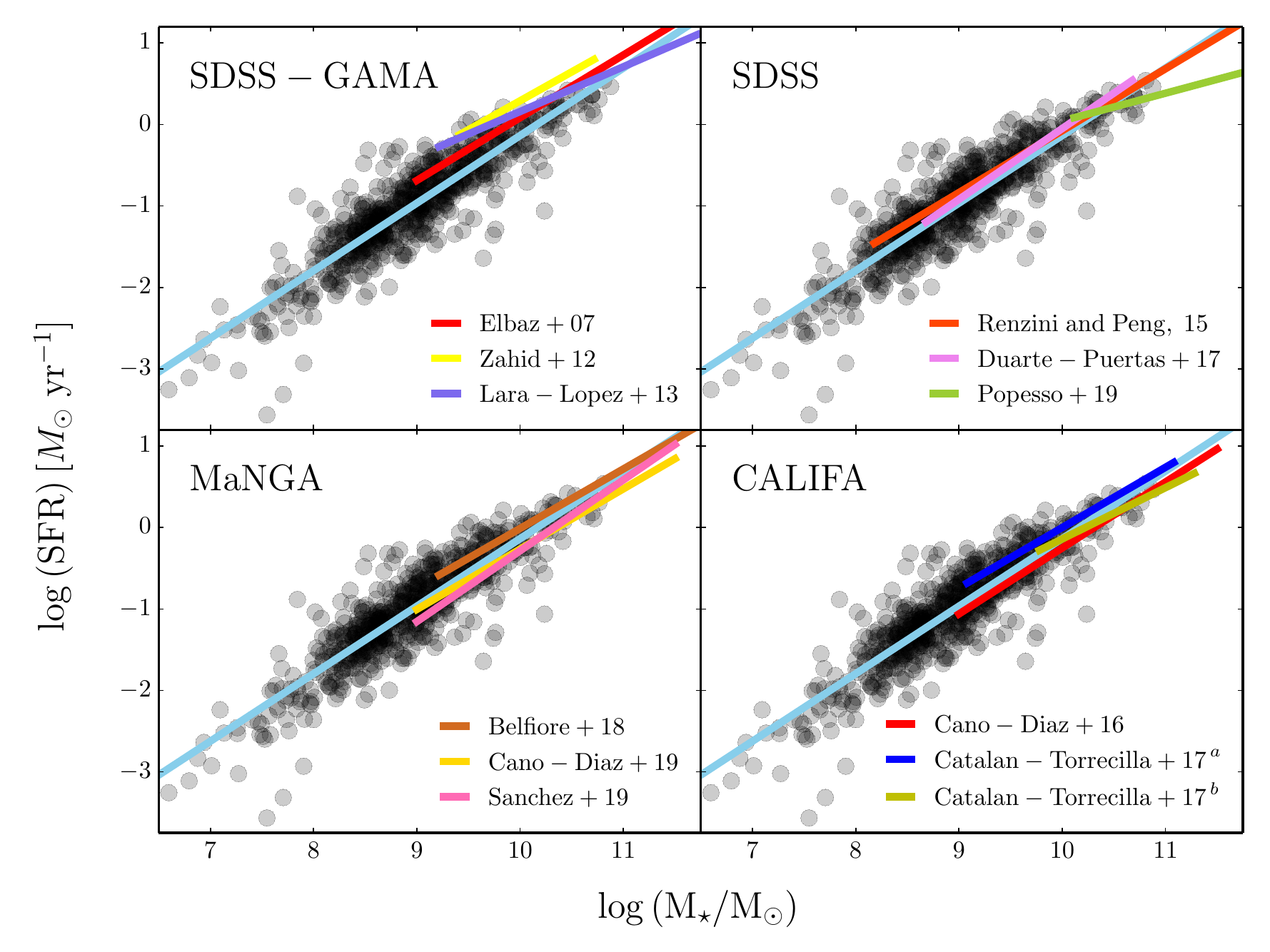}
    \caption{Star-formation main sequence in J-PLUS (bullets and cyan line) and in previous work in the literature. The source papers are labeled in the panels and summarized in Table.~\ref{tab:SFMS_literatura}. The draw lines cover the stellar mass range spanned by each study.}
    \label{fig:panel_SFMS}
\end{figure*}

\begin{table*}[]
    \caption{Compilation of local ($\mathrm{z} < 0.1$) SFMS based on H$\alpha$ as star formation tracer.}
    \centering
        \begin{tabular}{llccccc}
    \hline\hline\noalign{\smallskip}
         Reference & Survey & Redshift range & $\langle \mathrm{z} \rangle$  & $a$ & $b$ & $\Delta b_{\mathrm{z}}$\tnote{$\dagger$} \\
    \noalign{\smallskip}\hline\noalign{\smallskip}    
        This Work, Blue Sample                &  J-PLUS - imaging              & $0.001<\mathrm{z}<0.017$  & 0.012  & $0.83$   & $-8.44$  & $--$    \\              
        \cite{Elbaz07}                        &  SDSS - fiber                  & $0.04<\mathrm{z}<0.10$    & 0.077  & $0.77$   & $-7.53$  & $-0.087$ \\
        \cite{Zahid2012}                      &  SDSS - fiber                  & $0.04<\mathrm{z}<0.10$    & 0.070  & $0.71$   & $-6.73$  & $-0.078$ \\   
        \cite{Gavazzi13,Gavazzi15}            &  H$\alpha$3 - imaging          & $0.001<\mathrm{z}<0.03$   & 0.020  & $--$     & $--$     & $-0.011$ \\
        \cite{LaraLopez2013}, V1              &  SDSS \& GAMA - fiber          & $0.04<\mathrm{z}<0.10$    & 0.077  & $0.55$   & $-5.25$  & $-0.087$ \\  
        \cite{RenziniPeng2015}                &  SDSS - fiber                  & $0.02<\mathrm{z}<0.085$   & 0.067  & $0.76$   & $-7.60$  & $-0.074$ \\  
        \cite{CanoDiaz2016}                   &  CALIFA - IFS                  & $0.005<\mathrm{z}<0.03$   & 0.020  & $0.81$   & $-8.34$  & $-0.011$ \\  
        \cite{DuartePuertas2017}              &  SDSS - fiber                  & $0.005<\mathrm{z}<0.05$   & 0.040  & $0.87$   & $-8.71$  & $-0.037$ \\   
        \cite{CatalanTorrecilla2017}          &  CALIFA - IFS - Sb/Sbc         & $0.005<\mathrm{z}<0.03$   & 0.020  & $0.74$   & $-7.39$  & $-0.011$ \\   
        \cite{CatalanTorrecilla2017}          &  CALIFA - IFS - Sc/Sdm         & $0.005<\mathrm{z}<0.03$   & 0.020  & $0.63$   & $-6.43$  & $-0.011$ \\  
        \cite{McGaugh17}                      &  LSB galaxies - imaging        & $0.001<\mathrm{z}<0.017$  & 0.012  & $1.04$   & $-10.77$ & $--$     \\  
        \cite{Belfiore2018}                   &  MaNGA - IFS                   & $0.01<\mathrm{z}<0.15$    & 0.030   & $0.73$   & $-7.29$  & $-0.025$ \\  
        \cite{Popesso2019a}                   &  SDSS - WISE + fiber           & $0.01<\mathrm{z}<0.085$   & 0.067   & $0.34$   & $-3.28$  & $-0.074$ \\  
        \cite{CanoDiaz2019}                   &  MaNGA - IFS                   & $0.005<\mathrm{z}<0.15$   & 0.030  & $0.74$   & $-7.64$  & $-0.025$ \\   
        \cite{Sanchez2019}                    &  MaNGA - IFS                   & $0.005<\mathrm{z}<0.15$   & 0.030  & $0.87$   & $-8.96$  & $-0.025$ \\   
    \hline 
        \end{tabular}
    \label{tab:SFMS_literatura}
\end{table*}

\subsection{Star formation main sequence}\label{sect:SFMS_discusion}
\begin{figure*}
    \centering
    \includegraphics{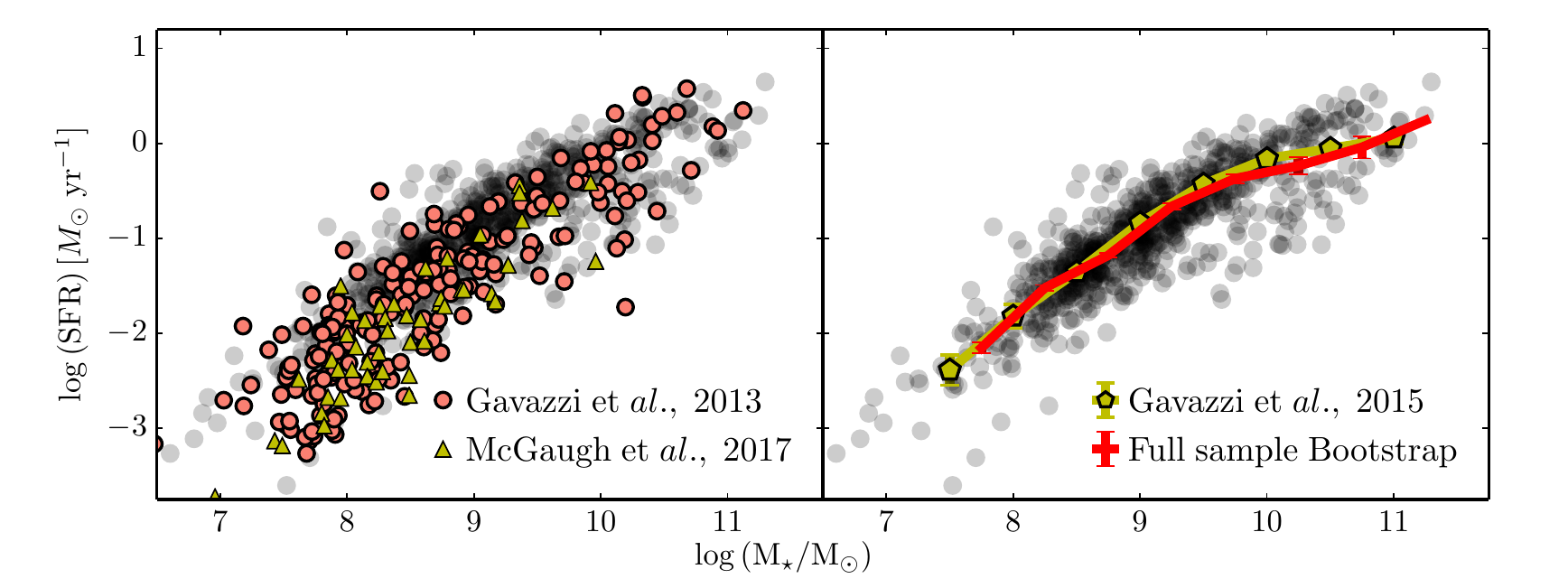}
    \caption{Star formation rate {\it vs}. stellar mass in the J-PLUS DR1 Full sample (bullets) in comparison with previous work based on H$\alpha$ imaging. {\it Left panel}: Results from H$\alpha$3 data in the Local Supercluster area (circles, \citealt{Gavazzi13}), and for LSB galaxies at $\mathrm{z} \leq 0.017$ (triangles, \citealt{McGaugh17}). {\it Right panel}: Median $\log \mathrm{SFR}$ as a function of stellar mass in the Local Supercluster and the Coma cluster derived by \citet{Gavazzi15}. The red solid line marks the median $\log \mathrm{SFR}$ estimated from the Full Sample.}
    \label{fig:Gavazzi_Bending}
\end{figure*}

We compare here the estimation of the Blue sample SFMS with prior determinations in literature. The number of studies regarding the SFMS at low redshift ($\mathrm{z} \lesssim 0.1$) is large, and we present the comparisons with spectroscopic work based on H$\alpha$ as SFR tracer in Fig.~\ref{fig:panel_SFMS}. The comparison with studies based on H$\alpha$ imaging is presented in Fig.~\ref{fig:Gavazzi_Bending}. The main characteristics of each study are summarized in Table~\ref{tab:SFMS_literatura}. For this comparison, stellar masses and SFRs were converted to a \citet{Salpeter55} IMF.

We also accounted for the time evolution in the normalization of the SFMS down to our median redshift, $\mathrm{z} = 0.012$, following the $(1+\mathrm{z})^{3.21}$ evolution found by \citet{Popesso2019b}. We denoted such evolution as $\Delta b_{\mathrm{z}}$, and the assumed values are reported in Table~\ref{tab:SFMS_literatura}. The use of other suggested evolution for the SFMS (i.e. \citealt{Speagle14}) does not alter the conclusions in this section.

\begin{figure}
    \centering
    \includegraphics{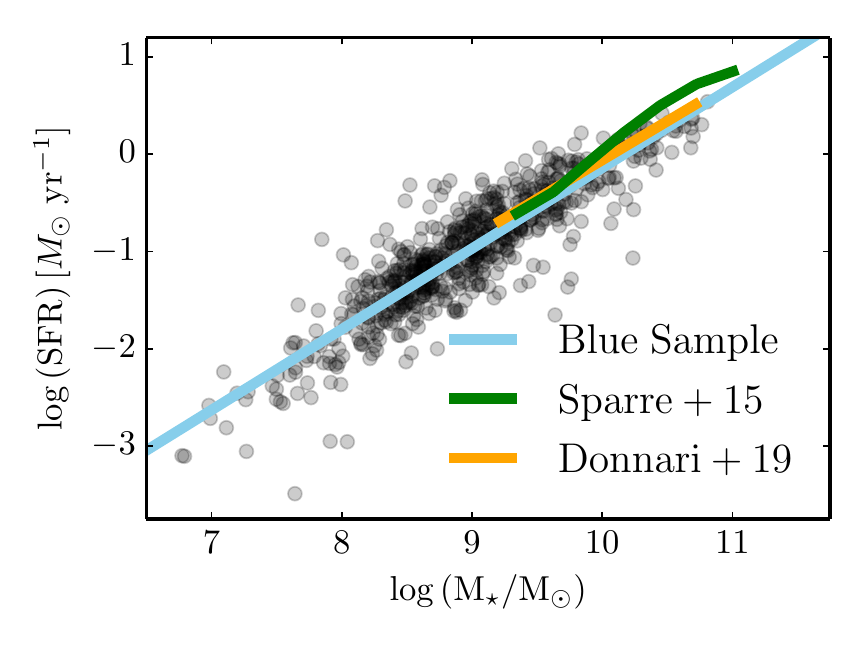}
    \caption{Star formation main sequence in J-PLUS (bullets and blue line) compared with the expectations from the cosmological hydro-dynamical simulations Illustris \citep[][green line]{Sparre2015} and IllustrisTNG \citep[][orange line]{Donnari2019}.}
    \label{fig:Illustris}
\end{figure}

We focus first on spectroscopic studies from SDSS, GAMA, CALIFA, and MaNGA surveys. On the one hand, regardless of the study we compare with, there is strong consistency between our slope, $a = 0.83$, and prior determinations of this value, well constrained between $a\sim0.7$ and $a\sim0.9$. This robustness is also pointed out in the study by \cite{CanoDiaz2016}. On the other hand, the normalization factor shows more variance. In particular, this is the main source of discrepancy between our work and some previous fiber-based spectroscopic surveys. We consider that the most probable explanation for this difference is the aperture corrections.

Our image-based measurements, as well as most IFS surveys, do not require aperture corrections to obtain the total H$\alpha$ flux of a galaxy. Those measurements based on fibers or long-slit spectroscopy demand an aperture correction to account for the missing H$\alpha$ flux outside the available aperture \citep[e.g.][]{Brinchmann04}. The comparison of our SFMS with those derived with CALIFA \citep{CanoDiaz2016, CatalanTorrecilla2017} and MaNGA \citep{Belfiore2017b, CanoDiaz2019, Sanchez2019} data is satisfactory, while some of the fiber-based measurements from SDSS provide larger SFRs than expected by our measurements even after accounting for time evolution \citep{Elbaz07, Zahid2012, LaraLopez2013}. We note that the average redshift from IFS studies ($\mathrm{z} \sim 0.02 - 0.03$) is closer to our probed volume than those based on fiber spectroscopy ($\mathrm{z} \sim 0.07$). The studies of \citet{RenziniPeng2015} and \citet{DuartePuertas2017} are also based on SDSS fiber measurements, and are in much better agreement with our results. We highlight the work of \citet{DuartePuertas2017}, that uses an improved aperture correction based on CALIFA data \citep{iglesias-Paramo2016} to recover the total flux from SDSS fiber spectroscopy. The excellent agreement with our SFMS supports their aperture correction.

We continue the discussion by comparing our results with the H$\alpha$ imaging estimations from \citet{Gavazzi13} and \citet{Gavazzi15}, as presented in Fig.~\ref{fig:Gavazzi_Bending}. They are based on H$\alpha$3 survey data, and we use their $g$ and $i$ magnitudes when possible to estimate the stellar mass of the galaxies following Eq.~(\ref{eq:mass}). For a detailed explanation about H$\alpha$3 observations and H$\alpha$ flux determinations, we refer the reader to \citet{Gavazzi12}. We find an excellent agreement with these studies, covering J-PLUS and H$\alpha$3 the same parameter space. It is apparent on the {\it left panel} in Fig.~\ref{fig:Gavazzi_Bending} that the Red Sample population is also present in the H$\alpha$3 data, which is HI-selected from the HI Arecibo legacy fast ALFA (ALFALFA, \citealt{alfalfa}) survey, so a prevalence of gas-rich, star-forming galaxies is expected. This supports that the H$\alpha$ emission in the Red Sample has a star-forming origin, as discussed in Sect.~\ref{sect:pAGB}. We also note that the results from \citet{Gavazzi15} on the {\it right panel} in Fig.~\ref{fig:Gavazzi_Bending} suggests a bending in the high-mass end ($\log M_{\star} \gtrsim 10$) of the SFMS. Such bending have been studied in several works \citep{Lee2015,Popesso2019a}, and it produces a lower slope ($a \sim 0.4$) at the high-mass end than on less massive samples ($a \sim 0.8$). The high-mass bending is also appreciable in our data when the Full Sample is analyzed, but it is not apparent from the Blue Sample alone. This implies that the Red Sample composed by lower SFR galaxies has a measurable impact in the SFMS at the high-mass end. We also found that the result from \citet{Popesso2019a}, estimated at $\log M_{\star} \geq 10$, provides similar SFRs in the shared mass range despite the large difference in the slope between both studies ($a = 0.83$ {\it vs}. $a = 0.34$), warning about the direct comparison between the SFMS parameters obtained from the fitting to the data. Finally, we analyse the morphological properties of the Red Sample in the forthcoming paper by \citet{Logrono2020} to get further clues on this regard.

\begin{figure*}
    \centering
    \includegraphics{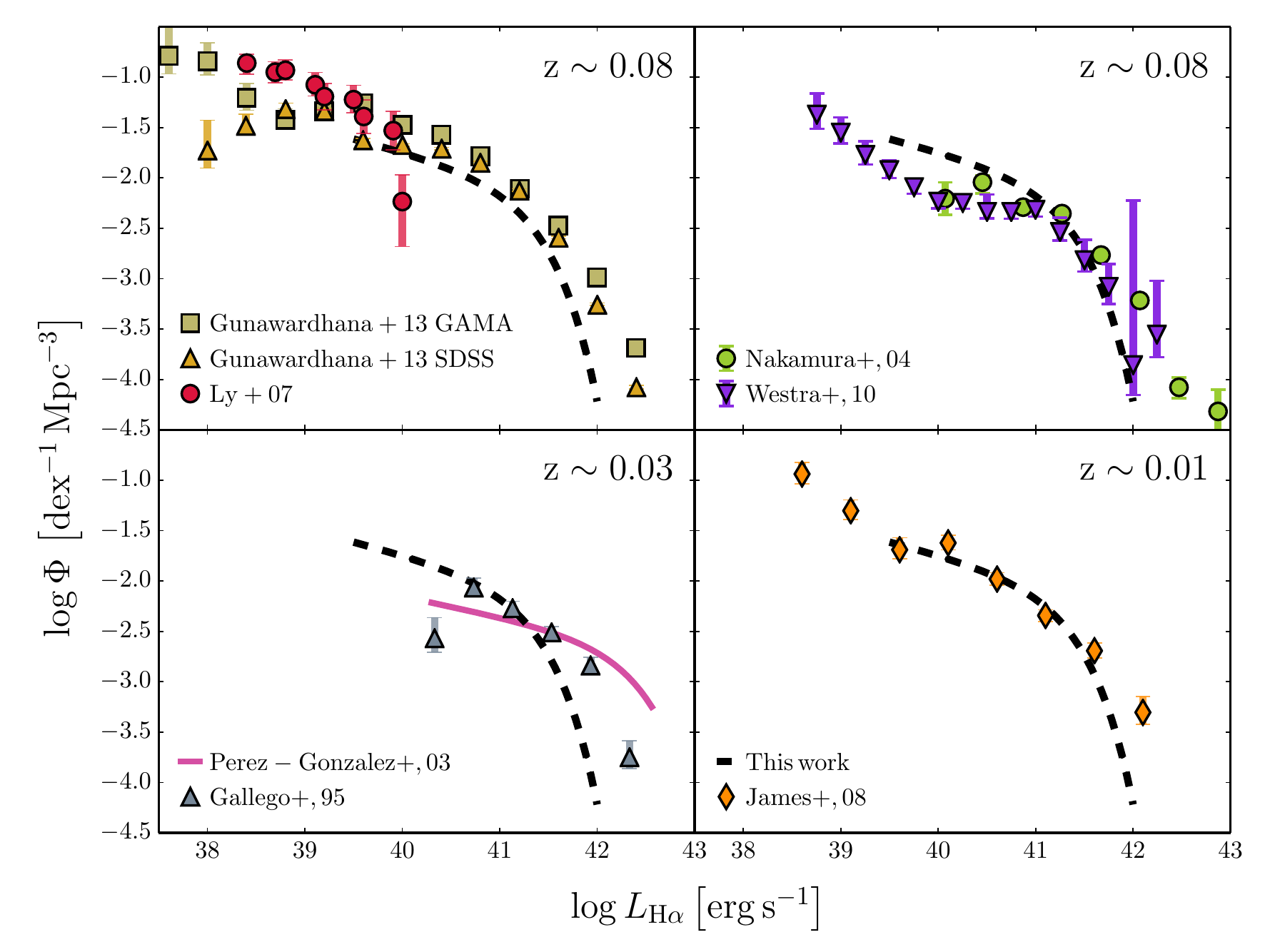}
    \caption{H$\alpha$ luminosity function in J-PLUS DR1 at $\mathrm{z} \leq 0.017$ (black dashed line) and previous estimations from the literature at $\mathrm{z} \leq 0.1$. The source papers and their legends are labeled in the corresponding panels.}
    \label{fig:panel_LF}
\end{figure*}

The data from \citet{McGaugh17} probes the low surface brightness (LSB) regime. Interestingly, the LSB galaxies studied by \citet{McGaugh17} cover the low SFR region of our SFMS. This suggests that our sample is not biased against LSB systems, reinforcing the surface brightness analysis presented in Sect.~\ref{sect:GlobalDistributions}.

Finally, in Fig.\ref{fig:Illustris}, we compare our observational SFMS in the local Universe with the expectations from the cosmological hydro-dynamical simulations Illustris \citep{Sparre2015} and IllustrisTNG \citep{Donnari2019}. These simulations probe galaxies with stellar masses larger than $\log M_{\star} \gtrsim 9$. We find an excellent agreement with the latest IllustrisTNG expectations. The predictions from the original Illustris suite over-predict the SFR at $\log M_{\star} \gtrsim 10$, as reported by \citet{Donnari2019}, and are in agreement with our observations at lower stellar masses.

As summary, the comparison with previous SFMS estimations presented in this section support our results. We highlight the good agreement with IFS surveys and with the H$\alpha$ imaging work of \citet{Gavazzi13} and \citet{Gavazzi15}. The comparison of the stellar mass function in our sample with the local estimation from GAMA \citep{Baldry2012} is detailed in Sect.~\ref{sect:SMF}, so we move to the H$\alpha$ luminosity function in the next Section. 

\begin{table*}[]
    \caption{Compilation of the Schechter parameters for local ($\mathrm{z} < 0.1$) H$\alpha$ luminosity functions}
    \centering
        \begin{tabular}{lcccc}
    \hline\hline\noalign{\smallskip}
        Refrence & $\log L_{\mathrm{H}\alpha}^{\star}$     & $\alpha$     & $\log \phi^{\star}$     \\ \noalign{\smallskip}
                 & [$\mathrm{erg\,s^{-1}}$]  &   &  [$\mathrm{Mpc^{-3}}$] \\
    \noalign{\smallskip}\hline\noalign{\smallskip}
        This work  &  $41.34 \pm^{0.12}_{0.10}$  &  $-1.25\pm^{0.07}_{0.07}$   & $-2.43 \pm^{0.11}_{0.13}$  \\
        \cite{Gallego1995}\tnote{a}  &  $41.87 \pm 0.08$  &  $-1.30 \pm 0.20$   & $-2.76\pm0.03$ \\ 
        \cite{PerezGonzalez2003}    &  $42.43 \pm 0.17$  &  $-1.20 \pm 0.20$   & $-3.00\pm0.20$  \\ 
        \cite{Nakamura2004}          &  $41.99 \pm 0.10$  &  $-1.43 \pm 0.10$   & $-3.02\pm0.17$ \\ 
        \cite{Westra2010}\tnote{b,$\dagger$}   &  $41.74 \pm 0.13$  &  $-1.22 \pm 0.06$   & $-2.90\pm0.10$ \\
        \noalign{\smallskip}\hline
        \end{tabular}
    \label{tab:LF_literatura}
\end{table*}

\subsection{H$\alpha$ luminosity function}\label{sect:HaLF_discussion}
We compare now our H$\alpha$LF with previous determinations. This is shown in Figure~\ref{fig:panel_LF}, where we plot the data of each study transformed to our assumed cosmology. The Schechter parameters provided by some studies are listed in Table~\ref{tab:LF_literatura}. For this comparison, we use the mass-weighted H$\alpha$LF derived from the Full Sample. 

The first comparison is with the work by \cite{gama13}. We find that our distribution lies below their data. Given that their effective redshift is larger than ours ($\mathrm{z} \sim 0.08$), the samples are not directly comparable, as some time evolution is expected. In such case, their estimation can be assumed as an upper limit to our local determination. The same argument is valid for the values measured at the faint end ($\log L_{\mathrm{H}\alpha} \lesssim 40$) by \citet{Ly2007}.

We compare now our results with the studies by \cite{Gallego1995}, based on the UCM Survey \citep{Zamorano1994}, and \cite{Nakamura2004}, based on SDSS. Both studies are based on spectroscopic measurements of the H$\alpha$ flux. We see in Table~\ref{tab:LF_literatura} that their best-fitting parameters show better consistency within them than with ours. On the one hand, we find our value of $\log L_{\mathrm{H}\alpha}^{\star}$ to be below theirs; on the other hand, our measurement of $\log \phi_{\star}$ is above. This behavior is expected due to the strong correlation that is found between both parameters.

Being both studies spectroscopic in origin, they require aperture corrections to account for the H$\alpha$ flux in the outer regions of the galaxy. The good agreement that we find with IFS surveys, as shown in Sect.~ \ref{sect:SFMS_discusion}, implies that our measurements are not biased on this regard. Other potential source of discrepancy is dust correction. The typical dust attenuation found by \citeauthor{Gallego1995} and \citeauthor{Nakamura2004} is $A_{\mathrm{H}\alpha} \sim 0.85$, significantly larger than the median extinction in our sample, $A_{\mathrm{H}\alpha} \sim 0.15$. However, we have shown in Sect.~\ref{sect:polvo} that the estimated H$\alpha$ attenuation in our sample follows the mass-to-extinction relation by \cite{DuartePuertas2017} in SDSS. Thus, we believe that our dust correction is appropriate.

We move to the study from \citet{James2008}, that estimate the H$\alpha$LF in a $d \lesssim 40$ Mpc sample using narrow-band imaging from the H$\alpha$ galaxy survey (H$\alpha$GS, \citealt{James2004}). The authors use the relation between $B-$band luminosity and SFR from H$\alpha$GS to weight the $B-$band luminosity function in the local volume, obtaining the H$\alpha$LF. They do not provide a Schechter fit to their data, shown in the {\it bottom right panel} at Fig.~\ref{fig:panel_LF}. Their values present an excellent agreement with our results within the common luminosity range, except for their highest luminosity point at $\log L_{\rm{H}\alpha} = 42.1$ with $\log \Phi = -3.3$. We note that to reach this number density in our surveyed volume, only two extra galaxies would be needed in our sample.

We find the strongest disagreement with the works by \cite{Westra2010} and \cite{PerezGonzalez2003}. Despite the determination by \cite{Westra2010} is drawn from a sample that covers a similar redshift range to the one of \cite{gama13}, in their case the H$\alpha$LF lies significantly below our distribution at $\log L_{\mathrm{H}\alpha} \lesssim 41$. However, it is worth noting that their work is based in observations of only $4$ deg\textsuperscript{2}, which points at cosmic variance as the main source of disagreement. 

\begin{figure*}
    \centering
    \includegraphics{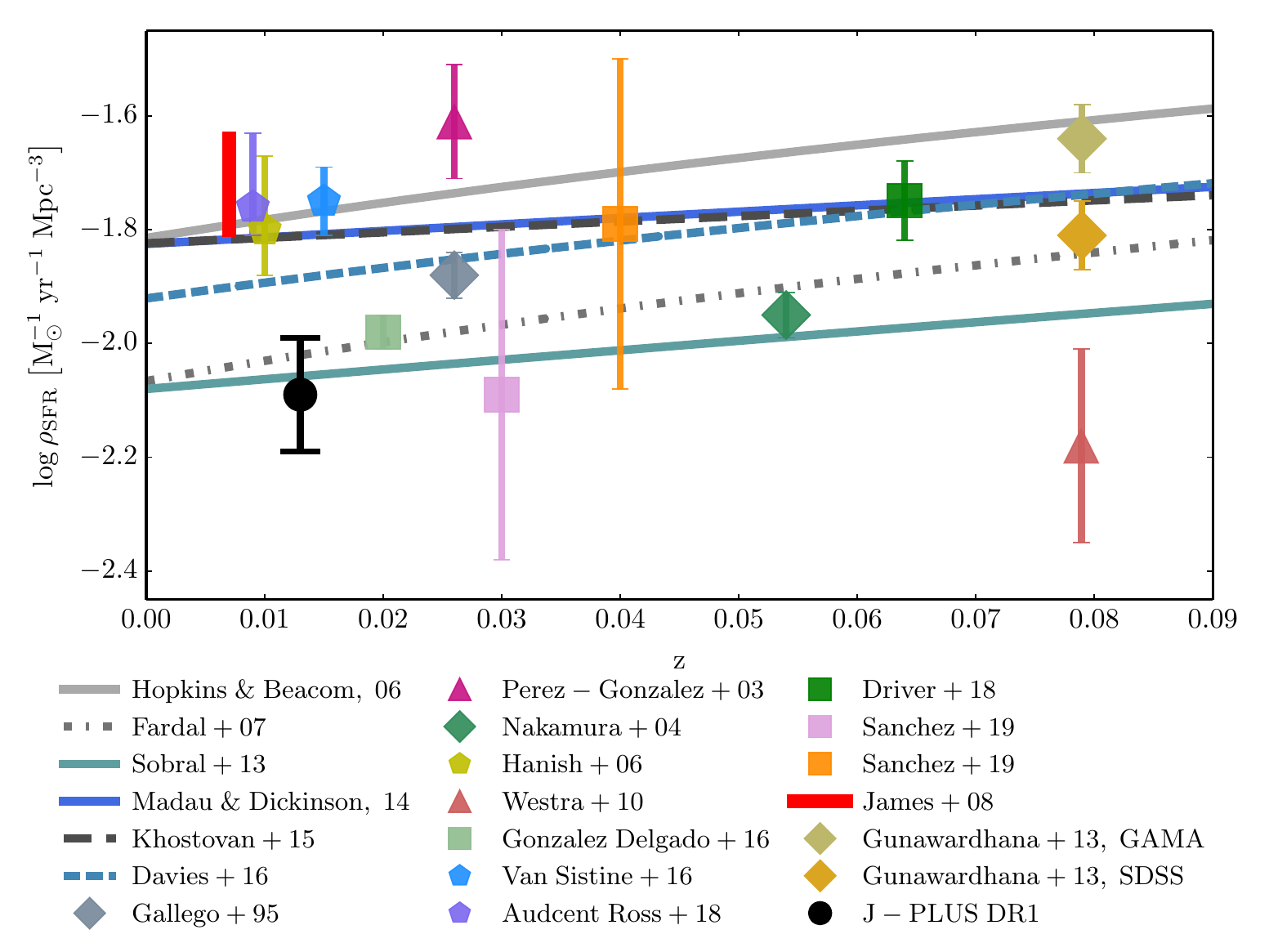}
    \caption{Star formation rate density at $\mathrm{z} \leq 0.1$. The blue dot shows the value from the local sample in J-PLUS DR1. The measurements (symbols) and the evolution parametrizations (lines) from the literature are depicted in the legend.}
    \label{fig:SFRD_lit}
\end{figure*}

\begin{table*}
    \centering
    \caption{Compilation of star formation rate densities at $\mathrm{z} \leq 0.1$}
    \label{tab:SFRD_literatura}
    \begin{tabular}{@{}lccc@{}}
        \hline\hline\noalign{\smallskip}
            Reference                        &   Redshift range               & $\langle \mathrm{z} \rangle$ & $\log \rho_{\mathrm{SFR}}$ \\
                             &   &  &  $\mathrm{[M_{\odot}\,yr^{-1}\,Mpc^{-3}]}$ \\
        \noalign{\smallskip}\hline\noalign{\smallskip}
        This work                    & $0.001\leq \mathrm{z} \leq0.017$      &    $ 0.012 $       &    $-2.10\pm0.11$                \\
        
        \cite{Gallego1995}           &  $0\leq \mathrm{z} \leq0.045$         &    $ 0.026 $       &    $-1.88\pm0.04$                \\ 
        \cite{PerezGonzalez2003}     &  $0\leq \mathrm{z} \leq0.045$         &    $ 0.026 $       &    $-1.61\pm0.10$                \\ 
        \cite{Nakamura2004}          &  $0.01\leq \mathrm{z} \leq0.12$       &    $ 0.054 $       &    $-1.95\pm0.04$                \\ 
        \cite{Hanish2006}            &  $0.001\leq \mathrm{z} \leq0.017$     &    $ 0.010 $       &    $-1.80\pm^{0.13}_{0.08}$      \\
        \cite{James2008}             &   $0.0\leq \mathrm{z} \leq0.009$      &    $ 0.007 $       &    [$-1.80,-1.64$]               \\
        \cite{Westra2010}            &  $0.01\leq \mathrm{z} \leq0.10$       &    $ 0.079 $       &    $-2.18\pm0.17$                \\ 
        \cite{gama13} - GAMA         &  $0\leq \mathrm{z}    \leq0.10$       &    $ 0.079 $       &    $-1.64\pm0.06$                \\
        \cite{gama13} - SDSS         &  $0\leq \mathrm{z}    \leq0.10$       &    $ 0.079 $       &    $-1.81\pm0.06$                \\
        \cite{RosaGD2016}            &  $0.005\leq \mathrm{z}\leq 0.03$      &    $ 0.020 $       &    $-1.98\pm0.03$                \\
        \cite{ALFALFA2016}           &  $0.005\leq \mathrm{z} \leq0.025$     &    $ 0.015 $       &    $-1.75\pm0.06$                \\
        \cite{AudcentRoss2018}       &  $0.001\leq \mathrm{z} \leq0.032$     &    $ 0.009 $       &    $-1.76\pm^{0.13}_{0.05}$      \\
        \cite{driver2018}            &  $0.02\leq \mathrm{z} \leq 0.08$      &    $ 0.064 $       &    $-1.75\pm0.07$                \\
        \cite{Sanchez2019}           &  $0.03\leq \mathrm{z} \leq 0.17$       &    $ 0.03 $        &    $-2.09\pm0.29$                \\
        \cite{Sanchez2019}           &  $0.03\leq \mathrm{z} \leq 0.17$       &    $ 0.04 $        &    $-1.79\pm0.29$                \\
        \hline
    \end{tabular}
\end{table*}

Conversely, the work by \citeauthor{PerezGonzalez2003} presents the most discrepant value of $L_{\mathrm{H}\alpha}^{*}$ when compared to ours and also a deficit of $\log L_{{\rm H}\alpha} \sim 40$ galaxies by 0.5 dex is noticeable. In their work, \citeauthor{PerezGonzalez2003} use a sample of $79$ galaxies drawn from a parent sample of $191$ UCM Survey galaxies. The H$\alpha$ fluxes of these galaxies are measured using narrow-band imaging, so no aperture correction is needed. However, we note that some of their galaxies present H$\alpha$ luminosities that we do not find in our volume ($\log L_{\mathrm{H}\alpha} \geq 42$, and even $\geq 42.5$). In their study, they emphasize that the UCM Survey galaxies present enhanced star formation when compared with normal, quiescent spiral galaxies. Hence, part of the discrepancies with \citet{PerezGonzalez2003} and \citet{Gallego1995} could be attributed to the target pre-selection with objective prism spectroscopy \citep{Zamorano1994}.

The H$\alpha$LF comparison performed in this section reveals tension with previous estimations in the literature. The origin of such discrepancies is not fully understood. The main discrepant feature is our lower $L_{\mathrm{H}\alpha}^{*}$ by $\sim 0.5$ dex. This is caused by the lack of high-luminosity (i.e., high star-forming) galaxies in our sample. On top of the possible impact of cosmic evolution and target pre-selection in other studies, our probed cosmological volume could be not large enough to find the most extreme, sparse H$\alpha$ emitters. As a reference, the cosmic volume covered by \citet{Gallego1995} is a factor of ten larger than ours. This factor increases to 85 in the case of \citet{Nakamura2004} and \citet{gama13}. However, we can not discard that such extreme emitters are indeed absent at $d \lesssim 75$ Mpc, suggesting a low star-formation rate environment for the Local Group.

Irrespective of the evident discrepancies pointed out, the good agreement with the estimated SFMS from IFS and H$\alpha$ imaging surveys (Sect.~\ref{sect:SFMS_discusion}), and with the stellar mass function derived by \citet{Baldry2012} in the GAMA survey (Sect.~\ref{sect:SMF}), suggest that our data are representative of the star-forming population at $d \lesssim 75$ Mpc.

 \subsection{Star formation rate density}\label{sect:SFRD_discussion}
 In Sect.~\ref{sect:SFRD}, we present our determination of the SFRD, obtained by integrating the H$\alpha$LF. In Figure~\ref{fig:SFRD_lit}, we present our measurement in the context of previous local determinations, as summarized in Table~\ref{tab:SFRD_literatura}. We focus on studies based on H$\alpha$ as SFR tracer, supplemented with the recent SED-fitting results from \citet{RosaGD2016}, \citet{driver2018}, and \citet{Sanchez2019}.

We found that our measured $\log \rho_{\mathrm{SFR}} = -2.10 \pm 0.11$ is at the lower envelope of previous estimations at $\mathrm{z} \leq 0.1$.

We start by comparing the J-PLUS DR1 value with the estimations from different SFRD evolution models (lines in Fig.~\ref{fig:SFRD_lit}). The functions from \citet{Hopkins2006}, \citet{Madau2014}, \citet{Khostovan2015}, and \citet{davies2016} yield $\log \rho_{\rm SFR} \sim -1.85$, while the functions from \citet{Fardal2007} and \citet{Sobral2013} provide $\log \rho_{\rm SFR} \sim -2.05$. Our measurement is in agreement with the latter, and it is $\sim 0.25$ dex below the former. Regarding individual observations, the values at $\mathrm{z} \gtrsim 0.05$, that cluster at $\log \rho_{\mathrm{SFR}} \sim -1.8$, can be reconciled within the expected evolution derived using the parametrization in \citet{Fardal2007}. 

Interestingly, the comparison with some of the most local values at $\mathrm{z} \lesssim 0.05$ can not be explained neither by cosmic evolution nor uncertainties in the measurements. The result from \citet{James2008} at $d \lesssim 40$ Mpc is $\sim 0.4$ dex larger than our value, $\log \rho_{\mathrm{SFR}} \in [-1.80,-1.64]$. The authors weights the observed $B-$band luminosity function by the SFR-to-luminosity ratio derived from a sub-sample of galaxies with H$\alpha$ imaging, being their confidence interval lead by dust extinction uncertainties. They also replicate the process with a SDSS sample of blue galaxies, finding a smaller value at a larger effective redshift ($\log \rho_{\mathrm{SFR}} = -1.89$ at $\mathrm{z} \sim 0.04$). As we already discussed in Sect.~\ref{sect:HaLF_discussion}, the H$\alpha$LFs in both works are similar in the range $\log L_{\rm{H}\alpha} \in [39.5,41.6]$, with a higher number density in \citet{James2008} above $\log  L_{\rm{H}\alpha} > 41.6$. Without making any assumptions for the analytical distribution of their H$\alpha$LF, we estimated that this excess could account for $\sim0.1$ dex, moving their lower limit in SFRD to $\rho_{\mathrm{SFR}} \sim -1.90$, still at $\sim 2\sigma$ of the J-PLUS estimation.

The results from \citet{Hanish2006}, \citet{ALFALFA2016} and \citet{AudcentRoss2018} suggests $\log \rho_{\rm SFR} \sim -1.75$, that is $0.3$ dex larger than our value. These studies follow a similar approach to estimate the SFRD: starting for a sample of HI~\!-\!~selected galaxies, a sub-sample is followed with H$\alpha$ imaging to estimate the relation between SFR and HI mass. Then, they integrate the HI mass function weighted with the previous SFR-HI mass relation to compute the SFRD. After carefully revising the data in \citet{ALFALFA2016}, we find a significant density of galaxies with $\log {\rm SFR} > 0.7$ (their Fig.~$13$) that are not present in our sample. We argue that the relation between SFR and HI mass from J-PLUS data could be less step than in \cite{ALFALFA2016}, mitigating the retrieved discrepancy. This is supported by the good agreement with \citet{Gavazzi13} and \citet{Gavazzi15} presented in Sec.~\ref{sect:SFMS_discusion}, as their sample is HI-selected from ALFALFA as well. A detailed study of the SFR {\it vs}. HI mass relation is beyond the scope of the present paper, and we plan to address this issue in a future work.

The higher value found by \citet{PerezGonzalez2003} is explained by their larger $\log L_{\mathrm{H}\alpha}^{*}$ when compared with our measurements. We refer to reader to Sect.~\ref{sect:HaLF_discussion} for a detailed discussion.

Finally, we highlight the agreement with the recent study of \citet{RosaGD2016} based on the SED-fitting analysis of CALIFA galaxies, that provides $\log \rho_{\rm SFR} = -1.98$ at $\mathrm{z} = 0.02$. The understanding of the reasons behind the current tension between our local measurement and those at similar redshift will benefit of the estimation of the H$\alpha$-based SFRD from IFS surveys such as CALIFA and MaNGA, a missing piece of information to our best knowledge.

Regarding the $\sim15$\% contribution of the Red Sample to the total SFRD in the local Universe, other studies in the literature also point out the significant contribution of red and morphological early-type galaxies to the star-forming population at low redshift \citep[e.g.][]{Guna2015, LopezFernandez2018, Sanchez2019}. We will analyze the morphology of the Blue and Red samples in the forthcoming paper by \citet{Logrono2020}.

We conclude that the star formation rate density from J-PLUS DR1 at $d \lesssim 75$ Mpc favors the SFRD evolution proposed by \citet{Fardal2007} and \citet{Sobral2013}. Further studies in the local Universe are needed to better understand the current discrepancies and provide a robust anchoring point for evolutionary studies.

\section{Summary and conclusions}\label{sect:Conclusions}
In this paper, we have used J-PLUS DR1 to obtain a representative collection of 805 H$\alpha$-emitting galaxies located at $\mathrm{z}\leq0.017$ ($d \lesssim 75$ Mpc). We have used this sample to study the star formation main sequence, and its two projections: the H$\alpha$ luminosity function and the stellar mass function. With this information, we have determined the star formation rate density in the local Universe. The most relevant aspects and results of this work are:
\begin{itemize}
    \item The local sample comprises $49$ AGN, $689$ blue galaxies, and $67$ red galaxies. The color classification was performed in the $(u-g)$ {\it vs}. $(g-z)$ color-color diagram, where our sample exhibits a bimodal distribution, computed from those pixels in J-PLUS images with signal-to-noise larger than three in H$\alpha$ emission.
    \item The star formation main sequence is clearly defined by the blue galaxies, with the red galaxies located below them. The SFMS is described as $\log \mathrm{SFR} = 0.83 \log M_{\star} - 8.44$. We find a good agreement with previous SFMSs in the literature, specially those based on integral field spectroscopy. 
    \item The stellar mass function of our blue galaxies closely resembles the one from blue galaxies in the GAMA survey, while our red galaxies follows a downgraded version by a factor of $0.35$ of the red stellar mass function presented in GAMA. This implies that our local sample is $95\%$ complete for $\log M_{\star} > 9$ galaxies. 
    \item The H$\alpha$ luminosity function is well described by a Schechter function with $\log L_{\mathrm{H}\alpha}^{*} = 41.34$, $\log \phi^{*} = -2.43$, and $\alpha = -1.25$. We find a lower characteristic luminosity that several previous work in the literature.
    \item The star formation rate density at $d \lesssim 75$ Mpc is $\log \rho_{\mathrm{SFR}}=-2.10 \pm 0.11$ for a Salpeter IMF. The red galaxies account for 15\% of the local SFRD. This value favors the evolutionary fittings presented in \citet{Fardal2007} and \citet{Sobral2013}.
\end{itemize}

We have computed these results in a consistent way, including in each step all the uncertainties, and accounting for the potential correlation between parameters.

The results presented along this paper show a good agreement with previous findings regarding the SFMS. However, the comparison of the H$\alpha$LF reveals a lower $L_{\mathrm{H}\alpha}^{*}$. We argue that this can be a real effect, being the local volume devoid of highly star-forming galaxies, or a sampling bias, being our surveyed volume unable to trace the sparse, high-luminosity population. Finally, our SFRD measurement is lower than most values at the same redshift. Further work is needed to understand the discrepancies and provide a robust anchoring point in the local Universe for evolutionary studies.

The analysis in this paper makes use of J-PLUS DR1 data. The covered area will be double in the future second data release, providing a test for the cosmic variance hypothesis about the lack of high-star forming systems in the current sample. This will be also possible with other large-area photometric surveys with narrow-band filters. We highlight the Census of the local Universe (CLU; \citealt{clu}) survey, covering $\sim 26\,700$ deg$^2$ down to $m \sim 19$ with four filters of $8$ nm width to explore H$\alpha$ at $\mathrm{z} \leq 0.047$; and the Javalambre Physics of the Accelerating Universe Astrophysical Survey \citep[J-PAS\footnote{http://www.j-pas.org/},][]{BenitezJPAS, minijpas}, that will cover $\sim 8\,500$ deg$^2$ down to $m \sim 22.5$ with $56$ filters of $14$ nm width to study H$\alpha$ emission at $\mathrm{z} \lesssim 0.4$. The combination of area and depth from J-PLUS and these surveys will provide important clues about the evolution of the SFRD in the last $2$ Gyr of the Universe.

\begin{acknowledgements}
Based on observations made with the JAST/T80 telescope for J-PLUS
project at the Observatorio Astrof\'{\i}sico de Javalambre in Teruel, a Spanish Infraestructura Cientifico-T\'ecnica Singular (ICTS) owned, managed and operated by the Centro de Estudios de F\'{\i}sica del Cosmos de Arag\'on (CEFCA). Data has been processed and provided by CEFCA's Unit of Processing and Archiving Data (UPAD). Funding for the J-PLUS Project has been provided by the Governments of Spain and Arag\'on through the Fondo de Inversiones de Teruel; the Arag\'on Government through the Research Groups E96, E103, and E16\_17R; the Spanish Ministry of Science, Innovation and Universities
(MCIU/AEI/FEDER, UE) with grants PGC2018-097585-B-C21 and PGC2018-097585-B-C22; the Spanish Ministry of Economy and Competitiveness (MINECO) under AYA2015-66211-C2-1-P, AYA2015-66211-C2-2, AYA2012-30789, and ICTS-2009-14; and European FEDER funding (FCDD10-4E-867, FCDD13-4E-2685). The Brazilian agencies FINEP, FAPESP, and the National Observatory of Brazil have also contributed to this project.

We thank the comments and suggestions of the referee, that improved the manuscript and the quality of the results.

G.~V.~R. wants to thank A. and K. Gavin O'Shaughnessy for the language corrections. 

J.~M.~V. acknowledges financial support from the State Agency for Research of the Spanish MCIU through the ''Center of Excellence Severo Ochoa'' award to the Instituto de Astrof\'{\i}sica de Andaluc\'{\i}a (SEV-2017-0709), and grant AYA2016-79724-C4-4P.

R.~A.~D. acknowledges support from the Conselho Nacional de Desenvolvimento Científico e Tecnológico – CNPq through BP grant 308105/2018-4, and the Financiadora de Estudos e Projetos – FINEP grants REF. 1217/13 – 01.13.0279.00 and REF 0859/10 – 01.10.0663.00 for hardware support for the J-PLUS project through the National Observatory of Brazil.

We thank the help of Prof. G. Gavazzi in the analysis and understanding of H$\alpha$3 data.

This research has made use of the NASA/IPAC Extragalactic Database (NED), which is operated by the Jet Propulsion Laboratory, California Institute of Technology, under contract with the National Aeronautics and Space Administration.

This research made use of \texttt{Astropy}, a community-developed core \texttt{Python} package for Astronomy \citep{astropy}, and \texttt{Matplotlib}, a 2D graphics package used for \texttt{Python} for publication-quality image generation across user interfaces and operating systems \citep{pylab}. Some of the data sets discussed in the manuscript were recovered with WebPlotDigitizer (\url{https://automeris.io/WebPlotDigitizer/index.html}), a free software developed to facilitate easy and accurate data extraction from a variety of plot types. 

\end{acknowledgements}

\bibliographystyle{aa}
\bibliography{Bibliography}

\begin{thebibliography}{121}
\expandafter\ifx\csname natexlab\endcsname\relax\def\natexlab#1{#1}\fi

\bibitem[{{Astropy Collaboration} {et~al.}(2013){Astropy Collaboration},
  {Robitaille}, {Tollerud}, {Greenfield}, {Droettboom}, {Bray}, {Aldcroft},
  {Davis}, {Ginsburg}, {Price-Whelan}, {Kerzendorf}, {Conley}, {Crighton},
  {Barbary}, {Muna}, {Ferguson}, {Grollier}, {Parikh}, {Nair}, {Unther},
  {Deil}, {Woillez}, {Conseil}, {Kramer}, {Turner}, {Singer}, {Fox}, {Weaver},
  {Zabalza}, {Edwards}, {Azalee Bostroem}, {Burke}, {Casey}, {Crawford},
  {Dencheva}, {Ely}, {Jenness}, {Labrie}, {Lim}, {Pierfederici}, {Pontzen},
  {Ptak}, {Refsdal}, {Servillat}, \& {Streicher}}]{astropy}
{Astropy Collaboration}, {Robitaille}, T.~P., {Tollerud}, E.~J., {et~al.} 2013,
  \aap, 558, A33

\bibitem[{{Audcent-Ross} {et~al.}(2018){Audcent-Ross}, {Meurer}, {Wong},
  {Zheng}, {Hanish}, {Zwaan}, {Bland-Hawthorn}, {Elagali}, {Meyer}, {Putman},
  {Ryan-Weber}, {Sweet}, {Thilker}, {Seibert}, {Allen}, {Dopita}, {Doyle-Pegg},
  {Drinkwater}, {Ferguson}, {Freeman}, {Heckman}, {Kennicutt}, {Kilborn},
  {Kim}, {Knezek}, {Koribalski}, {Smith}, {Staveley-Smith}, {Webster}, \&
  {Werk}}]{AudcentRoss2018}
{Audcent-Ross}, F.~M., {Meurer}, G.~R., {Wong}, O.~I., {et~al.} 2018, \mnras,
  480, 119

\bibitem[{{Baldry} {et~al.}(2012){Baldry}, {Driver}, {Loveday}, {Taylor},
  {Kelvin}, {Liske}, {Norberg}, {Robotham}, {Brough}, {Hopkins}, {Bamford},
  {Peacock}, {Bland-Hawthorn}, {Conselice}, {Croom}, {Jones}, {Parkinson},
  {Popescu}, {Prescott}, {Sharp}, \& {Tuffs}}]{Baldry2012}
{Baldry}, I.~K., {Driver}, S.~P., {Loveday}, J., {et~al.} 2012, \mnras, 421,
  621

\bibitem[{{Baldwin} {et~al.}(1981){Baldwin}, {Phillips}, \&
  {Terlevich}}]{Baldwin1981}
{Baldwin}, J.~A., {Phillips}, M.~M., \& {Terlevich}, R. 1981, \pasp, 93, 5

\bibitem[{{Belfiore} {et~al.}(2018){Belfiore}, {Maiolino}, {Bundy}, {Masters},
  {Bershady}, {Oyarz{\'u}n}, {Lin}, {Cano-Diaz}, {Wake}, {Spindler}, {Thomas},
  {Brownstein}, {Drory}, \& {Yan}}]{Belfiore2018}
{Belfiore}, F., {Maiolino}, R., {Bundy}, K., {et~al.} 2018, \mnras, 477, 3014

\bibitem[{{Belfiore} {et~al.}(2017){Belfiore}, {Maiolino}, {Tremonti},
  {S{\'a}nchez}, {Bundy}, {Bershady}, {Westfall}, {Lin}, {Drory}, {Boquien},
  {Thomas}, \& {Brinkmann}}]{Belfiore2017b}
{Belfiore}, F., {Maiolino}, R., {Tremonti}, C., {et~al.} 2017, \mnras, 469, 151

\bibitem[{{Ben\'{\i}tez} {et~al.}(2014){Ben\'{\i}tez}, {Dupke}, {Moles},
  {Sodre}, {Cenarro}, {Marin-Franch}, {Taylor}, {Cristobal}, {Fernandez-Soto},
  {Mendes de Oliveira}, {Cepa-Nogue}, {Abramo}, {Alcaniz}, {Overzier},
  {Hernandez-Monteagudo}, {Alfaro}, {Kanaan}, {Carvano}, {Reis}, {Martinez
  Gonzalez}, {Ascaso}, {Ballesteros}, {Xavier}, {Varela}, {Ederoclite},
  {Vazquez Ramio}, {Broadhurst}, {Cypriano}, {Angulo}, {Diego}, {Zandivarez},
  {Diaz}, {Melchior}, {Umetsu}, {Spinelli}, {Zitrin}, {Coe}, {Yepes}, {Vielva},
  {Sahni}, {Marcos-Caballero}, {Shu Kitaura}, {Maroto}, {Masip}, {Tsujikawa},
  {Carneiro}, {Gonzalez Nuevo}, {Carvalho}, {Reboucas}, {Carvalho}, {Abdalla},
  {Bernui}, {Pigozzo}, {Ferreira}, {Chandrachani Devi}, {Bengaly}, {Campista},
  {Amorim}, {Asari}, {Bongiovanni}, {Bonoli}, {Bruzual}, {Cardiel}, {Cava},
  {Cid Fernandes}, {Coelho}, {Cortesi}, {Delgado}, {Diaz Garcia}, {Espinosa},
  {Galliano}, {Gonzalez-Serrano}, {Falcon-Barroso}, {Fritz}, {Fernandes},
  {Gorgas}, {Hoyos}, {Jimenez-Teja}, {Lopez-Aguerri}, {Lopez-San Juan},
  {Mateus}, {Molino}, {Novais}, {OMill}, {Oteo}, {Perez-Gonzalez}, {Poggianti},
  {Proctor}, {Ricciardelli}, {Sanchez-Blazquez}, {Storchi-Bergmann}, {Telles},
  {Schoennell}, {Trujillo}, {Vazdekis}, {Viironen}, {Daflon},
  {Aparicio-Villegas}, {Rocha}, {Ribeiro}, {Borges}, {Martins}, {Marcolino},
  {Martinez-Delgado}, {Perez-Torres}, {Siffert}, {Calvao}, {Sako}, {Kessler},
  {Alvarez-Candal}, {De Pra}, {Roig}, {Lazzaro}, {Gorosabel}, {Lopes de
  Oliveira}, {Lima-Neto}, {Irwin}, {Liu}, {Alvarez}, {Balmes}, {Chueca},
  {Costa-Duarte}, {da Costa}, {Dantas}, {Diaz}, {Fabregat}, {Ferrari},
  {Gavela}, {Gracia}, {Gruel}, {Gutierrez}, {Guzman}, {Hernandez-Fernandez},
  {Herranz}, {Hurtado-Gil}, {Jablonsky}, {Laporte}, {Le Tiran}, {Licandro},
  {Lima}, {Martin}, {Martinez}, {Montero}, {Penteado}, {Pereira}, {Peris},
  {Quilis}, {Sanchez-Portal}, {Soja}, {Solano}, {Torra}, \&
  {Valdivielso}}]{BenitezJPAS}
{Ben\'{\i}tez}, N., {Dupke}, R., {Moles}, M., {et~al.} 2014, ArXiv e-prints
  [\eprint[arXiv]{1403.5237}]

\bibitem[{{Bertin} \& {Arnouts}(1996)}]{Bertin1996}
{Bertin}, E. \& {Arnouts}, S. 1996, \aaps, 117, 393

\bibitem[{{Bonatto} {et~al.}(2019){Bonatto}, {Chies-Santos}, {Coelho},
  {Varela}, {Larsen}, {Javier Cenarro}, {San Roman}, {Mar{\'\i}n-Franch},
  {Mendes de Oliveira}, {Molino}, {Ederoclite}, {Cortesi}, {L{\'o}pez-Sanjuan},
  {Crist{\'o}bal-Hornillos}, {V{\'a}zquez Rami{\'o}}, {Sodr{\'e}}, {Sampedro},
  {Costa-Duarte}, {Novais}, {Dupke}, {Overzier}, {Ribeiro}, {Santos}, \&
  {Schoennell}}]{Bonatto2019}
{Bonatto}, C., {Chies-Santos}, A.~L., {Coelho}, P. R.~T., {et~al.} 2019, \aap,
  622, A179

\bibitem[{{Bonoli} {et~al.}(2020){Bonoli}, {Mar{\'\i}n-Franch}, {Varela},
  {V{\'a}zquez Rami{\'o}}, {Abramo}, {Cenarro}, {Dupke}, {V{\'\i}lchez},
  {Crist{\'o}bal-Hornillos}, {Gonz{\'a}lez Delgado},
  {Hern{\'a}ndez-Monteagudo}, {L{\'o}pez-Sanjuan}, {Muniesa}, {Civera},
  {Ederoclite}, {Hern{\'a}n-Caballero}, {Marra}, {Baqui}, {Cortesi},
  {Cypriano}, {Daflon}, {de Amorim}, {D{\'\i}az-Garc{\'\i}a}, {Diego},
  {Mart{\'\i}nez-Solaeche}, {P{\'e}rez}, {Placco}, {Prada}, {Queiroz},
  {Alcaniz}, {Alvarez-Candal}, {Cepa}, {Maroto}, {Roig}, {Siffert}, {Taylor},
  {Benitez}, {Moles}, {Sodr{\'e}}, {Carneiro}, {Mendes de Oliveira}, {Abdalla},
  {Angulo}, {Aparicio Resco}, {Balaguera-Antol{\'\i}nez}, {Ballesteros},
  {Brito-Silva}, {Broadhurst}, {Carrasco}, {Castro}, {Cid Fernandes}, {Coelho},
  {de Melo}, {Doubrawa}, {Fernandez-Soto}, {Ferrari}, {Finoguenov},
  {Garc{\'\i}a-Benito}, {Iglesias-P{\'a}ramo}, {Jim{\'e}nez-Teja}, {Kitaura},
  {Laur}, {Lopes}, {Lucatelli}, {Mart{\'\i}nez}, {Maturi}, {Quartin},
  {Pigozzo}, {Rodr{\`\i}guez-Mart{\`\i}n}, {Salzano}, {Tamm}, {Tempel},
  {Umetsu}, {Valdivielso}, {von Marttens}, {Zitrin}, {D{\'\i}az-Mart{\'\i}n},
  {L{\'o}pez-Alegre}, {L{\'o}pez-Sainz}, {Yanes-D{\'\i}az}, {Rueda-Teruel},
  {Rueda-Teruel}, {Abril Iba{\~n}ez}, {Ant{\'o}n Bravo}, {Bello Ferrer},
  {Bielsa}, {Casino}, {Castillo}, {Chueca}, {Cuesta}, {Garzar{\'a}n Calderaro},
  {Iglesias-Marzoa}, {{\'I}niguez}, {Lamadrid Gutierrez}, {Lopez-Martinez},
  {Lozano-P{\'e}rez}, {Ma{\'\i}cas Sacrist{\'a}n}, {Molina-Ib{\'a}{\~n}ez},
  {Moreno-Signes}, {Rodr{\'\i}guez Llano}, {Royo Navarro}, {Tilve Rua},
  {Andrade}, {Alfaro}, {Akras}, {Arnalte-Mur}, {Ascaso}, {Barbosa},
  {Beltr{\'a}n Jim{\'e}nez}, {Benetti}, {Bengaly}, {Bernui}, {Blanco-Pillado},
  {Borges Fernandes}, {Bregman}, {Bruzual}, {Calderone}, {Carvano}, {Casarini},
  {Chies-Santos}, {Coutinho de Carvalho}, {Dimauro}, {Duarte Puertas},
  {Figueruelo}, {Gonz{\'a}lez-Serrano}, {Guerrero}, {Gurung-L{\'o}pez},
  {Herranz}, {Huertas-Company}, {Irwin}, {Izquierdo-Villalba}, {Kanaan},
  {Kehrig}, {Kirkpatrick}, {Lim}, {Lopes}, {Lopes de Oliveira},
  {Marcos-Caballero}, {Mart{\'\i}nez-Delgado}, {Mart{\'\i}nez-Gonz{\'a}lez},
  {Mart{\'\i}nez-Somonte}, {Oliveira}, {Orsi}, {Overzier}, {Penna-Lima},
  {Reis}, {Spinoso}, {Tsujikawa}, {Vielva}, {Vitorelli}, {Xia}, {Yuan},
  {Arroyo-Polonio}, {Dantas}, {Galarza}, {Gon{\c{c}}alves}, {Gon{\c{c}}alves},
  {Gonzalez}, {Gonzalez}, {Greisel}, {Land im}, {Lazzaro}, {Magris},
  {Monteiro-Oliveira}, {Pereira}, {Rebou{\c{c}}as}, {Rodriguez-Espinosa},
  {Santos da Costa}, \& {Telles}}]{minijpas}
{Bonoli}, S., {Mar{\'\i}n-Franch}, A., {Varela}, J., {et~al.} 2020, \aap,
  submitted [\eprint[arXiv]{2007.01910}]

\bibitem[{{Bothwell} {et~al.}(2011){Bothwell}, {Kenicutt}, {Johnson}, {Wu},
  {Lee}, {Dale}, {Engelbracht}, {Calzetti}, \& {Skillman}}]{Bothwell2011}
{Bothwell}, M.~S., {Kenicutt}, R.~C., {Johnson}, B.~D., {et~al.} 2011, \mnras,
  415, 1815

\bibitem[{{Brinchmann} {et~al.}(2004){Brinchmann}, {Charlot}, {White},
  {Tremonti}, {Kauffmann}, {Heckman}, \& {Brinkmann}}]{Brinchmann04}
{Brinchmann}, J., {Charlot}, S., {White}, S.~D.~M., {et~al.} 2004, \mnras, 351,
  1151

\bibitem[{{Bundy} {et~al.}(2015){Bundy}, {Bershady}, {Law}, {Yan}, {Drory},
  {MacDonald}, {Wake}, {Cherinka}, {S{\'a}nchez-Gallego}, {Weijmans}, {Thomas},
  {Tremonti}, {Masters}, {Coccato}, {Diamond-Stanic}, {Arag{\'o}n-Salamanca},
  {Avila-Reese}, {Badenes}, {Falc{\'o}n-Barroso}, {Belfiore}, {Bizyaev},
  {Blanc}, {Bland-Hawthorn}, {Blanton}, {Brownstein}, {Byler}, {Cappellari},
  {Conroy}, {Dutton}, {Emsellem}, {Etherington}, {Frinchaboy}, {Fu}, {Gunn},
  {Harding}, {Johnston}, {Kauffmann}, {Kinemuchi}, {Klaene}, {Knapen},
  {Leauthaud}, {Li}, {Lin}, {Maiolino}, {Malanushenko}, {Malanushenko}, {Mao},
  {Maraston}, {McDermid}, {Merrifield}, {Nichol}, {Oravetz}, {Pan}, {Parejko},
  {Sanchez}, {Schlegel}, {Simmons}, {Steele}, {Steinmetz}, {Thanjavur},
  {Thompson}, {Tinker}, {van den Bosch}, {Westfall}, {Wilkinson}, {Wright},
  {Xiao}, \& {Zhang}}]{manga}
{Bundy}, K., {Bershady}, M.~A., {Law}, D.~R., {et~al.} 2015, \apj, 798, 7

\bibitem[{{Bunker} {et~al.}(1995){Bunker}, {Warren}, {Hewett}, \&
  {Clements}}]{Bunker1995}
{Bunker}, A.~J., {Warren}, S.~J., {Hewett}, P.~C., \& {Clements}, D.~L. 1995,
  \mnras, 273, 513

\bibitem[{{Calzetti}(2013)}]{Calzetti2013}
{Calzetti}, D. 2013, {Star Formation Rate Indicators}, ed.
  J.~{Falc{\'o}n-Barroso} \& J.~H. {Knapen}, 419

\bibitem[{{Calzetti} {et~al.}(2000){Calzetti}, {Armus}, {Bohlin}, {Kinney},
  {Koornneef}, \& {Storchi-Bergmann}}]{Calzetti2000}
{Calzetti}, D., {Armus}, L., {Bohlin}, R.~C., {et~al.} 2000, \apj, 533, 682

\bibitem[{{Cano-D{\'\i}az} {et~al.}(2019){Cano-D{\'\i}az}, {{\'A}vila-Reese},
  {S{\'a}nchez}, {Hern{\'a}ndez-Toledo}, {Rodr{\'\i}guez-Puebla}, {Boquien}, \&
  {Ibarra-Medel}}]{CanoDiaz2019}
{Cano-D{\'\i}az}, M., {{\'A}vila-Reese}, V., {S{\'a}nchez}, S.~F., {et~al.}
  2019, \mnras, 488, 3929

\bibitem[{{Cano-D{\'{\i}}az} {et~al.}(2016){Cano-D{\'{\i}}az}, {S{\'a}nchez},
  {Zibetti}, {Ascasibar}, {Bland-Hawthorn}, {Ziegler}, {Gonz{\'a}lez Delgado},
  {Walcher}, {Garc{\'{\i}}a-Benito}, {Mast}, {Mendoza-P{\'e}rez},
  {Falc{\'o}n-Barroso}, {Galbany}, {Husemann}, {Kehrig}, {Marino},
  {S{\'a}nchez-Bl{\'a}zquez}, {L{\'o}pez-Cob{\'a}}, {L{\'o}pez-S{\'a}nchez}, \&
  {Vilchez}}]{CanoDiaz2016}
{Cano-D{\'{\i}}az}, M., {S{\'a}nchez}, S.~F., {Zibetti}, S., {et~al.} 2016,
  \apjl, 821, L26

\bibitem[{{Catal{\'a}n-Torrecilla} {et~al.}(2017){Catal{\'a}n-Torrecilla}, {Gil
  de Paz}, {Castillo-Morales}, {M{\'e}ndez-Abreu}, {Falc{\'o}n-Barroso},
  {Bekeraite}, {Costantin}, {de Lorenzo-C{\'a}ceres}, {Florido},
  {Garc{\'{\i}}a-Benito}, {Husemann}, {Iglesias-P{\'a}ramo}, {Kennicutt},
  {Mast}, {Pascual}, {Ruiz-Lara}, {S{\'a}nchez-Menguiano}, {S{\'a}nchez},
  {Walcher}, {Bland-Hawthorn}, {Duarte Puertas}, {Marino}, {Masegosa},
  {S{\'a}nchez-Bl{\'a}zquez}, \& {CALIFA
  Collaboration}}]{CatalanTorrecilla2017}
{Catal{\'a}n-Torrecilla}, C., {Gil de Paz}, A., {Castillo-Morales}, A.,
  {et~al.} 2017, \apj, 848, 87

\bibitem[{{Cenarro} {et~al.}(2019){Cenarro}, {Moles},
  {Crist{\'o}bal-Hornillos}, {Mar{\'{\i}}n-Franch}, {Ederoclite}, {Varela},
  {L{\'o}pez-Sanjuan}, {Hern{\'a}ndez-Monteagudo}, {Angulo}, {V{\'a}zquez
  Rami{\'o}}, {Viironen}, {Bonoli}, {Orsi}, {Hurier}, {San Roman}, {Greisel},
  {Vilella-Rojo}, {D{\'{\i}}az-Garc{\'{\i}}a}, {Logro{\~n}o-Garc{\'{\i}}a},
  {Gurung-L{\'o}pez}, {Spinoso}, {Izquierdo-Villalba}, {Aguerri}, {Allende
  Prieto}, {Bonatto}, {Carvano}, {Chies-Santos}, {Daflon}, {Dupke},
  {Falc{\'o}n-Barroso}, {Gon{\c c}alves}, {Jim{\'e}nez-Teja}, {Molino},
  {Placco}, {Solano}, {Whitten}, {Abril}, {Ant{\'o}n}, {Bello}, {Bielsa de
  Toledo}, {Castillo-Ram{\'{\i}}rez}, {Chueca}, {Civera},
  {D{\'{\i}}az-Mart{\'{\i}}n}, {Dom{\'{\i}}nguez-Mart{\'{\i}}nez},
  {Garzar{\'a}n-Calderaro}, {Hern{\'a}ndez-Fuertes}, {Iglesias-Marzoa},
  {I{\~n}iguez}, {Jim{\'e}nez Ruiz}, {Kruuse}, {Lamadrid}, {Lasso-Cabrera},
  {L{\'o}pez-Alegre}, {L{\'o}pez-Sainz}, {Ma{\'{\i}}cas}, {Moreno-Signes},
  {Muniesa}, {Rodr{\'{\i}}guez-Llano}, {Rueda-Teruel}, {Rueda-Teruel},
  {Soriano-Lagu{\'{\i}}a}, {Tilve}, {Valdivielso}, {Yanes-D{\'{\i}}az},
  {Alcaniz}, {Mendes de Oliveira}, {Sodr{\'e}}, {Coelho}, {Lopes de Oliveira},
  {Tamm}, {Xavier}, {Abramo}, {Akras}, {Alfaro}, {Alvarez-Candal}, {Ascaso},
  {Beasley}, {Beers}, {Borges Fernandes}, {Bruzual}, {Buzzo}, {Carrasco},
  {Cepa}, {Cortesi}, {Costa-Duarte}, {De Pr{\'a}}, {Favole}, {Galarza},
  {Galbany}, {Garcia}, {Gonz{\'a}lez Delgado}, {Gonz{\'a}lez-Serrano},
  {Guti{\'e}rrez-Soto}, {Hernandez-Jimenez}, {Kanaan}, {Kuncarayakti},
  {Landim}, {Laur}, {Licandro}, {Lima Neto}, {Lyman}, {Ma{\'{\i}}z
  Apell{\'a}niz}, {Miralda-Escud{\'e}}, {Morate}, {Nogueira-Cavalcante},
  {Novais}, {Oncins}, {Oteo}, {Overzier}, {Pereira}, {Rebassa-Mansergas},
  {Reis}, {Roig}, {Sako}, {Salvador-Rusi{\~n}ol}, {Sampedro},
  {S{\'a}nchez-Bl{\'a}zquez}, {Santos}, {Schmidtobreick}, {Siffert}, {Telles},
  \& {Vilchez}}]{Cenarro2019}
{Cenarro}, A.~J., {Moles}, M., {Crist{\'o}bal-Hornillos}, D., {et~al.} 2019,
  \aap, 622, A176

\bibitem[{{Cenarro} {et~al.}(2014){Cenarro}, {Moles}, {Mar{\'\i}n-Franch},
  {Crist{\'o}bal-Hornillos}, {Yanes D{\'\i}az}, {Ederoclite}, {Varela},
  {V{\'a}zquez Rami{\'o}}, {Valdivielso}, {Ben{\'\i}tez}, {Cepa}, {Dupke},
  {Fern{\'a}ndez-Soto}, {Mendes de Oliveira}, {Sodr{\'e}}, {Taylor},
  {Rueda-Teruel}, {Rueda-Teruel}, {Luis-Simoes}, {Chueca}, {Ant{\'o}n},
  {Bello}, {D{\'\i}az-Mart{\'\i}n}, {Guill{\'e}n-Civera},
  {Hern{\'a}ndez-Fuertes}, {Iglesias-Marzoa}, {Jim{\'e}nez-Mej{\'\i}as},
  {Lasso-Cabrera}, {L{\'o}pez-Alegre}, {L{\'o}pez-Sainz},
  {Rodr{\'\i}guez-Hern{\'a}ndez}, {Su{\'a}rez}, {Lamadrid}, {Ma{\'\i}cas},
  {Abril-Iba{\~n}ez}, {Tilve}, \& {Rodr{\'\i}guez-Llano}}]{cenarro14}
{Cenarro}, A.~J., {Moles}, M., {Mar{\'\i}n-Franch}, A., {et~al.} 2014, in
  Society of Photo-Optical Instrumentation Engineers (SPIE) Conference Series,
  Vol. 9149, \procspie, 91491I

\bibitem[{{Cid Fernandes} {et~al.}(2011){Cid Fernandes}, {Stasi{\'n}ska},
  {Mateus}, \& {Vale Asari}}]{CidFernandez2011}
{Cid Fernandes}, R., {Stasi{\'n}ska}, G., {Mateus}, A., \& {Vale Asari}, N.
  2011, \mnras, 413, 1687

\bibitem[{{Cook} {et~al.}(2019){Cook}, {Kasliwal}, {Van Sistine}, {Kaplan},
  {Sutter}, {Kupfer}, {Shupe}, {Laher}, {Masci}, {Dale}, {Sesar}, {Brady},
  {Yan}, {Ofek}, {Reitze}, \& {Kulkarni}}]{clu}
{Cook}, D.~O., {Kasliwal}, M.~M., {Van Sistine}, A., {et~al.} 2019, \apj, 880,
  7

\bibitem[{{Cucciati} {et~al.}(2012){Cucciati}, {Tresse}, {Ilbert}, {Le
  F{\`e}vre}, {Garilli}, {Le Brun}, {Cassata}, {Franzetti}, {Maccagni},
  {Scodeggio}, {Zucca}, {Zamorani}, {Bardelli}, {Bolzonella}, {Bielby},
  {McCracken}, {Zanichelli}, \& {Vergani}}]{cucciati12}
{Cucciati}, O., {Tresse}, L., {Ilbert}, O., {et~al.} 2012, \aap, 539, A31

\bibitem[{{Davies} {et~al.}(2016){Davies}, {Driver}, {Robotham}, {Grootes},
  {Popescu}, {Tuffs}, {Hopkins}, {Alpaslan}, {Andrews}, {Bland -Hawthorn},
  {Bremer}, {Brough}, {Brown}, {Cluver}, {Croom}, {da Cunha}, {Dunne},
  {Lara-L{\'o}pez}, {Liske}, {Loveday}, {Moffett}, {Owers}, {Phillipps},
  {Sansom}, {Taylor}, {Michalowski}, {Ibar}, {Smith}, \& {Bourne}}]{davies2016}
{Davies}, L.~J.~M., {Driver}, S.~P., {Robotham}, A.~S.~G., {et~al.} 2016,
  \mnras, 461, 458

\bibitem[{{Donnari} {et~al.}(2019){Donnari}, {Pillepich}, {Nelson},
  {Vogelsberger}, {Genel}, {Weinberger}, {Marinacci}, {Springel}, \&
  {Hernquist}}]{Donnari2019}
{Donnari}, M., {Pillepich}, A., {Nelson}, D., {et~al.} 2019, \mnras, 485, 4817

\bibitem[{{Driver} {et~al.}(2018){Driver}, {Andrews}, {da Cunha}, {Davies},
  {Lagos}, {Robotham}, {Vinsen}, {Wright}, {Alpaslan}, {Bland -Hawthorn},
  {Bourne}, {Brough}, {Bremer}, {Cluver}, {Colless}, {Conselice}, {Dunne},
  {Eales}, {Gomez}, {Holwerda}, {Hopkins}, {Kafle}, {Kelvin}, {Loveday},
  {Liske}, {Maddox}, {Phillipps}, {Pimbblet}, {Rowlands}, {Sansom}, {Taylor},
  {Wang}, \& {Wilkins}}]{driver2018}
{Driver}, S.~P., {Andrews}, S.~K., {da Cunha}, E., {et~al.} 2018, \mnras, 475,
  2891

\bibitem[{{Driver} {et~al.}(2011){Driver}, {Hill}, {Kelvin}, {Robotham},
  {Liske}, {Norberg}, {Baldry}, {Bamford}, {Hopkins}, {Loveday}, {Peacock},
  {Andrae}, {Bland-Hawthorn}, {Brough}, {Brown}, {Cameron}, {Ching}, {Colless},
  {Conselice}, {Croom}, {Cross}, {de Propris}, {Dye}, {Drinkwater}, {Ellis},
  {Graham}, {Grootes}, {Gunawardhana}, {Jones}, {van Kampen}, {Maraston},
  {Nichol}, {Parkinson}, {Phillipps}, {Pimbblet}, {Popescu}, {Prescott},
  {Roseboom}, {Sadler}, {Sansom}, {Sharp}, {Smith}, {Taylor}, {Thomas},
  {Tuffs}, {Wijesinghe}, {Dunne}, {Frenk}, {Jarvis}, {Madore}, {Meyer},
  {Seibert}, {Staveley-Smith}, {Sutherland}, \& {Warren}}]{gama}
{Driver}, S.~P., {Hill}, D.~T., {Kelvin}, L.~S., {et~al.} 2011, \mnras, 413,
  971

\bibitem[{{Driver} \& {Robotham}(2010)}]{driver10}
{Driver}, S.~P. \& {Robotham}, A. S.~G. 2010, \mnras, 407, 2131

\bibitem[{{Duarte Puertas} {et~al.}(2017){Duarte Puertas}, {Vilchez},
  {Iglesias-P{\'a}ramo}, {Kehrig}, {P{\'e}rez-Montero}, \&
  {Rosales-Ortega}}]{DuartePuertas2017}
{Duarte Puertas}, S., {Vilchez}, J.~M., {Iglesias-P{\'a}ramo}, J., {et~al.}
  2017, \aap, 599, A71

\bibitem[{{Efstathiou} {et~al.}(1988){Efstathiou}, {Ellis}, \&
  {Peterson}}]{Efstathiou1988}
{Efstathiou}, G., {Ellis}, R.~S., \& {Peterson}, B.~A. 1988, \mnras, 232, 431

\bibitem[{{Elbaz} {et~al.}(2007){Elbaz}, {Daddi}, {Le Borgne}, {Dickinson},
  {Alexander}, {Chary}, {Starck}, {Brandt}, {Kitzbichler}, {MacDonald},
  {Nonino}, {Popesso}, {Stern}, \& {Vanzella}}]{Elbaz07}
{Elbaz}, D., {Daddi}, E., {Le Borgne}, D., {et~al.} 2007, \aap, 468, 33

\bibitem[{{Fardal} {et~al.}(2007){Fardal}, {Katz}, {Weinberg}, \&
  {Dav{\'e}}}]{Fardal2007}
{Fardal}, M.~A., {Katz}, N., {Weinberg}, D.~H., \& {Dav{\'e}}, R. 2007, \mnras,
  379, 985

\bibitem[{{Foreman-Mackey} {et~al.}(2013){Foreman-Mackey}, {Hogg}, {Lang}, \&
  {Goodman}}]{emcee}
{Foreman-Mackey}, D., {Hogg}, D.~W., {Lang}, D., \& {Goodman}, J. 2013, \pasp,
  125, 306

\bibitem[{{Gaia Collaboration} {et~al.}(2018){Gaia Collaboration}, {Brown},
  {Vallenari}, {Prusti}, {de Bruijne}, {Babusiaux}, {Bailer-Jones}, {Biermann},
  {Evans}, {Eyer}, {Jansen}, {Jordi}, {Klioner}, {Lammers}, {Lindegren},
  {Luri}, {Mignard}, {Panem}, {Pourbaix}, {Randich}, {Sartoretti}, {Siddiqui},
  {Soubiran}, {van Leeuwen}, {Walton}, {Arenou}, {Bastian}, {Cropper},
  {Drimmel}, {Katz}, {Lattanzi}, {Bakker}, {Cacciari}, {Casta{\~n}eda},
  {Chaoul}, {Cheek}, {De Angeli}, {Fabricius}, {Guerra}, {Holl}, {Masana},
  {Messineo}, {Mowlavi}, {Nienartowicz}, {Panuzzo}, {Portell}, {Riello},
  {Seabroke}, {Tanga}, {Th{\'e}venin}, {Gracia-Abril}, {Comoretto},
  {Garcia-Reinaldos}, {Teyssier}, {Altmann}, {Andrae}, {Audard},
  {Bellas-Velidis}, {Benson}, {Berthier}, {Blomme}, {Burgess}, {Busso},
  {Carry}, {Cellino}, {Clementini}, {Clotet}, {Creevey}, {Davidson}, {De
  Ridder}, {Delchambre}, {Dell'Oro}, {Ducourant},
  {Fern{\'a}ndez-Hern{\'a}ndez}, {Fouesneau}, {Fr{\'e}mat}, {Galluccio},
  {Garc{\'\i}a-Torres}, {Gonz{\'a}lez-N{\'u}{\~n}ez}, {Gonz{\'a}lez-Vidal},
  {Gosset}, {Guy}, {Halbwachs}, {Hambly}, {Harrison}, {Hern{\'a}ndez},
  {Hestroffer}, {Hodgkin}, {Hutton}, {Jasniewicz}, {Jean-Antoine-Piccolo},
  {Jordan}, {Korn}, {Krone-Martins}, {Lanzafame}, {Lebzelter}, {L{\"o}ffler},
  {Manteiga}, {Marrese}, {Mart{\'\i}n-Fleitas}, {Moitinho}, {Mora}, {Muinonen},
  {Osinde}, {Pancino}, {Pauwels}, {Petit}, {Recio-Blanco}, {Richards},
  {Rimoldini}, {Robin}, {Sarro}, {Siopis}, {Smith}, {Sozzetti}, {S{\"u}veges},
  {Torra}, {van Reeven}, {Abbas}, {Abreu Aramburu}, {Accart}, {Aerts},
  {Altavilla}, {{\'A}lvarez}, {Alvarez}, {Alves}, {Anderson}, {Andrei},
  {Anglada Varela}, {Antiche}, {Antoja}, {Arcay}, {Astraatmadja}, {Bach},
  {Baker}, {Balaguer-N{\'u}{\~n}ez}, {Balm}, {Barache}, {Barata}, {Barbato},
  {Barblan}, {Barklem}, {Barrado}, {Barros}, {Barstow}, {Bartholom{\'e}
  Mu{\~n}oz}, {Bassilana}, {Becciani}, {Bellazzini}, {Berihuete}, {Bertone},
  {Bianchi}, {Bienaym{\'e}}, {Blanco-Cuaresma}, {Boch}, {Boeche}, {Bombrun},
  {Borrachero}, {Bossini}, {Bouquillon}, {Bourda}, {Bragaglia}, {Bramante},
  {Breddels}, {Bressan}, {Brouillet}, {Br{\"u}semeister}, {Brugaletta},
  {Bucciarelli}, {Burlacu}, {Busonero}, {Butkevich}, {Buzzi}, {Caffau},
  {Cancelliere}, {Cannizzaro}, {Cantat-Gaudin}, {Carballo}, {Carlucci},
  {Carrasco}, {Casamiquela}, {Castellani}, {Castro-Ginard}, {Charlot},
  {Chemin}, {Chiavassa}, {Cocozza}, {Costigan}, {Cowell}, {Crifo}, {Crosta},
  {Crowley}, {Cuypers}, {Dafonte}, {Damerdji}, {Dapergolas}, {David}, {David},
  {de Laverny}, {De Luise}, {De March}, {de Martino}, {de Souza}, {de Torres},
  {Debosscher}, {del Pozo}, {Delbo}, {Delgado}, {Delgado}, {Di Matteo},
  {Diakite}, {Diener}, {Distefano}, {Dolding}, {Drazinos}, {Dur{\'a}n},
  {Edvardsson}, {Enke}, {Eriksson}, {Esquej}, {Eynard Bontemps}, {Fabre},
  {Fabrizio}, {Faigler}, {Falc{\~a}o}, {Farr{\`a}s Casas}, {Federici},
  {Fedorets}, {Fernique}, {Figueras}, {Filippi}, {Findeisen}, {Fonti},
  {Fraile}, {Fraser}, {Fr{\'e}zouls}, {Gai}, {Galleti}, {Garabato},
  {Garc{\'\i}a-Sedano}, {Garofalo}, {Garralda}, {Gavel}, {Gavras}, {Gerssen},
  {Geyer}, {Giacobbe}, {Gilmore}, {Girona}, {Giuffrida}, {Glass}, {Gomes},
  {Granvik}, {Gueguen}, {Guerrier}, {Guiraud}, {Guti{\'e}rrez-S{\'a}nchez},
  {Haigron}, {Hatzidimitriou}, {Hauser}, {Haywood}, {Heiter}, {Helmi}, {Heu},
  {Hilger}, {Hobbs}, {Hofmann}, {Holland}, {Huckle}, {Hypki}, {Icardi},
  {Jan{\ss}en}, {Jevardat de Fombelle}, {Jonker}, {Juh{\'a}sz}, {Julbe},
  {Karampelas}, {Kewley}, {Klar}, {Kochoska}, {Kohley}, {Kolenberg},
  {Kontizas}, {Kontizas}, {Koposov}, {Kordopatis}, {Kostrzewa-Rutkowska},
  {Koubsky}, {Lambert}, {Lanza}, {Lasne}, {Lavigne}, {Le Fustec}, {Le
  Poncin-Lafitte}, {Lebreton}, {Leccia}, {Leclerc}, {Lecoeur-Taibi},
  {Lenhardt}, {Leroux}, {Liao}, {Licata}, {Lindstr{\o}m}, {Lister}, {Livanou},
  {Lobel}, {L{\'o}pez}, {Managau}, {Mann}, {Mantelet}, {Marchal}, {Marchant},
  {Marconi}, {Marinoni}, {Marschalk{\'o}}, {Marshall}, {Martino}, {Marton},
  {Mary}, {Massari}, {Matijevi{\v{c}}}, {Mazeh}, {McMillan}, {Messina},
  {Michalik}, {Millar}, {Molina}, {Molinaro}, {Moln{\'a}r}, {Montegriffo},
  {Mor}, {Morbidelli}, {Morel}, {Morris}, {Mulone}, {Muraveva}, {Musella},
  {Nelemans}, {Nicastro}, {Noval}, {O'Mullane}, {Ord{\'e}novic},
  {Ord{\'o}{\~n}ez-Blanco}, {Osborne}, {Pagani}, {Pagano}, {Pailler},
  {Palacin}, {Palaversa}, {Panahi}, {Pawlak}, {Piersimoni}, {Pineau}, {Plachy},
  {Plum}, {Poggio}, {Poujoulet}, {Pr{\v{s}}a}, {Pulone}, {Racero}, {Ragaini},
  {Rambaux}, {Ramos-Lerate}, {Regibo}, {Reyl{\'e}}, {Riclet}, {Ripepi}, {Riva},
  {Rivard}, {Rixon}, {Roegiers}, {Roelens}, {Romero-G{\'o}mez}, {Rowell},
  {Royer}, {Ruiz-Dern}, {Sadowski}, {Sagrist{\`a} Sell{\'e}s}, {Sahlmann},
  {Salgado}, {Salguero}, {Sanna}, {Santana-Ros}, {Sarasso}, {Savietto},
  {Schultheis}, {Sciacca}, {Segol}, {Segovia}, {S{\'e}gransan}, {Shih},
  {Siltala}, {Silva}, {Smart}, {Smith}, {Solano}, {Solitro}, {Sordo}, {Soria
  Nieto}, {Souchay}, {Spagna}, {Spoto}, {Stampa}, {Steele},
  {Steidelm{\"u}ller}, {Stephenson}, {Stoev}, {Suess}, {Surdej}, {Szabados},
  {Szegedi-Elek}, {Tapiador}, {Taris}, {Tauran}, {Taylor}, {Teixeira},
  {Terrett}, {Teyssand ier}, {Thuillot}, {Titarenko}, {Torra Clotet}, {Turon},
  {Ulla}, {Utrilla}, {Uzzi}, {Vaillant}, {Valentini}, {Valette}, {van Elteren},
  {Van Hemelryck}, {van Leeuwen}, {Vaschetto}, {Vecchiato}, {Veljanoski},
  {Viala}, {Vicente}, {Vogt}, {von Essen}, {Voss}, {Votruba}, {Voutsinas},
  {Walmsley}, {Weiler}, {Wertz}, {Wevers}, {Wyrzykowski}, {Yoldas},
  {{\v{Z}}erjal}, {Ziaeepour}, {Zorec}, {Zschocke}, {Zucker}, {Zurbach}, \&
  {Zwitter}}]{gaia_dr2}
{Gaia Collaboration}, {Brown}, A.~G.~A., {Vallenari}, A., {et~al.} 2018, \aap,
  616, A1

\bibitem[{{Gallego} {et~al.}(1995){Gallego}, {Zamorano}, {Aragon-Salamanca}, \&
  {Rego}}]{Gallego1995}
{Gallego}, J., {Zamorano}, J., {Aragon-Salamanca}, A., \& {Rego}, M. 1995,
  ApJL, 455, L1

\bibitem[{{Garn} \& {Best}(2010)}]{GarnBest2010}
{Garn}, T. \& {Best}, P.~N. 2010, \mnras, 409, 421

\bibitem[{{Gavazzi} {et~al.}(2015){Gavazzi}, {Consolandi}, {Dotti}, {Fanali},
  {Fossati}, {Fumagalli}, {Viscardi}, {Savorgnan}, {Boselli}, {Guti{\'e}rrez},
  {Hern{\'a}ndez Toledo}, {Giovanelli}, \& {Haynes}}]{Gavazzi15}
{Gavazzi}, G., {Consolandi}, G., {Dotti}, M., {et~al.} 2015, \aap, 580, A116

\bibitem[{{Gavazzi} {et~al.}(2013){Gavazzi}, {Fumagalli}, {Fossati}, {Galardo},
  {Grossetti}, {Boselli}, {Giovanelli}, \& {Haynes}}]{Gavazzi13}
{Gavazzi}, G., {Fumagalli}, M., {Fossati}, M., {et~al.} 2013, \aap, 553, A89

\bibitem[{{Gavazzi} {et~al.}(2012){Gavazzi}, {Fumagalli}, {Galardo},
  {Grossetti}, {Boselli}, {Giovanelli}, {Haynes}, \& {Fabello}}]{Gavazzi12}
{Gavazzi}, G., {Fumagalli}, M., {Galardo}, V., {et~al.} 2012, \aap, 545, A16

\bibitem[{{Gilbank} {et~al.}(2010){Gilbank}, {Baldry}, {Balogh}, {Glazebrook},
  \& {Bower}}]{Gilbank2010}
{Gilbank}, D.~G., {Baldry}, I.~K., {Balogh}, M.~L., {Glazebrook}, K., \&
  {Bower}, R.~G. 2010, \mnras, 405, 2594

\bibitem[{{Giovanelli} {et~al.}(2005){Giovanelli}, {Haynes}, {Kent},
  {Perillat}, {Saintonge}, {Brosch}, {Catinella}, {Hoffman}, {Stierwalt},
  {Spekkens}, {Lerner}, {Masters}, {Momjian}, {Rosenberg}, {Springob},
  {Boselli}, {Charmand aris}, {Darling}, {Davies}, {Garcia Lambas}, {Gavazzi},
  {Giovanardi}, {Hardy}, {Hunt}, {Iovino}, {Karachentsev}, {Karachentseva},
  {Koopmann}, {Marinoni}, {Minchin}, {Muller}, {Putman}, {Pantoja}, {Salzer},
  {Scodeggio}, {Skillman}, {Solanes}, {Valotto}, {van Driel}, \& {van
  Zee}}]{alfalfa}
{Giovanelli}, R., {Haynes}, M.~P., {Kent}, B.~R., {et~al.} 2005, \aj, 130, 2598

\bibitem[{{Gomes} {et~al.}(2016){Gomes}, {Papaderos}, {Kehrig}, {V{\'\i}lchez},
  {Lehnert}, {S{\'a}nchez}, {Ziegler}, {Breda}, {Dos Reis},
  {Iglesias-P{\'a}ramo}, {Bland-Hawthorn}, {Galbany}, {Bomans},
  {Rosales-Ortega}, {Cid Fernandes}, {Walcher}, {Falc{\'o}n-Barroso},
  {Garc{\'\i}a-Benito}, {M{\'a}rquez}, {Del Olmo}, {Masegosa}, {Moll{\'a}},
  {Marino}, {Gonz{\'a}lez Delgado}, {L{\'o}pez-S{\'a}nchez}, \& {CALIFA
  Collaboration}}]{gomes16}
{Gomes}, J.~M., {Papaderos}, P., {Kehrig}, C., {et~al.} 2016, \aap, 588, A68

\bibitem[{{Gonz{\'a}lez Delgado} {et~al.}(2016){Gonz{\'a}lez Delgado}, {Cid
  Fernandes}, {P{\'e}rez}, {Garc{\'{\i}}a-Benito}, {L{\'o}pez Fern{\'a}ndez},
  {Lacerda}, {Cortijo-Ferrero}, {de Amorim}, {Vale Asari}, {S{\'a}nchez},
  {Walcher}, {Wisotzki}, {Mast}, {Alves}, {Ascasibar}, {Bland-Hawthorn},
  {Galbany}, {Kennicutt}, {M{\'a}rquez}, {Masegosa}, {Moll{\'a}},
  {S{\'a}nchez-Bl{\'a}zquez}, \& {V{\'{\i}}lchez}}]{RosaGD2016}
{Gonz{\'a}lez Delgado}, R.~M., {Cid Fernandes}, R., {P{\'e}rez}, E., {et~al.}
  2016, \aap, 590, A44

\bibitem[{{Green} {et~al.}(2018){Green}, {Schlafly}, {Finkbeiner}, {Rix},
  {Martin}, {Burgett}, {Draper}, {Flewelling}, {Hodapp}, {Kaiser}, {Kudritzki},
  {Magnier}, {Metcalfe}, {Tonry}, {Wainscoat}, \& {Waters}}]{bayestar17}
{Green}, G.~M., {Schlafly}, E.~F., {Finkbeiner}, D., {et~al.} 2018, \mnras,
  478, 651

\bibitem[{{Gunawardhana} {et~al.}(2013){Gunawardhana}, {Hopkins},
  {Bland-Hawthorn}, {Brough}, {Sharp}, {Loveday}, {Taylor}, {Jones},
  {Lara-L{\'o}pez}, {Bauer}, {Colless}, {Owers}, {Baldry},
  {L{\'o}pez-S{\'a}nchez}, {Foster}, {Bamford}, {Brown}, {Driver},
  {Drinkwater}, {Liske}, {Meyer}, {Norberg}, {Robotham}, {Ching}, {Cluver},
  {Croom}, {Kelvin}, {Prescott}, {Steele}, {Thomas}, \& {Wang}}]{gama13}
{Gunawardhana}, M.~L.~P., {Hopkins}, A.~M., {Bland-Hawthorn}, J., {et~al.}
  2013, \mnras, 433, 2764

\bibitem[{{Gunawardhana} {et~al.}(2015){Gunawardhana}, {Hopkins}, {Taylor},
  {Bland-Hawthorn}, {Norberg}, {Baldry}, {Loveday}, {Owers}, {Wilkins},
  {Colless}, {Brown}, {Driver}, {Alpaslan}, {Brough}, {Cluver}, {Croom},
  {Kelvin}, {Lara-L{\'o}pez}, {Liske}, {L{\'o}pez-S{\'a}nchez}, \&
  {Robotham}}]{Guna2015}
{Gunawardhana}, M.~L.~P., {Hopkins}, A.~M., {Taylor}, E.~N., {et~al.} 2015,
  \mnras, 447, 875

\bibitem[{{Hampel}(1974)}]{nmad}
{Hampel}, F.~R. 1974, Journal of the American Statistical Association, 69, 383

\bibitem[{{Hanish} {et~al.}(2006){Hanish}, {Meurer}, {Ferguson}, {Zwaan},
  {Heckman}, {Staveley-Smith}, {Bland-Hawthorn}, {Kilborn}, {Koribalski},
  {Putman}, {Ryan-Weber}, {Oey}, {Kennicutt}, {Knezek}, {Meyer}, {Smith},
  {Webster}, {Dopita}, {Doyle}, {Drinkwater}, {Freeman}, \&
  {Werk}}]{Hanish2006}
{Hanish}, D.~J., {Meurer}, G.~R., {Ferguson}, H.~C., {et~al.} 2006, \apj, 649,
  150

\bibitem[{{Hopkins} \& {Beacom}(2006)}]{Hopkins2006}
{Hopkins}, A.~M. \& {Beacom}, J.~F. 2006, \apj, 651, 142

\bibitem[{{Hsieh} {et~al.}(2017){Hsieh}, {Lin}, {Lin}, {Pan}, {Hsu},
  {S{\'a}nchez}, {Cano-D{\'\i}az}, {Zhang}, {Yan}, {Barrera-Ballesteros},
  {Boquien}, {Riffel}, {Brownstein}, {Cruz-Gonz{\'a}lez}, {Hagen}, {Ibarra},
  {Pan}, {Bizyaev}, {Oravetz}, \& {Simmons}}]{Hsieh2017}
{Hsieh}, B.~C., {Lin}, L., {Lin}, J.~H., {et~al.} 2017, \apjl, 851, L24

\bibitem[{{Huchra} \& {Sargent}(1973)}]{Huchra1973}
{Huchra}, J. \& {Sargent}, W.~L.~W. 1973, \apj, 186, 433

\bibitem[{Hunter(2007)}]{pylab}
Hunter, J.~D. 2007, Computing In Science \& Engineering, 9, 90

\bibitem[{{Iglesias-P{\'a}ramo} {et~al.}(2016){Iglesias-P{\'a}ramo},
  {V{\'\i}lchez}, {Rosales-Ortega}, {S{\'a}nchez}, {Duarte Puertas},
  {Petropoulou}, {Gil de Paz}, {Galbany}, {Moll{\'a}},
  {Catal{\'a}n-Torrecilla}, {Castillo Morales}, {Mast}, {Husemann},
  {Garc{\'\i}a-Benito}, {Mendoza}, {Kehrig}, {P{\'e}rez-Montero}, {Papaderos},
  {Gomes}, {Walcher}, {Gonz{\'a}lez Delgado}, {Marino},
  {L{\'o}pez-S{\'a}nchez}, {Ziegler}, {Flores}, \&
  {Alves}}]{iglesias-Paramo2016}
{Iglesias-P{\'a}ramo}, J., {V{\'\i}lchez}, J.~M., {Rosales-Ortega}, F.~F.,
  {et~al.} 2016, \apj, 826, 71

\bibitem[{{Ilbert} {et~al.}(2015){Ilbert}, {Arnouts}, {Le Floc'h}, {Aussel},
  {Bethermin}, {Capak}, {Hsieh}, {Kajisawa}, {Karim}, {Le F{\`e}vre}, {Lee},
  {Lilly}, {McCracken}, {Michel-Dansac}, {Moutard}, {Renzini}, {Salvato},
  {Sanders}, {Scoville}, {Sheth}, {Silverman}, {Smol{\v{c}}i{\'c}},
  {Taniguchi}, \& {Tresse}}]{Ilbert2015}
{Ilbert}, O., {Arnouts}, S., {Le Floc'h}, E., {et~al.} 2015, \aap, 579, A2

\bibitem[{{Izquierdo-Villalba} {et~al.}(2019){Izquierdo-Villalba}, {Angulo},
  {Orsi}, {Hurier}, {Vilella-Rojo}, {Bonoli}, {L{\'o}pez-Sanjuan}, {Alcaniz},
  {Cenarro}, {Crist{\'o}bal-Hornillos}, {Dupke}, {Ederoclite},
  {Hern{\'a}ndez-Monteagudo}, {Mar{\'\i}n-Franch}, {Moles}, {Mendes de
  Oliveira}, {Sodr{\'e}}, {Varela}, \& {V{\'a}zquez
  Rami{\'o}}}]{IzquierdoVillalba2019}
{Izquierdo-Villalba}, D., {Angulo}, R.~E., {Orsi}, A., {et~al.} 2019, \aap,
  631, A82

\bibitem[{{James} {et~al.}(2008){James}, {Knapen}, {Shane}, {Baldry}, \& {de
  Jong}}]{James2008}
{James}, P.~A., {Knapen}, J.~H., {Shane}, N.~S., {Baldry}, I.~K., \& {de Jong},
  R.~S. 2008, \aap, 482, 507

\bibitem[{{James} {et~al.}(2004){James}, {Shane}, {Beckman}, {Cardwell},
  {Collins}, {Etherton}, {de Jong}, {Fathi}, {Knapen}, {Peletier}, {Percival},
  {Pollacco}, {Seigar}, {Stedman}, \& {Steele}}]{James2004}
{James}, P.~A., {Shane}, N.~S., {Beckman}, J.~E., {et~al.} 2004, \aap, 414, 23

\bibitem[{{Jim{\'e}nez-Teja} {et~al.}(2019){Jim{\'e}nez-Teja}, {Dupke}, {Lopes
  de Oliveira}, {Xavier}, {Coelho}, {Chies-Santos}, {L{\'o}pez-Sanjuan},
  {Alvarez-Candal}, {Costa-Duarte}, {Telles}, {Hernandez-Jimenez},
  {Ben{\'\i}tez}, {Alcaniz}, {Cenarro}, {Crist{\'o}bal-Hornillos},
  {Ederoclite}, {Mar{\'\i}n-Franch}, {Mendes de Oliveira}, {Moles},
  {Sodr{\'e}}, {Varela}, \& {V{\'a}zquez Rami{\'o}}}]{JimenezTeja2019}
{Jim{\'e}nez-Teja}, Y., {Dupke}, R.~A., {Lopes de Oliveira}, R., {et~al.} 2019,
  \aap, 622, A183

\bibitem[{{Kawinwanichakij} {et~al.}(2020){Kawinwanichakij}, {Papovich},
  {Ciardullo}, {Finkelstein}, {Stevans}, {Wold}, {Jogee}, {Sherman}, {Florez},
  \& {Gronwall}}]{kawin20}
{Kawinwanichakij}, L., {Papovich}, C., {Ciardullo}, R., {et~al.} 2020, \apj,
  892, 7

\bibitem[{{Kehrig} {et~al.}(2012){Kehrig}, {Monreal-Ibero}, {Papaderos},
  {V{\'\i}lchez}, {Gomes}, {Masegosa}, {S{\'a}nchez}, {Lehnert}, {Cid
  Fernandes}, {Bland-Hawthorn}, {Bomans}, {Marquez}, {Mast}, {Aguerri},
  {L{\'o}pez-S{\'a}nchez}, {Marino}, {Pasquali}, {Perez}, {Roth},
  {S{\'a}nchez-Bl{\'a}zquez}, \& {Ziegler}}]{Kehrig2012}
{Kehrig}, C., {Monreal-Ibero}, A., {Papaderos}, P., {et~al.} 2012, \aap, 540,
  A11

\bibitem[{{Kennicutt}(1998)}]{Kennicutt1998}
{Kennicutt}, Jr., R.~C. 1998, ARA\&A, 36, 189

\bibitem[{{Khostovan} {et~al.}(2015){Khostovan}, {Sobral}, {Mobasher}, {Best},
  {Smail}, {Stott}, {Hemmati}, \& {Nayyeri}}]{Khostovan2015}
{Khostovan}, A.~A., {Sobral}, D., {Mobasher}, B., {et~al.} 2015, \mnras, 452,
  3948

\bibitem[{{Lara-L{\'o}pez} {et~al.}(2013){Lara-L{\'o}pez}, {Hopkins},
  {L{\'o}pez-S{\'a}nchez}, {Brough}, {Gunawardhana}, {Colless}, {Robotham},
  {Bauer}, {Bland-Hawthorn}, {Cluver}, {Driver}, {Foster}, {Kelvin}, {Liske},
  {Loveday}, {Owers}, {Ponman}, {Sharp}, {Steele}, {Taylor}, \&
  {Thomas}}]{LaraLopez2013}
{Lara-L{\'o}pez}, M.~A., {Hopkins}, A.~M., {L{\'o}pez-S{\'a}nchez}, A.~R.,
  {et~al.} 2013, \mnras, 434, 451

\bibitem[{{Lee} {et~al.}(2015){Lee}, {Sanders}, {Casey}, {Toft}, {Scoville},
  {Hung}, {Le Floc'h}, {Ilbert}, {Zahid}, {Aussel}, {Capak}, {Kartaltepe},
  {Kewley}, {Li}, {Schawinski}, {Sheth}, \& {Xiao}}]{Lee2015}
{Lee}, N., {Sanders}, D.~B., {Casey}, C.~M., {et~al.} 2015, \apj, 801, 80

\bibitem[{{Leslie} {et~al.}(2020){Leslie}, {Schinnerer}, {Liu}, {Magnelli},
  {Algera}, {Karim}, {Davidzon}, {Gozaliasl}, {Jim{\'e}nez-Andrade}, {Lang},
  {Sargent}, {Novak}, {Groves}, {Smol{\v{c}}i{\'c}}, {Zamorani}, {Vaccari},
  {Battisti}, {Vardoulaki}, {Peng}, \& {Kartaltepe}}]{Leslie2020}
{Leslie}, S.~K., {Schinnerer}, E., {Liu}, D., {et~al.} 2020, \apj, 899, 58

\bibitem[{{Lilly} {et~al.}(1996){Lilly}, {Le Fevre}, {Hammer}, \&
  {Crampton}}]{Lilly1996}
{Lilly}, S.~J., {Le Fevre}, O., {Hammer}, F., \& {Crampton}, D. 1996, ApJL,
  460, L1

\bibitem[{{Logro{\~n}o-Garc{\'{\i}}a}
  {et~al.}(2021){Logro{\~n}o-Garc{\'{\i}}a}, {Vilella-Rojo},
  {L{\'o}pez-Sanjuan}, {Varela}, {Viironen}, \& collaboration}]{Logrono2020}
{Logro{\~n}o-Garc{\'{\i}}a}, R., {Vilella-Rojo}, G., {L{\'o}pez-Sanjuan}, C.,
  {et~al.} 2021, in preparation

\bibitem[{{Logro{\~n}o-Garc{\'{\i}}a}
  {et~al.}(2019){Logro{\~n}o-Garc{\'{\i}}a}, {Vilella-Rojo},
  {L{\'o}pez-Sanjuan}, {Varela}, {Viironen}, {Muniesa}, {Cenarro},
  {Crist{\'o}bal-Hornillos}, {Ederoclite}, {Mar{\'{\i}}n-Franch}, {Moles},
  {V{\'a}zquez Rami{\'o}}, {Bonoli}, {D{\'{\i}}az-Garc{\'{\i}}a}, {Orsi}, {San
  Roman}, {Akras}, {Chies-Santos}, {Coelho}, {Daflon}, {Costa-Duarte}, {Dupke},
  {Galbany}, {Gonz{\'a}lez Delgado}, {Hernandez-Jimenez}, {Lopes de Oliveira},
  {Mendes de Oliveira}, {Oteo}, {Gon{\c c}alves}, {S{\'a}nchez-Portal},
  {Schmidtobreick}, \& {Sodr{\'e}}}]{Logrono2019}
{Logro{\~n}o-Garc{\'{\i}}a}, R., {Vilella-Rojo}, G., {L{\'o}pez-Sanjuan}, C.,
  {et~al.} 2019, \aap, 622, A180

\bibitem[{{L{\'o}pez Fern{\'a}ndez} {et~al.}(2018){L{\'o}pez Fern{\'a}ndez},
  {Gonz{\'a}lez Delgado}, {P{\'e}rez}, {Garc{\'\i}a-Benito}, {Cid Fernandes},
  {Schoenell}, {S{\'a}nchez}, {Gallazzi}, {S{\'a}nchez-Bl{\'a}zquez}, {Vale
  Asari}, \& {Walcher}}]{LopezFernandez2018}
{L{\'o}pez Fern{\'a}ndez}, R., {Gonz{\'a}lez Delgado}, R.~M., {P{\'e}rez}, E.,
  {et~al.} 2018, \aap, 615, A27

\bibitem[{{L{\'o}pez-Sanjuan} {et~al.}(2015){L{\'o}pez-Sanjuan}, {Cenarro},
  {Hern{\'a}ndez-Monteagudo}, {Arnalte-Mur}, {Varela}, {Viironen},
  {Fern{\'a}ndez-Soto}, {Mart{\'\i}nez}, {Alfaro}, {Ascaso}, {del Olmo},
  {D{\'\i}az-Garc{\'\i}a}, {Hurtado-Gil}, {Moles}, {Molino}, {Perea},
  {Povi{\'c}}, {Aguerri}, {Aparicio-Villegas}, {Ben{\'\i}tez}, {Broadhurst},
  {Cabrera-Ca{\~n}o}, {Castander}, {Cepa}, {Cervi{\~n}o},
  {Crist{\'o}bal-Hornillos}, {Gonz{\'a}lez Delgado}, {Husillos}, {Infante},
  {M{\'a}rquez}, {Masegosa}, {Prada}, \& {Quintana}}]{clsj15cv}
{L{\'o}pez-Sanjuan}, C., {Cenarro}, A.~J., {Hern{\'a}ndez-Monteagudo}, C.,
  {et~al.} 2015, \aap, 582, A16

\bibitem[{{L{\'o}pez-Sanjuan} {et~al.}(2019{\natexlab{a}}){L{\'o}pez-Sanjuan},
  {D{\'{\i}}az-Garc{\'{\i}}a}, {Cenarro}, {Fern{\'a}ndez-Soto}, {Viironen},
  {Molino}, {Ben{\'{\i}}tez}, {Crist{\'o}bal-Hornillos}, {Moles}, {Varela},
  {Arnalte-Mur}, {Ascaso}, {Castander}, {Cervi{\~n}o}, {Gonz{\'a}lez Delgado},
  {Husillos}, {M{\'a}rquez}, {Masegosa}, {Del Olmo}, {Povi{\'c}}, \&
  {Perea}}]{carlinhos2019masas}
{L{\'o}pez-Sanjuan}, C., {D{\'{\i}}az-Garc{\'{\i}}a}, L.~A., {Cenarro}, A.~J.,
  {et~al.} 2019{\natexlab{a}}, \aap, 622, A51

\bibitem[{{L{\'o}pez-Sanjuan} {et~al.}(2017){L{\'o}pez-Sanjuan}, {Tempel},
  {Ben{\'{\i}}tez}, {Molino}, {Viironen}, {D{\'{\i}}az-Garc{\'{\i}}a},
  {Fern{\'a}ndez-Soto}, {Santos}, {Varela}, {Cenarro}, {Moles}, {Arnalte-Mur},
  {Ascaso}, {Montero-Dorta}, {Povi{\'c}}, {Mart{\'{\i}}nez}, {Nieves-Seoane},
  {Stefanon}, {Hurtado-Gil}, {M{\'a}rquez}, {Perea}, {Aguerri}, {Alfaro},
  {Aparicio-Villegas}, {Broadhurst}, {Cabrera-Ca{\~n}o}, {Castander}, {Cepa},
  {Cervi{\~n}o}, {Crist{\'o}bal-Hornillos}, {Gonz{\'a}lez Delgado}, {Husillos},
  {Infante}, {Masegosa}, {del Olmo}, {Prada}, \& {Quintana}}]{clsj2017}
{L{\'o}pez-Sanjuan}, C., {Tempel}, E., {Ben{\'{\i}}tez}, N., {et~al.} 2017,
  \aap, 599, A62

\bibitem[{{L{\'o}pez-Sanjuan} {et~al.}(2019{\natexlab{b}}){L{\'o}pez-Sanjuan},
  {Varela}, {Crist{\'o}bal-Hornillos}, {V{\'a}zquez Rami{\'o}}, {Carrasco},
  {Tremblay}, {Whitten}, {Placco}, {Mar{\'\i}n-Franch}, {Cenarro},
  {Ederoclite}, {Alfaro}, {Coelho}, {Civera}, {Hern{\'a}ndez-Fuertes},
  {Jim{\'e}nez-Esteban}, {Jim{\'e}nez-Teja}, {Ma{\'\i}z Apell{\'a}niz},
  {Sobral}, {V{\'\i}lchez}, {Alcaniz}, {Angulo}, {Dupke},
  {Hern{\'a}ndez-Monteagudo}, {Mendes de Oliveira}, {Moles}, \&
  {Sodr{\'e}}}]{Carlinhos_Calibracion}
{L{\'o}pez-Sanjuan}, C., {Varela}, J., {Crist{\'o}bal-Hornillos}, D., {et~al.}
  2019{\natexlab{b}}, \aap, 631, A119

\bibitem[{{L{\'o}pez-Sanjuan} {et~al.}(2019{\natexlab{c}}){L{\'o}pez-Sanjuan},
  {V{\'a}zquez Rami{\'o}}, {Varela}, {Spinoso}, {Angulo}, {Muniesa},
  {Viironen}, {Crist{\'o}bal-Hornillos}, {Cenarro}, {Ederoclite},
  {Mar{\'{\i}}n-Franch}, {Moles}, {Ascaso}, {Bonoli}, {Chies-Santos}, {Coelho},
  {Costa-Duarte}, {Cortesi}, {D{\'{\i}}az-Garc{\'{\i}}a}, {Dupke}, {Galbany},
  {Hern{\'a}ndez-Monteagudo}, {Logro{\~n}o-Garc{\'{\i}}a}, {Molino}, {Orsi},
  {Placco}, {Sampedro}, {San Roman}, {Vilella-Rojo}, {Whitten}, {Mendes de
  Oliveira}, \& {Sodr{\'e}}}]{Carlinhos2019}
{L{\'o}pez-Sanjuan}, C., {V{\'a}zquez Rami{\'o}}, H., {Varela}, J., {et~al.}
  2019{\natexlab{c}}, \aap, 622, A177

\bibitem[{{Ly} {et~al.}(2011){Ly}, {Lee}, {Dale}, {Momcheva}, {Salim},
  {Staudaher}, {Moore}, \& {Finn}}]{Ly2011}
{Ly}, C., {Lee}, J.~C., {Dale}, D.~A., {et~al.} 2011, \apj, 726, 109

\bibitem[{{Ly} {et~al.}(2007){Ly}, {Malkan}, {Kashikawa}, {Shimasaku}, {Doi},
  {Nagao}, {Iye}, {Kodama}, {Morokuma}, \& {Motohara}}]{Ly2007}
{Ly}, C., {Malkan}, M.~A., {Kashikawa}, N., {et~al.} 2007, \apj, 657, 738

\bibitem[{{Madau} \& {Dickinson}(2014)}]{Madau2014}
{Madau}, P. \& {Dickinson}, M. 2014, \araa, 52, 415

\bibitem[{{Maniyar} {et~al.}(2018){Maniyar}, {B{\'e}thermin}, \&
  {Lagache}}]{Maniyar2018}
{Maniyar}, A.~S., {B{\'e}thermin}, M., \& {Lagache}, G. 2018, \aap, 614, A39

\bibitem[{{Marin-Franch} {et~al.}(2015){Marin-Franch}, {Taylor}, {Cenarro},
  {Cristobal-Hornillos}, \& {Moles}}]{t80cam}
{Marin-Franch}, A., {Taylor}, K., {Cenarro}, J., {Cristobal-Hornillos}, D., \&
  {Moles}, M. 2015, in IAU General Assembly, Vol.~29, 2257381

\bibitem[{{McGaugh} {et~al.}(2017){McGaugh}, {Schombert}, \&
  {Lelli}}]{McGaugh17}
{McGaugh}, S.~S., {Schombert}, J.~M., \& {Lelli}, F. 2017, \apj, 851, 22

\bibitem[{{Molino} {et~al.}(2019){Molino}, {Costa-Duarte}, {Mendes de
  Oliveira}, {Cenarro}, {Lima Neto}, {Cypriano}, {Sodr{\'e}}, {Coelho},
  {Chow-Mart{\'\i}nez}, {Monteiro-Oliveira}, {Sampedro}, {Cristobal-Hornillos},
  {Varela}, {Ederoclite}, {Chies-Santos}, {Schoenell}, {Ribeiro},
  {Mar{\'\i}n-Franch}, {L{\'o}pez-Sanjuan}, {Hern{\'a}ndez-Fern{\'a}ndez},
  {Cortesi}, {V{\'a}zquez Rami{\'o}}, {Santos}, {Cibirka}, {Novais}, {Pereira},
  {Hern{\'a}ndez-Jimenez}, {Jimenez-Teja}, {Moles}, {Ben{\'\i}tez}, \&
  {Dupke}}]{Molino2019}
{Molino}, A., {Costa-Duarte}, M.~V., {Mendes de Oliveira}, C., {et~al.} 2019,
  \aap, 622, A178

\bibitem[{{Moster} {et~al.}(2011){Moster}, {Somerville}, {Newman}, \&
  {Rix}}]{moster11}
{Moster}, B.~P., {Somerville}, R.~S., {Newman}, J.~A., \& {Rix}, H.-W. 2011,
  \apj, 731, 113

\bibitem[{{Nakamura} {et~al.}(2004){Nakamura}, {Fukugita}, {Brinkmann}, \&
  {Schneider}}]{Nakamura2004}
{Nakamura}, O., {Fukugita}, M., {Brinkmann}, J., \& {Schneider}, D.~P. 2004,
  \aj, 127, 2511

\bibitem[{{Noeske} {et~al.}(2007){Noeske}, {Weiner}, {Faber}, {Papovich},
  {Koo}, {Somerville}, {Bundy}, {Conselice}, {Newman}, {Schiminovich}, {Le
  Floc'h}, {Coil}, {Rieke}, {Lotz}, {Primack}, {Barmby}, {Cooper}, {Davis},
  {Ellis}, {Fazio}, {Guhathakurta}, {Huang}, {Kassin}, {Martin}, {Phillips},
  {Rich}, {Small}, {Willmer}, \& {Wilson}}]{Noeske2007}
{Noeske}, K.~G., {Weiner}, B.~J., {Faber}, S.~M., {et~al.} 2007, \apjl, 660,
  L43

\bibitem[{{Nogueira-Cavalcante} {et~al.}(2019){Nogueira-Cavalcante}, {Dupke},
  {Coelho}, {Dantas}, {Gon{\c{c}}alves}, {Men{\'e}ndez-Delmestre}, {Lopes de
  Oliveira}, {Jim{\'e}nez-Teja}, {L{\'o}pez-Sanjuan}, {Alcaniz}, {Angulo},
  {Cenarro}, {Crist{\'o}bal-Hornillos}, {Hern{\'a}ndez-Monteagudo},
  {Ederoclite}, {Mar{\'\i}n-Franch}, {Mendes de Oliveira}, {Moles},
  {Sodr{\'e}}, {Varela}, {V{\'a}zquez Rami{\'o}}, {Alvarez-Candal},
  {Chies-Santos}, {D{\'\i}az-Garc{\'\i}a}, {Galbany}, {Hernand ez-Jimenez},
  {S{\'a}nchez-Bl{\'a}zquez}, {S{\'a}nchez-Portal}, {Sobral}, {Telles}, \&
  {Tempel}}]{Nogueira-Cavalcante19}
{Nogueira-Cavalcante}, J.~P., {Dupke}, R., {Coelho}, P., {et~al.} 2019, \aap,
  630, A88

\bibitem[{{Novak} {et~al.}(2017){Novak}, {Smol{\v{c}}i{\'c}}, {Delhaize},
  {Delvecchio}, {Zamorani}, {Baran}, {Bondi}, {Capak}, {Carilli}, {Ciliegi},
  {Civano}, {Ilbert}, {Karim}, {Laigle}, {Le F{\`e}vre}, {Marchesi},
  {McCracken}, {Miettinen}, {Salvato}, {Sargent}, {Schinnerer}, \&
  {Tasca}}]{novak17}
{Novak}, M., {Smol{\v{c}}i{\'c}}, V., {Delhaize}, J., {et~al.} 2017, \aap, 602,
  A5

\bibitem[{{Oke} \& {Gunn}(1983)}]{Oke83}
{Oke}, J.~B. \& {Gunn}, J.~E. 1983, \apj, 266, 713

\bibitem[{{P{\'e}rez-Gonz{\'a}lez} {et~al.}(2003){P{\'e}rez-Gonz{\'a}lez},
  {Zamorano}, {Gallego}, {Arag{\'o}n-Salamanca}, \& {Gil de
  Paz}}]{PerezGonzalez2003}
{P{\'e}rez-Gonz{\'a}lez}, P.~G., {Zamorano}, J., {Gallego}, J.,
  {Arag{\'o}n-Salamanca}, A., \& {Gil de Paz}, A. 2003, \apj, 591, 827

\bibitem[{{Popesso} {et~al.}(2019{\natexlab{a}}){Popesso}, {Concas},
  {Morselli}, {Schreiber}, {Rodighiero}, {Cresci}, {Belli}, {Erfanianfar},
  {Mancini}, {Inami}, {Dickinson}, {Ilbert}, {Pannella}, \&
  {Elbaz}}]{Popesso2019a}
{Popesso}, P., {Concas}, A., {Morselli}, L., {et~al.} 2019{\natexlab{a}},
  \mnras, 483, 3213

\bibitem[{{Popesso} {et~al.}(2019{\natexlab{b}}){Popesso}, {Morselli},
  {Concas}, {Schreiber}, {Rodighiero}, {Cresci}, {Belli}, {Ilbert},
  {Erfanianfar}, {Mancini}, {Inami}, {Dickinson}, {Pannella}, \&
  {Elbaz}}]{Popesso2019b}
{Popesso}, P., {Morselli}, L., {Concas}, A., {et~al.} 2019{\natexlab{b}},
  \mnras, 490, 5285

\bibitem[{{Renzini} \& {Peng}(2015)}]{RenziniPeng2015}
{Renzini}, A. \& {Peng}, Y.-j. 2015, \apjl, 801, L29

\bibitem[{{Robertson}(2010)}]{robertson10}
{Robertson}, B.~E. 2010, \apjl, 716, L229

\bibitem[{{Salmon} {et~al.}(2015){Salmon}, {Papovich}, {Finkelstein}, {Tilvi},
  {Finlator}, {Behroozi}, {Dahlen}, {Dav{\'e}}, {Dekel}, {Dickinson},
  {Ferguson}, {Giavalisco}, {Long}, {Lu}, {Mobasher}, {Reddy}, {Somerville}, \&
  {Wechsler}}]{Salmon2015}
{Salmon}, B., {Papovich}, C., {Finkelstein}, S.~L., {et~al.} 2015, \apj, 799,
  183

\bibitem[{{Salpeter}(1955)}]{Salpeter55}
{Salpeter}, E.~E. 1955, \apj, 121, 161

\bibitem[{{San Roman} {et~al.}(2019){San Roman}, {S{\'a}nchez-Bl{\'a}zquez},
  {Cenarro}, {D{\'\i}az-Garc{\'\i}a}, {L{\'o}pez-Sanjuan}, {Varela},
  {Vilella-Rojo}, {Akras}, {Bonoli}, {Chies Santos}, {Coelho}, {Cortesi},
  {Ederoclite}, {Jim{\'e}nez-Teja}, {Logro{\~n}o-Garc{\'\i}a}, {Lopes de
  Oliveira}, {Nogueira-Cavalcante}, {Orsi}, {V{\'a}zquez Rami{\'o}},
  {Viironen}, {Crist{\'o}bal-Hornillos}, {Dupke}, {Mar{\'\i}n-Franch}, {Mendes
  de Oliveira}, {Moles}, \& {Sodr{\'e}}}]{SanRoman2019}
{San Roman}, I., {S{\'a}nchez-Bl{\'a}zquez}, P., {Cenarro}, A.~J., {et~al.}
  2019, \aap, 622, A181

\bibitem[{{S{\'a}nchez} {et~al.}(2019){S{\'a}nchez}, {Avila-Reese},
  {Rodr{\'\i}guez-Puebla}, {Ibarra-Medel}, {Calette}, {Bershady},
  {Hern{\'a}ndez-Toledo}, {Pan}, \& {Bizyaev}}]{Sanchez2019}
{S{\'a}nchez}, S.~F., {Avila-Reese}, V., {Rodr{\'\i}guez-Puebla}, A., {et~al.}
  2019, \mnras, 482, 1557

\bibitem[{{S{\'a}nchez} {et~al.}(2012){S{\'a}nchez}, {Rosales-Ortega},
  {Marino}, {Iglesias-P{\'a}ramo}, {V{\'{\i}}lchez}, {Kennicutt},
  {D{\'{\i}}az}, {Mast}, {Monreal-Ibero}, {Garc{\'{\i}}a-Benito},
  {Bland-Hawthorn}, {P{\'e}rez}, {Gonz{\'a}lez Delgado}, {Husemann},
  {L{\'o}pez-S{\'a}nchez}, {Cid Fernandes}, {Kehrig}, {Walcher}, {Gil de Paz},
  \& {Ellis}}]{Sanchez2012}
{S{\'a}nchez}, S.~F., {Rosales-Ortega}, F.~F., {Marino}, R.~A., {et~al.} 2012,
  \aap, 546, A2

\bibitem[{{Santini} {et~al.}(2017){Santini}, {Fontana}, {Castellano}, {Di
  Criscienzo}, {Merlin}, {Amorin}, {Cullen}, {Daddi}, {Dickinson}, {Dunlop},
  {Grazian}, {Lamastra}, {McLure}, {Micha{\l}owski}, {Pentericci}, \&
  {Shu}}]{Santini2017}
{Santini}, P., {Fontana}, A., {Castellano}, M., {et~al.} 2017, \apj, 847, 76

\bibitem[{{Schechter}(1976)}]{Schechter76}
{Schechter}, P. 1976, \apj, 203, 297

\bibitem[{{Schmidt}(1968)}]{Schmidt1968}
{Schmidt}, M. 1968, \apj, 151, 393

\bibitem[{{Smith}(2012)}]{Smith2012}
{Smith}, R.~E. 2012, \mnras, 426, 531

\bibitem[{{Sobral} {et~al.}(2011){Sobral}, {Best}, {Smail}, {Geach},
  {Cirasuolo}, {Garn}, \& {Dalton}}]{Sobral2011}
{Sobral}, D., {Best}, P.~N., {Smail}, I., {et~al.} 2011, \mnras, 411, 675

\bibitem[{{Sobral} {et~al.}(2013){Sobral}, {Smail}, {Best}, {Geach}, {Matsuda},
  {Stott}, {Cirasuolo}, \& {Kurk}}]{Sobral2013}
{Sobral}, D., {Smail}, I., {Best}, P.~N., {et~al.} 2013, \mnras, 428, 1128

\bibitem[{{Solano} {et~al.}(2019){Solano}, {Mart{\'\i}n}, {Caballero},
  {Rodrigo}, {Angulo}, {Alcaniz}, {Borges Fernand es}, {Cenarro},
  {Crist{\'o}bal-Hornillos}, {Dupke}, {Alfaro}, {Ederoclite},
  {Jim{\'e}nez-Esteban}, {Hernandez-Jimenez}, {Hern{\'a}ndez-Monteagudo},
  {Lopes de Oliveira}, {L{\'o}pez-Sanjuan}, {Mar{\'\i}n-Franch}, {Mendes de
  Oliveira}, {Moles}, {Orsi}, {Schmidtobreick}, {Sobral}, {Sodr{\'e}},
  {Varela}, \& {V{\'a}zquez Rami{\'o}}}]{Solano19}
{Solano}, E., {Mart{\'\i}n}, E.~L., {Caballero}, J.~A., {et~al.} 2019, \aap,
  627, A29

\bibitem[{{Somerville} {et~al.}(2004){Somerville}, {Lee}, {Ferguson},
  {Gardner}, {Moustakas}, \& {Giavalisco}}]{somerville04}
{Somerville}, R.~S., {Lee}, K., {Ferguson}, H.~C., {et~al.} 2004, \apjl, 600,
  L171

\bibitem[{{Sparre} {et~al.}(2015){Sparre}, {Hayward}, {Springel},
  {Vogelsberger}, {Genel}, {Torrey}, {Nelson}, {Sijacki}, \&
  {Hernquist}}]{Sparre2015}
{Sparre}, M., {Hayward}, C.~C., {Springel}, V., {et~al.} 2015, \mnras, 447,
  3548

\bibitem[{{Speagle} {et~al.}(2014){Speagle}, {Steinhardt}, {Capak}, \&
  {Silverman}}]{Speagle14}
{Speagle}, J.~S., {Steinhardt}, C.~L., {Capak}, P.~L., \& {Silverman}, J.~D.
  2014, \apjs, 214, 15

\bibitem[{{Spinoso} {et~al.}(2020){Spinoso}, {Orsi}, {L{\'o}pez-Sanjuan},
  {Bonoli}, {Viironen}, {Izquierdo-Villalba}, {Sobral}, {Gurung-L{\'o}pez},
  {Hern{\'a}n-Caballero}, {Ederoclite}, {Varela}, {Overzier},
  {Miralda-Escud{\'e}}, {Muniesa}, {Alcaniz}, {Angulo}, {Cenarro},
  {Crist{\'o}bal-Hornillos}, {Dupke}, {Hern{\'a}ndez-Monteagudo},
  {Mar{\'\i}n-Franch}, {Moles}, {Sodr{\'e}}, \&
  {V{\'a}zquez-Rami{\'o}}}]{Spinoso20}
{Spinoso}, D., {Orsi}, A., {L{\'o}pez-Sanjuan}, C., {et~al.} 2020, \aap,
  submitted [\eprint[arXiv]{2006.15084}]

\bibitem[{{Taylor} {et~al.}(2011){Taylor}, {Hopkins}, {Baldry}, {Brown},
  {Driver}, {Kelvin}, {Hill}, {Robotham}, {Bland-Hawthorn}, {Jones}, {Sharp},
  {Thomas}, {Liske}, {Loveday}, {Norberg}, {Peacock}, {Bamford}, {Brough},
  {Colless}, {Cameron}, {Conselice}, {Croom}, {Frenk}, {Gunawardhana},
  {Kuijken}, {Nichol}, {Parkinson}, {Phillipps}, {Pimbblet}, {Popescu},
  {Prescott}, {Sutherland}, {Tuffs}, {van Kampen}, \&
  {Wijesinghe}}]{Taylor2011}
{Taylor}, E.~N., {Hopkins}, A.~M., {Baldry}, I.~K., {et~al.} 2011, \mnras, 418,
  1587

\bibitem[{{Tomczak} {et~al.}(2016){Tomczak}, {Quadri}, {Tran}, {Labb{\'e}},
  {Straatman}, {Papovich}, {Glazebrook}, {Allen}, {Brammer}, {Cowley},
  {Dickinson}, {Elbaz}, {Inami}, {Kacprzak}, {Morrison}, {Nanayakkara},
  {Persson}, {Rees}, {Salmon}, {Schreiber}, {Spitler}, \&
  {Whitaker}}]{Tomczak2016}
{Tomczak}, A.~R., {Quadri}, R.~F., {Tran}, K.-V.~H., {et~al.} 2016, \apj, 817,
  118

\bibitem[{{Tully} \& {Fisher}(1977)}]{TullyFisher1977}
{Tully}, R.~B. \& {Fisher}, J.~R. 1977, \aap, 54, 661

\bibitem[{{Van Sistine} {et~al.}(2016){Van Sistine}, {Salzer}, {Sugden},
  {Giovanelli}, {Haynes}, {Janowiecki}, {Jaskot}, \& {Wilcots}}]{ALFALFA2016}
{Van Sistine}, A., {Salzer}, J.~J., {Sugden}, A., {et~al.} 2016, \apj, 824, 25

\bibitem[{{Vilella-Rojo} {et~al.}(2015){Vilella-Rojo}, {Viironen},
  {L{\'o}pez-Sanjuan}, {Cenarro}, {Varela}, {D{\'{\i}}az-Garc{\'{\i}}a},
  {Crist{\'o}bal-Hornillos}, {Ederoclite}, {Mar{\'{\i}}n-Franch}, \&
  {Moles}}]{mio}
{Vilella-Rojo}, G., {Viironen}, K., {L{\'o}pez-Sanjuan}, C., {et~al.} 2015,
  \aap, 580, A47

\bibitem[{{Wenger} {et~al.}(2000){Wenger}, {Ochsenbein}, {Egret}, {Dubois},
  {Bonnarel}, {Borde}, {Genova}, {Jasniewicz}, {Lalo{\"e}}, {Lesteven}, \&
  {Monier}}]{simbad}
{Wenger}, M., {Ochsenbein}, F., {Egret}, D., {et~al.} 2000, \aaps, 143, 9

\bibitem[{{Westra} {et~al.}(2010){Westra}, {Geller}, {Kurtz}, {Fabricant}, \&
  {Dell'Antonio}}]{Westra2010}
{Westra}, E., {Geller}, M.~J., {Kurtz}, M.~J., {Fabricant}, D.~G., \&
  {Dell'Antonio}, I. 2010, \apj, 708, 534

\bibitem[{{Whitaker} {et~al.}(2012){Whitaker}, {van Dokkum}, {Brammer}, \&
  {Franx}}]{Whitaker2012}
{Whitaker}, K.~E., {van Dokkum}, P.~G., {Brammer}, G., \& {Franx}, M. 2012,
  \apjl, 754, L29

\bibitem[{{Whitten} {et~al.}(2019){Whitten}, {Placco}, {Beers}, {Chies-Santos},
  {Bonatto}, {Varela}, {Crist{\'o}bal-Hornillos}, {Ederoclite}, {Masseron},
  {Lee}, {Akras}, {Borges Fernandes}, {Caballero}, {Cenarro}, {Coelho},
  {Costa-Duarte}, {Daflon}, {Dupke}, {Lopes de Oliveira}, {L{\'o}pez-Sanjuan},
  {Mar{\'\i}n-Franch}, {Mendes de Oliveira}, {Moles}, {Orsi}, {Rossi},
  {Sodr{\'e}}, \& {V{\'a}zquez Rami{\'o}}}]{Whitten2019}
{Whitten}, D.~D., {Placco}, V.~M., {Beers}, T.~C., {et~al.} 2019, \aap, 622,
  A182

\bibitem[{{York} {et~al.}(2000){York}, {Adelman}, {Anderson}, {Anderson},
  {Annis}, {Bahcall}, {Bakken}, {Barkhouser}, {Bastian}, {Berman}, {Boroski},
  {Bracker}, {Briegel}, {Briggs}, {Brinkmann}, {Brunner}, {Burles}, {Carey},
  {Carr}, {Castander}, {Chen}, {Colestock}, {Connolly}, {Crocker}, {Csabai},
  {Czarapata}, {Davis}, {Doi}, {Dombeck}, {Eisenstein}, {Ellman}, {Elms},
  {Evans}, {Fan}, {Federwitz}, {Fiscelli}, {Friedman}, {Frieman}, {Fukugita},
  {Gillespie}, {Gunn}, {Gurbani}, {de Haas}, {Haldeman}, {Harris}, {Hayes},
  {Heckman}, {Hennessy}, {Hindsley}, {Holm}, {Holmgren}, {Huang}, {Hull},
  {Husby}, {Ichikawa}, {Ichikawa}, {Ivezi{\'c}}, {Kent}, {Kim}, {Kinney},
  {Klaene}, {Kleinman}, {Kleinman}, {Knapp}, {Korienek}, {Kron}, {Kunszt},
  {Lamb}, {Lee}, {Leger}, {Limmongkol}, {Lindenmeyer}, {Long}, {Loomis},
  {Loveday}, {Lucinio}, {Lupton}, {MacKinnon}, {Mannery}, {Mantsch}, {Margon},
  {McGehee}, {McKay}, {Meiksin}, {Merelli}, {Monet}, {Munn}, {Narayanan},
  {Nash}, {Neilsen}, {Neswold}, {Newberg}, {Nichol}, {Nicinski}, {Nonino},
  {Okada}, {Okamura}, {Ostriker}, {Owen}, {Pauls}, {Peoples}, {Peterson},
  {Petravick}, {Pier}, {Pope}, {Pordes}, {Prosapio}, {Rechenmacher}, {Quinn},
  {Richards}, {Richmond}, {Rivetta}, {Rockosi}, {Ruthmansdorfer}, {Sand ford},
  {Schlegel}, {Schneider}, {Sekiguchi}, {Sergey}, {Shimasaku}, {Siegmund},
  {Smee}, {Smith}, {Snedden}, {Stone}, {Stoughton}, {Strauss}, {Stubbs},
  {SubbaRao}, {Szalay}, {Szapudi}, {Szokoly}, {Thakar}, {Tremonti}, {Tucker},
  {Uomoto}, {Vanden Berk}, {Vogeley}, {Waddell}, {Wang}, {Watanabe},
  {Weinberg}, {Yanny}, {Yasuda}, \& {SDSS Collaboration}}]{sdss}
{York}, D.~G., {Adelman}, J., {Anderson}, John~E., J., {et~al.} 2000, \aj, 120,
  1579

\bibitem[{{Zahid} {et~al.}(2012){Zahid}, {Dima}, {Kewley}, {Erb}, \&
  {Dav{\'e}}}]{Zahid2012}
{Zahid}, H.~J., {Dima}, G.~I., {Kewley}, L.~J., {Erb}, D.~K., \& {Dav{\'e}}, R.
  2012, \apj, 757, 54

\bibitem[{{Zamorano} {et~al.}(1994){Zamorano}, {Rego}, {Gallego}, {Vitores},
  {Gonzalez-Riestra}, \& {Rodriguez-Caderot}}]{Zamorano1994}
{Zamorano}, J., {Rego}, M., {Gallego}, J.~G., {et~al.} 1994, \apjs, 95, 387

\end{thebibliography}

\begin{appendix}

\section{Estimation of redshift-independent distances}\label{sect:A_Distance_errors}
The purpose of this Appendix is to explain how we assign uncertainties to redshift-independent distances. These uncertainties are the ones that we use in the Monte-Carlo sampling process, leading to the values of H$\alpha$ luminosity and stellar masses upon which we derive all the results. 

As we mentioned in Sect.~\ref{sect:Distancias}, we first check the NED for redshift-independent distances. In the NED, these are given by the distance modulus, $(m-M)$. In the case that a galaxy has multiple determinations of $(m-M)$, we compile all of them, and their corresponding uncertainties, $\epsilon\left(m-M\right)$ . Most of these are obtained using the Tully-Fischer relation \citep{TullyFisher1977}, but a small percentage of them use other indicators, such as the tip of the Red Giant Branch, Cepheid variables, or Type Ia Super Novae. 

After gathering the information from NED, each galaxy may have a different number of $\left(m-M\right)$ measurements. Thus, for each galaxy we compute three statistics:
\begin{enumerate}
    \item the median $(m-M)$, which we refer to as $\langle \left(m-M\right) \rangle$,
    \item the standard deviation of all the $(m-M)$ values, referred to as $\sigma \left(m-M\right)$,
    \item the median uncertainty $\epsilon (m-M)$, which we refer to as $\langle \epsilon \left(m-M\right) \rangle$.
\end{enumerate}

To assign a final uncertainty to a distance measurement, we combine both the standard deviation and the median uncertainty, such that
\begin{equation}\label{eq:delta_distancias}
    \delta \left(m-M\right) = \sqrt{ \langle\epsilon \left( m-M \right)\rangle^{2} + \left[  \sigma \left( m-M \right)\right]^{2}  } \,.
\end{equation}
We do this because the typical uncertainty in $\left(m-M\right)$ is significantly larger than the standard deviation of a collection of $\left(m-M\right) $ measurements. By compiling around $400$ measurements from the NED\footnote{As mentioned before, some galaxies have more than one measurement} we find that, regardless of the distance modulus, the median uncertainty converges to a value of $\delta \left( m-M \right) =0.42$. 

Now, to convert distance moduli $\left(m-M\right)$ to luminosity distances, $d_{L}$, we use:
\begin{equation}\label{dist_modulus_dist}
    \left(m-M\right)=5\,\left( \log\,d_{L} - 1 \right) \,,
\end{equation}

An expression for the error in distance computed with the distance moduli is obtained using the standard procedure of error propagation in Eq.~(\ref{dist_modulus_dist}), which leads to:
\begin{equation}\label{err_dist_modulus_dist}
    \delta\left( d_{L} \right) = \frac{1}{5}\,\mathrm{ln}\left(10\right)\,d_{L}\,\delta\left( m-M \right) = \,0.461\cdot d \cdot\delta\left( m-M \right).
\end{equation}
This means that the relative error budget of the distance obtained with this method is around a $20\%$:
\begin{equation}\label{rel_err_dist_modulus}
    \frac{\delta\left( d_{L} \right)}{d_{L}} = \,0.461\cdot\delta\left( m-M \right) \approx 0.461\cdot0.42\approx 0.2\,.
\end{equation}

In the end, we use $\langle \left(m-M\right) \rangle $ and Eq.~(\ref{dist_modulus_dist}) to compute the redshift-independent luminosity distance, and $\delta \left(m-M\right)$ and Eq.~(\ref{err_dist_modulus_dist}) to compute its uncertainty.

\section{Impact of the distance algorithm in the results}\label{sect:A_Distance_Algorithm}
In Section~\ref{sect:Distancias}, we explain the algorithm that we use to assign distances and their uncertainties to our sample of galaxies. In this Appendix, we want to quantify the impact of the algorithm that we use, when compared to other possible choices for the distance. 

\subsection{Definition of the distance Runs}
For this approach, we analyzed the exact same data in several different ways, that we will refer to as Runs. Some aspects are common to all the runs. These are:
\begin{enumerate}
    \item All the Runs consist of $300$ realizations to sample the parameter space following a Monte-Carlo approach.
    \item In each of them, H$\alpha$ fluxes are always perturbed with a random component drawn from a Gaussian distribution with $\sigma=\delta F_{\mathrm{H}\alpha}$.
    \item In all Runs, galaxies from Sample~$G0$ (no information of the distance whatsoever) are assigned a random distance according to a volume prior.
    \item Sample $G1$ galaxies are assigned a distance according to their spectroscopic redshift. This is perturbed with a term of peculiar velocity $v'$ that is drawn from a Gaussian distribution with $\mu=0$ and $\sigma = v_{\mathrm{peculiar}}$, so each galaxy is perturbed with a different $v'$. 
     \begin{equation*}
         z_{\mathrm{perturbed}} = \left( 1+z_{\mathrm{obs}} \right)\cdot \left( 1 + \frac{v'}{c} \right) -1\,.
     \end{equation*}    
     
    \item Galaxies with a redshift-independent distance, and below a certain distance limit (which we refer to as $d_{\mathrm{lim}}$ in this Section), are assigned their redshift-independent distance if $d<d_{\mathrm{lim}}$, and are perturbed with a term of noise that is drawn from a Gaussian distribution, with $\mu=0$ and $\sigma = \delta d$. The way we compute the uncertainty is explained in Appendix~\ref{sect:A_Distance_errors}.
    \item Galaxies with a redshift-independent distance, but with $d>d_{\mathrm{lim}}$, are assigned a distance using their spectroscopic redshift, and perturbed like it is described in Step $4$.
\end{enumerate}

The main difference between each run is the way we assign distances to galaxies, and their uncertainties. We summarize the properties of each run in Table~\ref{tab:propiedades_runs}, and briefly describe the motivation of each set of simulations.

\begin{table}[]
    \centering
    \caption{Summary of the properties of each Run.}
    \begin{tabular}{cccccc}
        \hline\hline\noalign{\smallskip}
    $v_{\rm peculiar}$      &       \multicolumn{5}{c}{$d_{\rm lim}\ \ {\rm [Mpc]}$}  \\
                                           \cline{2-6}\noalign{\smallskip}
      $\rm{[km\,s^{-1}]}$   & $0$    &   $40$  &    $50$    &  $60$            &   $\infty$    \\ 
                    \noalign{\smallskip}\hline\noalign{\smallskip}
$0$    & Run $1$   &      -    &   -       &    -                &  Run $7$      \\
$500$  & Run $2$   &   Run $5$    &   Run $6$    &    -                &  Run$8$      \\
$750$  & Run $3$   &      -    &       -   &   \textbf{Run $0$}  &  Run $9$      \\
$1000$ & Run $4$   &      -    &       -   &   -                 &  Run $10$     \\
\noalign{\smallskip}\hline
\end{tabular}
\label{tab:propiedades_runs}
\end{table}

\begin{table*}
\caption{Impact of the distance assignment on the main parameters derived along this paper}\label{tab:resumen_runs}
\centering
\begin{tabular}{lllllll} 
\hline\hline\noalign{\smallskip}
Parameter & Run $0$ & Run $1$ & Run $2$ & Run $3$ & Run $4$ & Run $5$ \\
\noalign{\smallskip}\hline\noalign{\smallskip}
$\log L_{\mathrm{H}\alpha}^{*}$\ \ $\mathrm{[erg\,s^{-1}]}$    & ${41.34}\pm^{0.12}_{0.1}$ &   ${41.37}\pm^{0.14}_{0.1}$ &   ${41.37}\pm^{0.14}_{0.1}$ &  ${41.4}\pm^{0.14}_{0.1}$ &    ${41.44}\pm^{0.14}_{0.11}$ & ${41.34}\pm^{0.12}_{0.1}$ \\
$\alpha$                             & ${-1.25}\pm^{0.07}_{0.07}$ &  ${-1.28}\pm^{0.07}_{0.07}$ &  ${-1.26}\pm^{0.07}_{0.07}$ & ${-1.27}\pm^{0.07}_{0.07}$ &  ${-1.28}\pm^{0.07}_{0.07}$ & ${-1.25}\pm^{0.07}_{0.07}$  \\
$\log \phi^{\star}$\ \ $\mathrm{[Mpc^{-3}]}$ & ${-2.43}\pm^{0.11}_{0.13}$ &  ${-2.47}\pm^{0.11}_{0.14}$ &  ${-2.45}\pm^{0.11}_{0.14}$ & ${-2.47}\pm^{0.11}_{0.14}$ &  ${-2.5}\pm^{0.12}_{0.14}$ &  ${-2.43}\pm^{0.11}_{0.13}$ \\
$\log \rho_{\star}$\ \ $\mathrm{[M_{\odot}\,yr^{-1}\,Mpc^{-3}]}$ & ${-2.1}\pm^{0.04}_{0.04}$ &   ${-2.09}\pm^{0.05}_{0.04}$ &  ${-2.09}\pm^{0.05}_{0.04}$ & ${-2.08}\pm^{0.05}_{0.04}$ &  ${-2.06}\pm^{0.05}_{0.04}$ & ${-2.1}\pm^{0.04}_{0.04}$  \\
$a$ (Blue Sample)             & ${0.83}\pm^{0.05}_{0.05}$ &   ${0.82}\pm^{0.05}_{0.05}$ &   ${0.83}\pm^{0.05}_{0.05}$ &  ${0.84}\pm^{0.05}_{0.05}$ &   ${0.84}\pm^{0.05}_{0.05}$ &  ${0.83}\pm^{0.05}_{0.05}$ \\
$b$ (Blue Sample)             & ${-8.44}\pm^{0.5}_{0.5}$ &    ${-8.38}\pm^{0.49}_{0.49}$ &  ${-8.46}\pm^{0.49}_{0.49}$ & ${-8.49}\pm^{0.49}_{0.49}$ &  ${-8.53}\pm^{0.49}_{0.49}$ & ${-8.43}\pm^{0.5}_{0.5}$\\
\noalign{\smallskip}\hline\hline\noalign{\smallskip}
       & Run $6$ & Run $7$ & Run $8$ & Run $9$ & Run $10$ &  \\
\noalign{\smallskip}\hline\noalign{\smallskip}
$\log L_{\mathrm{H}\alpha}^{*}$\ \ $\mathrm{[erg\,s^{-1}]}$    &  ${41.33}\pm^{0.12}_{0.1}$ &    ${41.37}\pm^{0.12}_{0.1}$ &   ${41.35}\pm^{0.12}_{0.1}$ &  ${41.36}\pm^{0.12}_{0.1}$ &  ${41.38}\pm^{0.13}_{0.11}$ \\
$\alpha$                             &  ${-1.25}\pm^{0.07}_{0.07}$ &   ${-1.28}\pm^{0.07}_{0.06}$ &  ${-1.25}\pm^{0.07}_{0.07}$ & ${-1.25}\pm^{0.07}_{0.07}$ & ${-1.25}\pm^{0.07}_{0.07}$ \\
$\log \phi^{\star}$\ \ $\mathrm{[Mpc^{-3}]}$ & ${-2.43}\pm^{0.11}_{0.13}$ & ${-2.47}\pm^{0.11}_{0.13}$ &  ${-2.43}\pm^{0.1}_{0.12}$ &  ${-2.44}\pm^{0.11}_{0.12}$ & ${-2.45}\pm^{0.11}_{0.14}$ \\
$\log \rho_{\star}$\ \ $\mathrm{[M_{\odot}\,yr^{-1}\,Mpc^{-3}]}$ &   ${-2.1}\pm^{0.04}_{0.04}$ &    ${-2.1}\pm^{0.04}_{0.04}$ &   ${-2.1}\pm^{0.04}_{0.04}$ &  ${-2.09}\pm^{0.04}_{0.04}$ & ${-2.08}\pm^{0.05}_{0.04}$ \\
$a$ (Blue Sample)             &  ${0.83}\pm^{0.05}_{0.05}$ &    ${0.82}\pm^{0.05}_{0.05}$ &   ${0.83}\pm^{0.05}_{0.05}$ &  ${0.83}\pm^{0.05}_{0.05}$ &  ${0.83}\pm^{0.05}_{0.05}$ \\
$b$ (Blue Sample)             &   ${-8.43}\pm^{0.5}_{0.5}$ &     ${-8.38}\pm^{0.49}_{0.49}$ &  ${-8.43}\pm^{0.5}_{0.5}$ &   ${-8.44}\pm^{0.5}_{0.5}$ &   ${-8.47}\pm^{0.49}_{0.49}$ \\
\noalign{\smallskip}\hline
\end{tabular}
\end{table*}

\paragraph{Run 0:}
This is the reference run, with which we obtain the values reported in the paper. We use $v_{\mathrm{peculiar}}=750\,\mathrm{km\,s^{-1}}$ and $d_{\mathrm{lim}} = 60$ Mpc. The choice of these fiducial values is justified in Sect.~$\ref{sect:Distancias}$.

\paragraph{Run $1$, Run $2$, Run $3$, and Run $4$:}
As can be seen in Table~\ref{tab:propiedades_runs}, these set of Runs have all in common $d_{\mathrm{lim}}=0$ Mpc, which means that we never use the redshift-independent distances. All galaxies are assigned a distance according to their spectroscopic redshift. The only difference between them is the $v_{\mathrm{peculiar}}$ that we use. In the case of Run $1$, we do not perturb distances at all, except for those that do not have either a redshift-independent measurement or a spectroscopic redshift, which are assigned a random distance each iteration. This will help us asses the importance of our assumption of $v_{\mathrm{peculiar}}$.

\paragraph{Run $7$, Run $8$, Run $9$, and Run $10$:}
Contrarily to the case of Runs $1$, $2$, $3$, and $4$, in these Runs we set $d_{\mathrm{lim}}=\infty$ Mpc,  which means that whenever a galaxy has a redshift-independent distance, we use it regardless of the distance. These set of Runs provide insights into the impact of using distances imposing any cut in $d_{\mathrm{lim}}$, disregarding the fact that at some point errors in redshift-independent distances can dominate over redshift-derived ones and introduce extra uncertainty.

\paragraph{Run $5$ and Run $6$:}
These runs are the middle point between the two other cases, slightly closer to the set up for Run $0$. 

\paragraph{Non-tested combinations:}
There are combinations that have not been tested as they lack physical sense, or would only provide redundant information. For instance, we have not considered the cases where $v_{\mathrm{peculiar}}=0\,\mathrm{km\,s^{-1}}$ and $d_{\mathrm{lim}}=40$ Mpc or $d_{\mathrm{lim}}=60$ Mpc. If we are going to assign errors to the sample with redshift-independent distance it had no point to not add errors to redshift-derived distances. On the other hand, if $v_{\mathrm{peculiar}}\geq1000\,\mathrm{km\,s^{-1}}$, the uncertainty that is introduced dominates over the error associated to redshift-independent distances. Hence, it has no sense to add a noise budget that is more likely to dominate over another source of uncertainty that is better constrained.  

\subsection{Results}
\begin{figure*}
    \centering
    \includegraphics{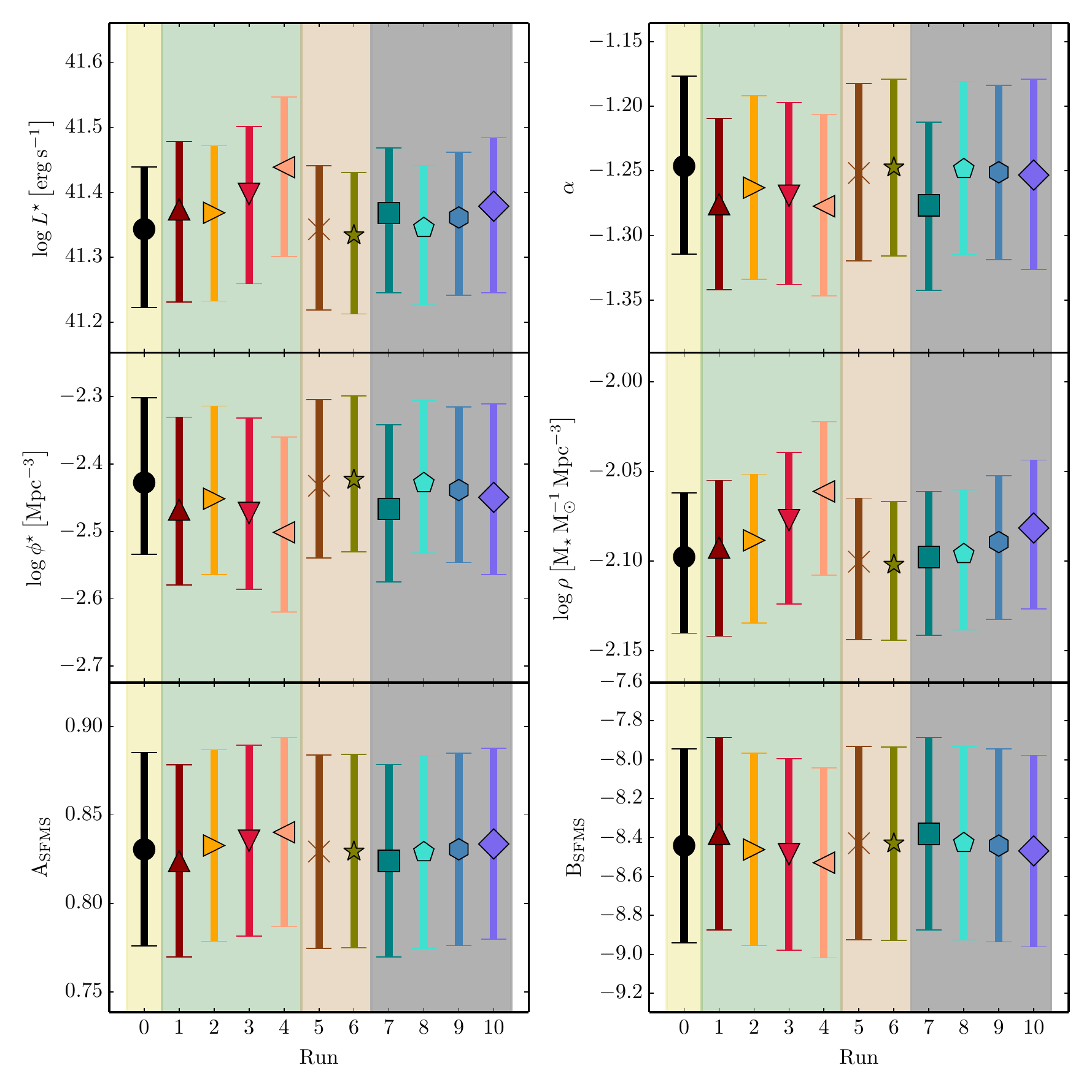}
    \caption{Summary of the best-fitting parameters for the Mass-weighted, Full Sample, H$\alpha$ LFs, and SFMS, as a function of the Run. For the properties of each Run, see Table~\ref{tab:resumen_runs}  }
    \label{fig:Resumen_runs}
\end{figure*}

We now compare the outcome of each run with the others to understand the impact of each assumption. We plot the values of $\log L_{\mathrm{H}\alpha}^{*}$, $\alpha$, $\log \phi^{*}$, and $\log \rho_{\mathrm{SFR}}$ in Figure~\ref{fig:Resumen_runs}, and present them in Table~\ref{tab:resumen_runs}.

We find that our algorithm to assign distances does not have a major impact on the values of the Schechter distribution, or on its integral. All of the eleven Runs are in good agreement, and their dispersion is well constrained by the error bars of each estimation.

If we consider the two extreme cases, which are Runs $1$ and $10$, we find that they are not the most dissimilar. In fact, the most discrepant values appear when we compare Runs $1$ and $4$, which both belong to the set of Runs that never use redshift-independent distances.  We see that the assumptions in $v_{\mathrm{peculiar}}$ are the ones that affect more the results, while the mixed methods (i.e., Runs $0$, $5$, and $6$) retrieve almost the same results for each parameter, being their discrepancies insignificant compared to their error bars. 

\subsection{Impact of the distance assignment}
In this Appendix, we have studied the impact of our assumptions when assigning distances to galaxies. To do so, we have performed eleven different measurements, changing each time the value of the two free parameters that we considered in our model: these are the distance when redshift-based distances have smaller uncertainty than redshift independent measurements ($d_{\mathrm{lim}}$) and the peculiar velocity field from which we draw a perturbation for the spectroscopic redshifts, $v_{\mathrm{peculiar}}$.

We find that all the values that we obtain are in good agreement within each other, leading us to conclude that our method to retrieve distances is not having a large impact in the results that have been presented in this paper.

\section{$V_{\rm int}/V_{\rm max}$ correction}\label{sect:VintVmax}
In Section~\ref{sect:HaLF}, we have computed the H$\alpha$LF using two different approaches: using the stellar mass function as a proxy for incompleteness, and assuming a correction based on the volume that we observe given our limit in H$\alpha$ flux. In this Appendix we provide more insight into this correction, which is usually referred to as $V_{\rm int}/V_{\rm max}$.

The core idea of this correction is to compensate for the Malmquist bias, which appears when one selects objects of fixed apparent magnitude (or, to put it in other words, with a cut in flux). The volume containing the more distant, intrinsically luminous objects is larger than the occupied by the nearer, intrinsically fainter ones. However, the most distant volume is poorly surveyed due to the the fact that intrinsically fainter objects at large distances will have fluxes below our limiting flux. The $V_{\rm int}/V_{\rm max}$ technique aims to correct this effect in a way in which one does not require any \emph{a priori} information, except the assumption that any sufficiently large sub-volume in your survey will be populated by objects with the same luminosity distribution. This is the weakest point of the method (see \citealt{Efstathiou1988}, for a detailed discussion on corrections). This technique is used in many studies to correct luminosity functions at different redshifts \citep{Gallego1995, PerezGonzalez2003, James2008, Ly2011, Bothwell2011}. Basically:

\begin{equation}\label{eq:vint}
 V_{\mathrm{int}} = \frac{1}{\Delta L} \int_{\mathrm{z}_{1}}^{\mathrm{z}_{2}} \int_{G(L)}^{L_{2}} dL d\mathrm{z} \frac{dV}{d\mathrm{z} d\Omega}\,,
\end{equation}

\noindent where $L_{1}$ and $L_{2}$ are the edges of the luminosity bin for which we want to compute the completeness, $\Delta L = L_{2}-L_{1}$, $G\left( L \right)$ is a function defined as $G\left(L\right) \equiv \max{[L_{1},\mathrm{F_{lim}}4\pi d^{2}(\mathrm{z})]}$, $\mathrm{z}_{1} = 0.001$, and $\mathrm{z}_{2}$ is the maximum redshift were we would be able to detect a galaxy with luminosity $L_{2}$ giving our limiting flux. More formally, and assuming a flat $\Lambda\mathrm{CDM}$ Universe (i.e., $\Omega_{k}=0$):

\begin{equation}\label{vint_dif}
    \frac{dV}{d\mathrm{z}d\Omega} = D_{H}^{3} \frac{1 }{ E\left(\mathrm{z}\right)} \left[ \int_{0}^{\mathrm{z}'} \frac{d\mathrm{z}'}{E\left(\mathrm{z}'\right)} \right]\,,
\end{equation}

\noindent where $D_{H} \equiv c/H_{0}$, and

\begin{equation}\label{vint_Ez}
    E\left(z\right) \equiv \sqrt{\Omega_{M}\left(1+z\right)^{3}+\Omega_{\Lambda}}.
\end{equation}

Finally, the completeness for each luminosity bin is estimated as $V_{\rm int}/V_{\rm max}$, where $V_{\rm max}$ is the maximum available volume and it is estimated with $G(L) = L_1$ and $\mathrm{z}_2 = 0.017$ in Eq.~(\ref{eq:vint}).

\begin{figure}
\center
\includegraphics[scale=0.45]{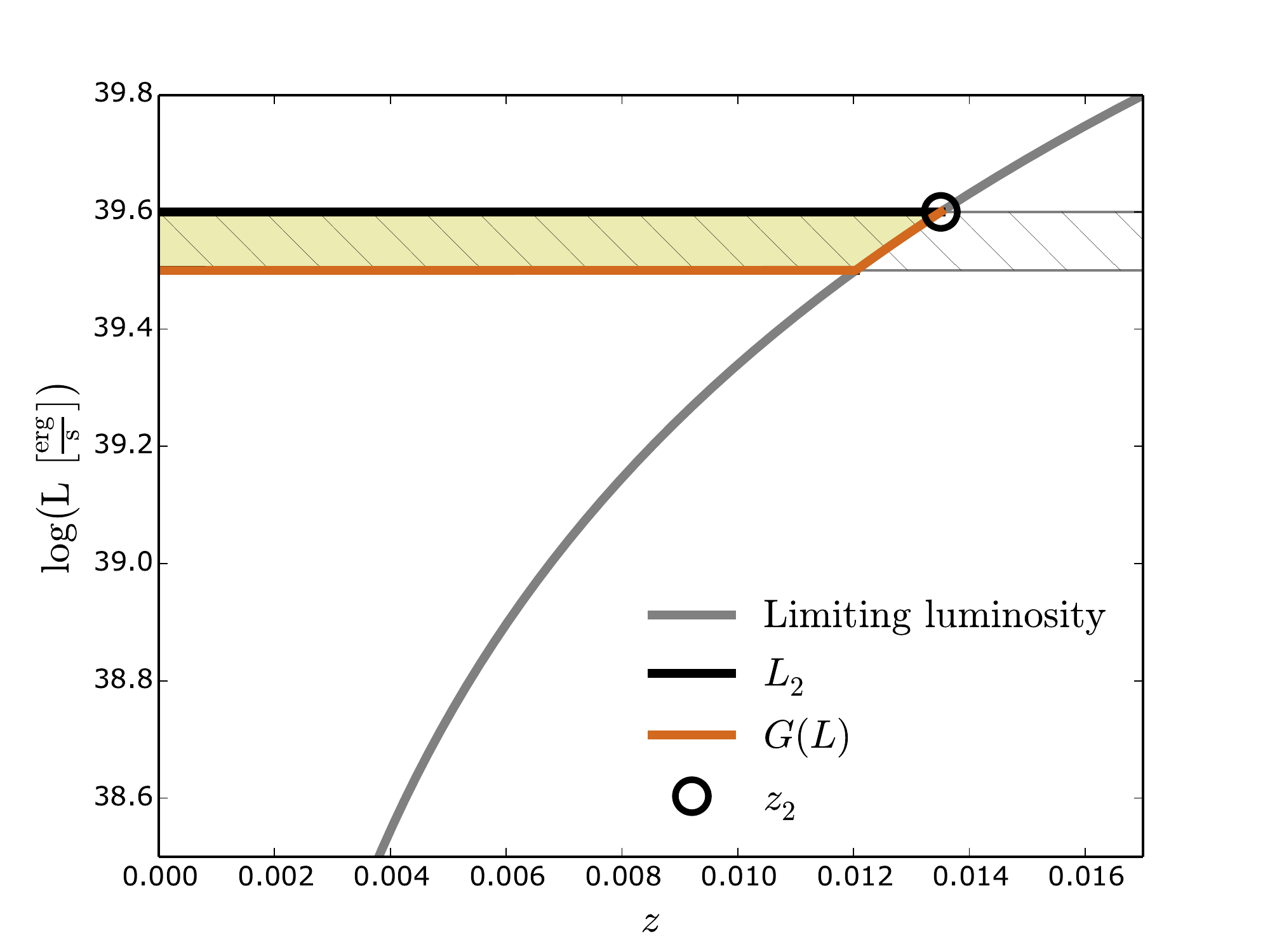} 
\caption{\textit{Solid curve}: luminosity associated to the limiting flux at each redshift. It represents the maximum luminosity that we are expected to see at each redshift. \textit{Yellow filled area}: volume traced by all the galaxies that have a luminosity with a given luminosity bin, enclosed between L1 and L2 (horizontal solid lines). We refer to this volume as $V_{\rm int}$ \textit{Diagonally filled area}: Total volume that would be traced without any limit in the minimum luminosity that we could observe. We refer to this as $V_{\rm max}$.}
\label{VintVmax}
\end{figure}

This correction is illustrated in Figure~\ref{VintVmax}. It provides the fraction of volume that we can trace given a limiting flux, at each redshift. Decreasing the limiting flux to fainter values (i.e., going deeper in flux) would shift the solid curve of limiting luminosity along the Y axis towards fainter luminosities.  Hence, a given luminosity bin would trace a larger volume.

\section{H$\alpha$ luminosity and stellar mass functions}\label{Sect:tablas}
In this Appendix, we provide the numerical values of the stellar mass function for the Blue and Red samples computed in Sect.~\ref{sect:SMF}, Table~\ref{tab:MF_values}, and the mass-weighted H$\alpha$ luminosity function computed form the Full Sample in Sect.~\ref{sect:HaLF}, Table~\ref{tab:FSMWHaLF}.

\begin{table}
\caption{Mass Function, divided into Blue and Red Sample, expressed in \cite{Salpeter55} IMF. Corresponds with values in Fig.~\ref{fig:Mass_function}}\label{tab:MF_values}
\centering
\begin{tabular}{@{\extracolsep{8pt}}ccccc@{}} 
\hline\hline\noalign{\smallskip}
 & \multicolumn{2}{c}{Blue Sample} & \multicolumn{2}{c}{Red Sample} \\
\noalign{\smallskip}\cline{2-3}\cline{4-5}\noalign{\smallskip}

$\log M_{\star}$ & $\log \mathrm{\Phi} $ & $\sigma$ & $\log \mathrm{\Phi} $ & $\sigma$ \\

$\left[ M_{\odot} \right]$ & $ \mathrm{\left[ dex^{-1}\,Mpc^{-3} \right]}$ & & $ \mathrm{\left[ dex^{-1}\,Mpc^{-3} \right]}$ & \\

\noalign{\smallskip}\hline\noalign{\smallskip}
$6.60$  & $-3.54$ & $0.31$  & $\ldots$& $\ldots$ \\
$6.80$  & $-3.54$ & $0.31$  & $\ldots$& $\ldots$ \\
$7.00$  & $-3.36$ & $0.25$  &$\ldots$ & $\ldots$ \\
$7.20$  & $-3.24$ & $0.22$  &$\ldots$ & $\ldots$ \\
$7.40$  & $-3.24$ & $0.22$  &$\ldots$ & $\ldots$ \\
$7.60$  & $-2.76$ & $0.13$  &$\ldots$ & $\ldots$ \\
$7.80$  & $-2.67$ & $0.11$  &$\ldots$ & $\ldots$ \\
$8.00$  & $-2.56$ & $0.10$   & $-3.54$ & $0.31$   \\
$8.20$  & $-2.23$ & $0.07$  & $-3.54$ & $0.31$   \\
$8.40$  & $-2.02$ & $0.05$  & $-3.54$ & $0.31$   \\
$8.60$  & $-1.91$ & $0.05$  & $-3.36$ & $0.25$   \\
$8.80$  & $-2.01$ & $0.05$  &$\ldots$ & $\ldots$ \\
$9.00$  & $-1.92$ & $0.05$  &$\ldots$ & $\ldots$ \\
$9.20$  & $-2.02$ & $0.05$  &$\ldots$ & $\ldots$ \\
$9.40$  & $-2.08$ & $0.06$  & $-3.36$ & $0.25$   \\
$9.60$  & $-2.14$ & $0.06$  & $-3.24$ & $0.22$   \\
$9.80$  & $-2.29$ & $0.07$  & $-3.14$ & $0.19$   \\
$10.0$ & $-2.50$  & $0.09$  & $-3.24$ & $0.22$   \\
$10.2$ & $-2.61$ & $0.11$  & $-2.84$ & $0.14$   \\
$10.4$ & $-2.64$ & $0.11$  & $-2.89$ & $0.14$   \\
$10.6$ & $-3.00$  & $0.16$  & $-2.94$ & $0.15$   \\
$10.8$ & $-3.00$  & $0.16$  & $-3.14$ & $0.19$   \\
$11.0$ & $\ldots$& $\ldots$& $-3.00$  & $0.16$   \\
$11.2$ & $\ldots$& $\ldots$& $-3.54$ & $0.31$   \\
$11.4$ & $\ldots$& $\ldots$& $-3.84$ & $0.43$   \\

\noalign{\smallskip}\hline
\end{tabular}
\end{table}

\begin{table}
\caption{Mass-weighted, Full Sample, H$\alpha$ luminosity function. These points correspond to the \textit{left panel} of Fig.~\ref{fig:HaLFs}}\label{tab:FSMWHaLF}
\centering
\begin{tabular}{cccccc} 
\hline\hline\noalign{\smallskip}
$\log\, L_{\mathrm{H}\alpha}$ & $\log \Phi$ &  \multicolumn{4}{c}{Percentiles} \\
 \cline{3-6}\noalign{\smallskip}
$\mathrm{\left[ erg\,s^{-1} \right]}$ & 
$\mathrm{\left[ Mpc^{-3}\,dex^{-1} \right]} $ & 
$\mathrm{16^{th}}$  & 
$\mathrm{84^{th}}$  & 
$\mathrm{2.5^{th}}$  & 
$\mathrm{97.5^{th}}$ 
\\
\noalign{\smallskip}\hline\noalign{\smallskip}
$39.56$ & $-1.68$ & $0.10$ & $0.09$ & $0.20$ & $0.15$ \\
$39.68$ & $-1.66$ & $0.08$ & $0.06$ & $0.14$ & $0.12$ \\
$39.80$ & $-1.66$ & $0.07$ & $0.07$ & $0.14$ & $0.12$ \\
$39.92$ & $-1.71$ & $0.07$ & $0.05$ & $0.13$ & $0.1$ \\
$40.04$ & $-1.76$ & $0.06$ & $0.06$ & $0.14$ & $0.12$ \\
$40.16$ & $-1.82$ & $0.07$ & $0.05$ & $0.14$ & $0.11$ \\
$40.28$ & $-1.87$ & $0.06$ & $0.06$ & $0.13$ & $0.12$ \\
$40.40$ & $-1.91$ & $0.06$ & $0.06$ & $0.13$ & $0.11$ \\
$40.52$ & $-1.94$ & $0.06$ & $0.06$ & $0.13$ & $0.12$ \\
$40.64$ & $-1.98$ & $0.06$ & $0.06$ & $0.13$ & $0.11$ \\
$40.76$ & $-2.05$ & $0.06$ & $0.06$ & $0.13$ & $0.11$ \\
$40.88$ & $-2.11$ & $0.07$ & $0.07$ & $0.16$ & $0.12$ \\
$41.00$ & $-2.20$ & $0.09$ & $0.08$ & $0.19$ & $0.14$ \\
$41.12$ & $-2.32$ & $0.07$ & $0.10$ & $0.19$ & $0.15$ \\
$41.24$ & $-2.42$ & $0.09$ & $0.10$ & $0.25$ & $0.16$ \\
$41.36$ & $-2.54$ & $0.12$ & $0.10$ & $0.30$ & $0.18$ \\
$41.48$ & $-2.72$ & $0.12$ & $0.14$ & $0.30$ & $0.21$ \\
$41.60$ & $-3.02$ & $0.30$ & $0.18$ & $0.60$ & $0.30$ \\
$41.72$ & $-3.32$ & $0.30$ & $0.18$ & $\ldots$ & $0.40$ \\
$41.84$ & $-3.62$ & $\ldots$ & $0.30$ & $\ldots$ & $0.48$ \\

\noalign{\smallskip}\hline
\end{tabular}
\end{table}

\end{appendix}

\end{document}